\documentclass[structabstract]{aa}
\usepackage{txfonts}
\usepackage{graphicx}
\usepackage{natbib}
\usepackage{array}
\bibpunct{(}{)}{;}{a}{}{,}

\begin{document}
\title{The precision of line position measurements of unresolved quasar absorption lines and its influence on the search for variations of fundamental constants}
\titlerunning{The precision of line position measurements of unresolved quasar absorption lines}
\author{N.\ Prause \and D.\ Reimers}
\institute{Hamburger Sternwarte, Gojenbergsweg 112, 21029 Hamburg, Germany}

\date{Received 1 November 2011/ Accepted 10 April 2013}

\abstract{}{Optical quasar spectra can be used to trace variations of the fine-structure constant $\alpha$. Controversial results that have been published in last years suggest that in addition to to wavelength calibration problems systematic errors might arise because of insufficient spectral resolution. The aim of this work is to estimate the impact of incorrect line decompositions in fitting procedures due to asymmetric line profiles. Methods are developed to distinguish between different sources of line position shifts and thus to minimize error sources in future work.}{To simulate asymmetric line profiles, two different methods were used. At first the profile was created as an unresolved blend of narrow lines and then, the profile was created using a macroscopic velocity field of the absorbing medium. The simulated spectra were analysed with standard methods to search for apparent shifts of line positions that would mimic a variation of fundamental constants. Differences between position shifts due to an incorrect line decomposition and a real variation of constants ẃere probed using methods that have been newly developed or adapted for this kind of analysis. The results were then applied to real data.}
{Apparent relative velocity shifts of several hundred meters per second are found in the analysis of simulated spectra with asymmetric line profiles. It was found that each system has to be analysed in detail to distinguish between different sources of line position shifts. A set of 16 \ion{Fe}{ii} systems in seven quasar spectra was analysed. With the methods developed, the mean $\alpha$ variation that appeared in these systems was reduced from the original $\Delta\alpha/\alpha=(2.1\pm2.0_\mathrm{stat})\cdot10^{-5}$ to $\Delta\alpha/\alpha=(0.1\pm0.8_\mathrm{stat})\cdot10^{-5}$. We thus conclude that incorrect line decompositions can be partly responsible for the conflicting results published so far.}{}


\maketitle

\section{Introduction}
The search for varying fundamental constants is a research field of ongoing interest in astronomy, laboratory experiments, and theory. As dimensionless constants, the electron to proton mass ratio $\mu=m_\mathrm{e}/m_\mathrm{p}$ and the fine-structure constant $\alpha=e^2/(\hbar c)$ are in the focus of astrophysical observations. High precision measurements have been made in laboratory experiments over the last years, giving an upper limit of \(\frac{\delta \ln\alpha}{\delta t}<10^{-17}\, yr^{-1}\) \citep{Karshenboim2008}. Although in astronomy the precision in estimating variations of fundamental constants is far lower, the time scales are typically higher by a factor of \(10^{10}\). Using high redshift quasar spectra, a look-back time of over 10 billion years can be observed. Assuming a linear variation with time, the methods are competitive in accuracy. However, there is no reason to believe that a change in fundamental constants would be linear in time, so astronomical observations trace a regime that cannot be tackled with laboratory experiments and are therefore a complementary research field. In the analysis of optical quasar spectra, many of the achievements of the last years have been, at least in part, conflicting. Results, ranging from $\Delta\alpha/\alpha=(-5.4\pm1.2)\cdot10^{-6}$ \citep{Murphy2003}, over \(\Delta\alpha/\alpha=(-0.4\pm3.3)\cdot10^{-6}\)\citep{Quast2004}, and up to $\Delta\alpha/\alpha=(5.4\pm2.5)\cdot10^{-6}$ \citep{Levshakov2007} have been reported in the literature. The reasons for these discrepancies are not yet fully understood. In addition to wavelength calibration difficulties \citep{Agafonova2011,Griest2010}, problems with methodology might be the cause. One of the problems is insufficient spectral resolution in present quasar spectra. It is known from very high resolution spectra (\(R=10^6\)) of galactic interstellar \ion{Na}{i} and \ion{Ca}{ii} absorption lines that the typical separation of subcomponents of interstellar lines is about \(1.2\,\mathrm{km\,s^{-1}}\) so that even at a resolution of \(0.5\,\mathrm{km\,s^{-1}}\), only \(\sim60\) \% of the individual subcomponents are detected \citep{Welty1994, Welty1996, Welty1998}. This means that even in the highest quality quasar spectra (\(R\approx80\,000 \sim 4\,\mathrm{km\,s^{-1}}\)), apparently single Doppler profiles may have many narrow, even saturated subcomponents that can be recognized only by line asymmetries. \citet{Murphy2001b} have simulated the impact of blends with single unidentified lines. Since they were mainly interested in effects that are statistically relevant for a high number of systems, they focussed on possible weak transitions that lie close to those used in their analysis. \citet{Chand2004} have probed the possibility of apparent position shifts due to unresolved line blends by simulating systems consisting of two closely blended components. They found that in these cases significant problems can arise for this kind of analysis and they restricted their work to systems with simple profiles.  \\
Small-scale velocity splittings become particularly important for quasar absorption systems formed in galactic discs. Even if more systems are formed in halos because of their larger cross sections, as argued by \citet{Murphy2003}, each individual absorption system has to be examined to detect possible sources of line position shifts that could mimic an \(\alpha\) variation. As long as lines of the same ion with similar transition strength \(f\lambda_0\) are compared (e.g. \ion{Fe}{ii} 1608 \AA\ with \ion{Fe}{ii} 2374 \AA), this has little impact on \(\alpha\) variation measurements (Sect.\ \ref{sect:simulations}). However, this was rarely the case in existing studies. To minimize systematic errors, a comparison of different ions formed possibly non-cospatially or of different transition strengths should be avoided (Sect.\ \ref{subsec:reg}).\\
In this work, possible apparent line position shifts that could be mimicked when using absorption lines with asymmetric profiles are discussed. While in previous works simulations have been done for simple blends of two components \citep{Murphy2001b,Chand2004}, in this work the line profiles are assumed to be more complex and are therefore a better representation of real data. Thus, simulated quasar spectra, including noise and the instrumental profile have been created to determine the influence of asymmetric line shapes on the results of the methods used to trace variations of the fine-structure constant $\alpha$ (Sect.\ \ref{sect:simulations}). The methods developed in the simulations are then applied to real data taken with UVES (Sect.\ \ref{sect:observations}). In Sect.\ \ref{sect:discussion} the results are discussed. Though this work concentrates on methods to detect possible variations of the fine-structure constant $\alpha$, most of the findings can also be used for related tasks.

\section{Simulations}
\label{sect:simulations}
Asymmetric line profiles can be formed by various mechanisms that in general, cannot be distinguished in real data. Usually they are treated as a simple blend of two or more Doppler profiles. If the real composition of the system is more complex, line position fits can be erroneous. The aim of this chapter is to show that this error source cannot be neglected when searching for varying fundamental constants and other related analyses where a very high precision in line positions is required. During fitting procedures, Doppler or Voigt profiles are usually used to simulate the line. In this work we use only Doppler profiles since the damping wings of the observed lines are negligible at low densities and/or low column densities. \\
Relativistic corrections to atomic fine structure transitions cause different sensitivities of the spectral line frequencies to a variation of the fine structure constant $\alpha$. This effect will result in a differential position shift of each transition i 
\begin{equation}
 z_i = z_0 + \kappa_{\alpha}Q_\mathrm{i},
\label{eq:reg}
\end{equation}
with the slope parameter 
\begin{equation}
 \kappa_{\alpha}=2(1+z_0)\frac{\Delta\alpha}{\alpha}
\label{eq:slope}
\end{equation}
and $Q_\mathrm{i}$ the dimensionless sensitivity coefficients \citep{Murphy2001, Levshakov2004}. When several transitions with different sensitivities are present, the slope parameter, and thus the \(\alpha\) variation, can be found with a regression analysis of Eq.\ \ref{eq:reg}. In many cases fitted line positions are not compatible with the regression model. It is thus important to probe which effects can cause shifts in line positions to identify the least affected transition.\\
The most important ion for this analysis is \ion{Fe}{ii} since it has a high sensitivity and is often found in quasar spectra. Sometimes \ion{Mg}{ii} lines are used as anchor lines. Table \ref{tab:values} shows the laboratory wavelength $\lambda_{0}$, oscillator strengths \(f\), and sensitivity coefficients \(Q\) used in this work. The factor \(f\lambda_0\) is a measure of the strength of each transition (see Eq.\ \ref{eq:opt_depth_2}).\\
The intensity $I_0$  of the light of a distant source is reduced by absorption of an intermediate gas with the optical depth $\tau$ by
\begin{equation}
\label{eq:intensity}
I=I_0\cdot\exp(-\tau).
\end{equation}
The optical depth is the integral of the opacity $\kappa$ times the number density $n$ over the spatial extension $s$ of the absorber,
\begin{equation}
\label{eq:opt_depth}
\tau_{\lambda}=\int\kappa n(s)\phi(\lambda,s)\rm{d}s,
\end{equation}
with $\kappa=\frac{\pi e^2}{m_\mathrm{e} c^2}\cdot f\lambda_0$, where $f$ and $\lambda_0$ are the oscillator strength and the laboratory wavelength of a specific transition, respectively, and $\phi$ is the profile function. When we consider thermal broadening as the dominant mechanism, a Doppler profile is used,
\begin{equation}
\label{eq:opt_depth_2}
\tau_{\lambda}=\frac{\sqrt{\pi} e^2}{m_\mathrm{e} c^2}f\lambda_0\cdot\int \frac{n(s)}{b(s)}\exp\left(-\left( c\frac{\lambda-\lambda_c}{b(s)\lambda_c}\right)^2\right)\mathrm{d}s,
\end{equation}
where $\lambda_\mathrm{c}=(1+z)(1+v(s)/c)\cdot\lambda_\mathrm{0}$ is the observed central wavelength of the line, $z$ the redshift, and \(v(s)\) the macroscopic velocity of the absorbing medium. The Doppler parameter $b$ is a measure of the line width, usually simplified as a combination of thermal broadening and turbulent velocity $b=\sqrt{b_\mathrm{th}^2+b_\mathrm{turb}^2}$, with $b_\mathrm{th}=\sqrt{\frac{2kT}{m}}$, where $m$ is the mass of the ion. Since the temperature, density distribution and turbulence of the absorbing system are not known, this integral cannot be solved analytically. Under the assumption of a constant temperature and turbulence, and no changes in the velocity field throughout the absorber, the optical depth can be written as
\begin{equation}
\label{eq:opt_depth_3}
\tau_{\lambda}=\frac{\sqrt{\pi} e^2}{m_\mathrm{e} c^2}\frac{f\lambda_0 N_\mathrm{c}}{b}\cdot\exp\left(-\left( c\frac{\Delta\lambda^2}{b^2}\right)\right),
\end{equation}
where $\Delta\lambda$ is the wavelength distance from the redshifted line centre $\Delta\lambda=\frac{\lambda-\lambda_\mathrm{c}}{\lambda_\mathrm{c}}$ and $N_\mathrm{c}$ is the column density, defined as the integral of the density over the length of the absorber $N_\mathrm{c}=\int n(s)\mathrm{d}s$. In the following the notation \(N:=\log\frac{N\mathrm{c}}{\mathrm{cm^{-2}}}\) will be used. This profile is generally accepted and will be used in Sect. \ref{subsec:blend}. In Sect. \ref{subsec:vel} the assumptions are abandoned to create more realistic line shapes.\\
The instrument measures the flux \(F_\lambda=\int I_\lambda\cos\vartheta\mathrm{d}\Omega\), which is the intensity integrated over the solid angle of the source. Since quasars are point sources, the behaviour of the flux and of the intensity are the same.\\
The flux is finally convoluted with the spectrograph point spread function \(P\),
\begin{equation}
\label{eq:flux}
F_\lambda=F_\mathrm{0}\cdot \exp(-\tau)\otimes P_\lambda,
\end{equation}
where $P$ is assumed to be a Gaussian with the width $\sigma_\lambda=\frac{\lambda}{2\sqrt{2ln2}R}$. The resolving power $R=\lambda/\Delta\lambda$ is defined as the smallest distance $\Delta\lambda$ at which two features can be separated.
A Poisson noise is added to the resulting spectrum. No additional white noise was included for it would not significantly affect the results. If not stated otherwise, the simulations are done with a high quality, that can be achieved in future observations for optical quasar spectra, namely a signal to noise ratio $S/N=150$ and a resolving power of \(R=60.000\) \((\mathrm{FWHM\sim4.6\,km\,s^{-1}})\). The pixel size is 0.0147\AA\ \((\sim2.2\,\mathrm{km\,s^{-1}-0.7\,km\,s^{-1}}\) at 2000 \AA\ - 6000 \AA). \\
To fit the simulated spectra as well as the real data, a minimization algorithm based on an evolution strategy developed by \citet{Quast2005} was used. For details we refer to \citet{Quast2005}. Absorption lines usually consist of several components. For the \(\alpha\) variation measurements we used relative positions of whole lines, rather than directly comparing positions of single components, i.e. in the fitting procedure the Doppler parameters \(b\), column densities \(N\), and distance between the components were assumed to be the same for all transitions, while the position of each system was individual. The positions of FeII transitions in the analysed system are calculated in velocity or redshift scale. Frequently the fine-structure constant \(\alpha\) is used as an additional fitting parameter. This approach assumes that any position shift between different transitions is necessarily created by a varying \(\alpha\). Since it is the aim of this paper to show that position shifts can have other reasons, this approach is not used.

\subsection{Narrow line blends}
\label{subsec:blend}

\begin{table}[ht]
  \caption{Laboratory wavelength $\lambda_{0}$, oscillator strength $f$, transition strength \(f\lambda_0\), and sensitivity coefficients $Q$ for \ion{Fe}{ii} and \ion{Mg}{ii} transitions.}
 \begin{tabular}{lllrr}
  Transition & $\lambda_{0}$ [\AA]& $f$\tablefootmark{3} & \(f\lambda_0\)  [\AA]& $Q$\tablefootmark{4} \\
  \hline
  \ion{Fe}{ii} 1608 & $1608.45081$\tablefootmark{1} & 0.0577 & 92.8 & -0.019  \\
  \ion{Fe}{ii} 2344 & $2344.2128$\tablefootmark{2} & 0.114 & 267.2 & 0.032 \\
  \ion{Fe}{ii} 2374 & $2374.4601$\tablefootmark{2} & 0.0313 & 74.3 & 0.039  \\
  \ion{Fe}{ii} 2382 & $2382.7641$\tablefootmark{2} & 0.320 & 762.5 & 0.036 \\
  \ion{Fe}{ii} 2586 & $2586.6494$\tablefootmark{2} & 0.0691 & 178.7 & 0.039 \\
  \ion{Fe}{ii} 2600 & $2600.1722$\tablefootmark{2} & 0.239 & 621.4 & 0.036 \\
 \end{tabular}
\tablebib{
(1)~\citet{Nave2011};
(2)~\citet{Aldenius2009};
(3)~\citet{Morton2003};
(4)~\citet{Berengut2010}.
}
\label{tab:values}
\end{table}

\subsubsection{Single ion}
\label{subsubsec:blend_sidam}

\begin{figure}
 \includegraphics[width=\columnwidth, bb=45 160 506 648]{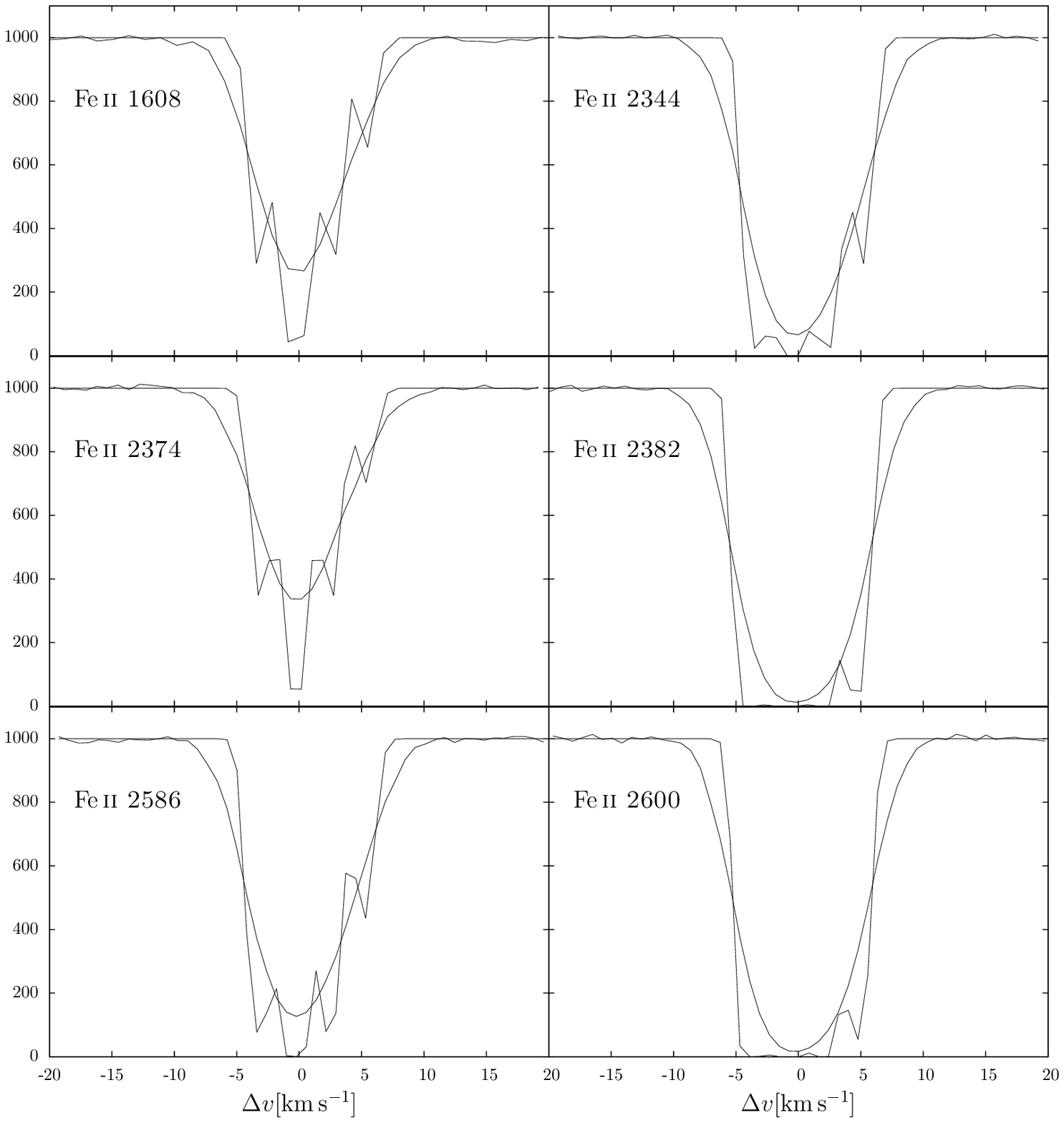}
 \caption{Simulated spectra of the first set-up. The original spectrum, prior to convolution with the instrument profile, is over-plotted by the final spectrum.}
\label{fig:spec_res_bl_130}
\end{figure}
\begin{figure}
 \includegraphics[width=\columnwidth, bb=45 160 506 648]{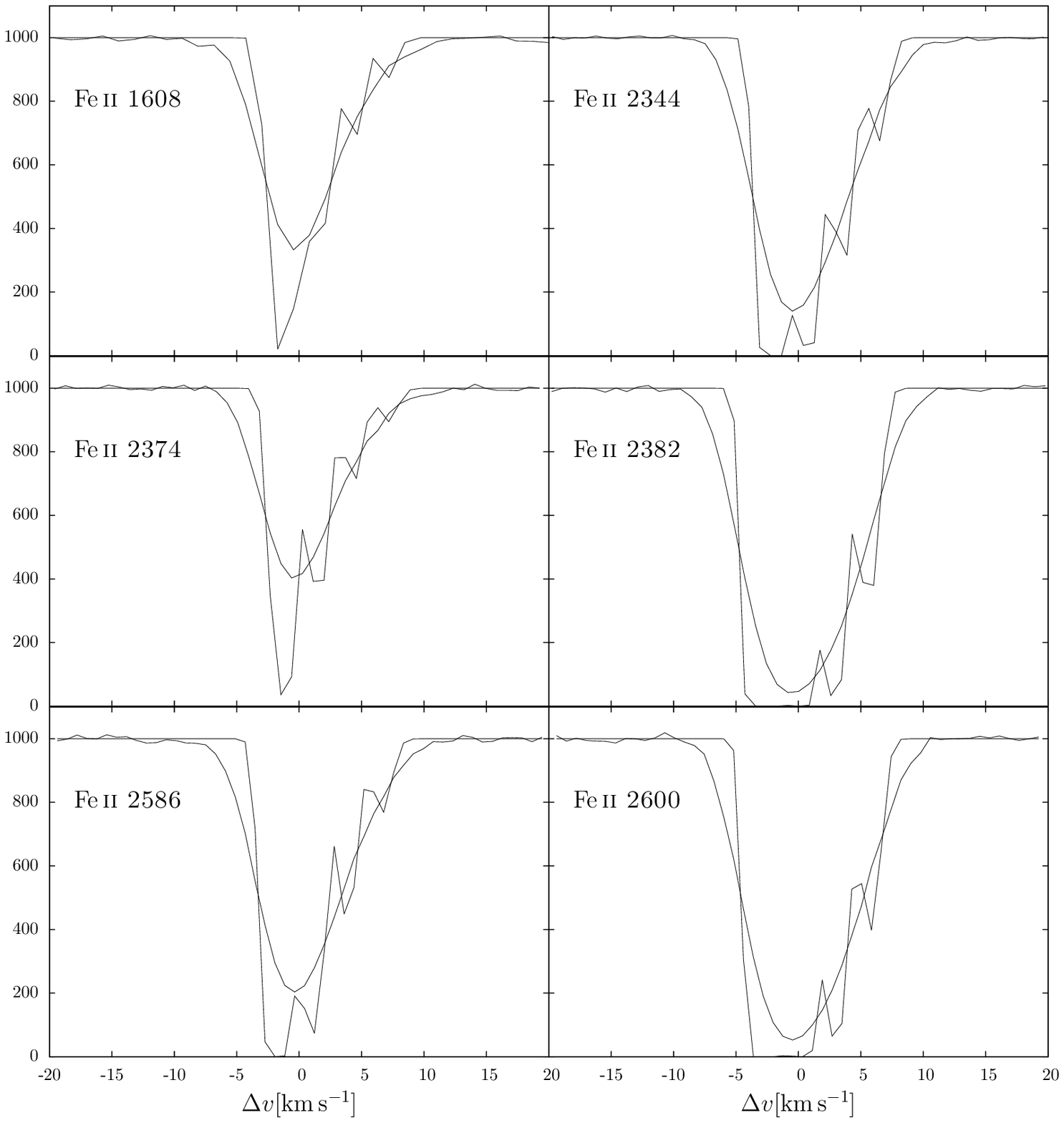}
 \caption{Simulated spectra of the second set-up. The original spectrum, prior to convolution with the instrument profile, is over-plotted by the final spectrum.}
\label{fig:spec_res_bl_135}
\end{figure}

 \begin{figure}
  \includegraphics[width=\columnwidth, bb=40 170 518 664]{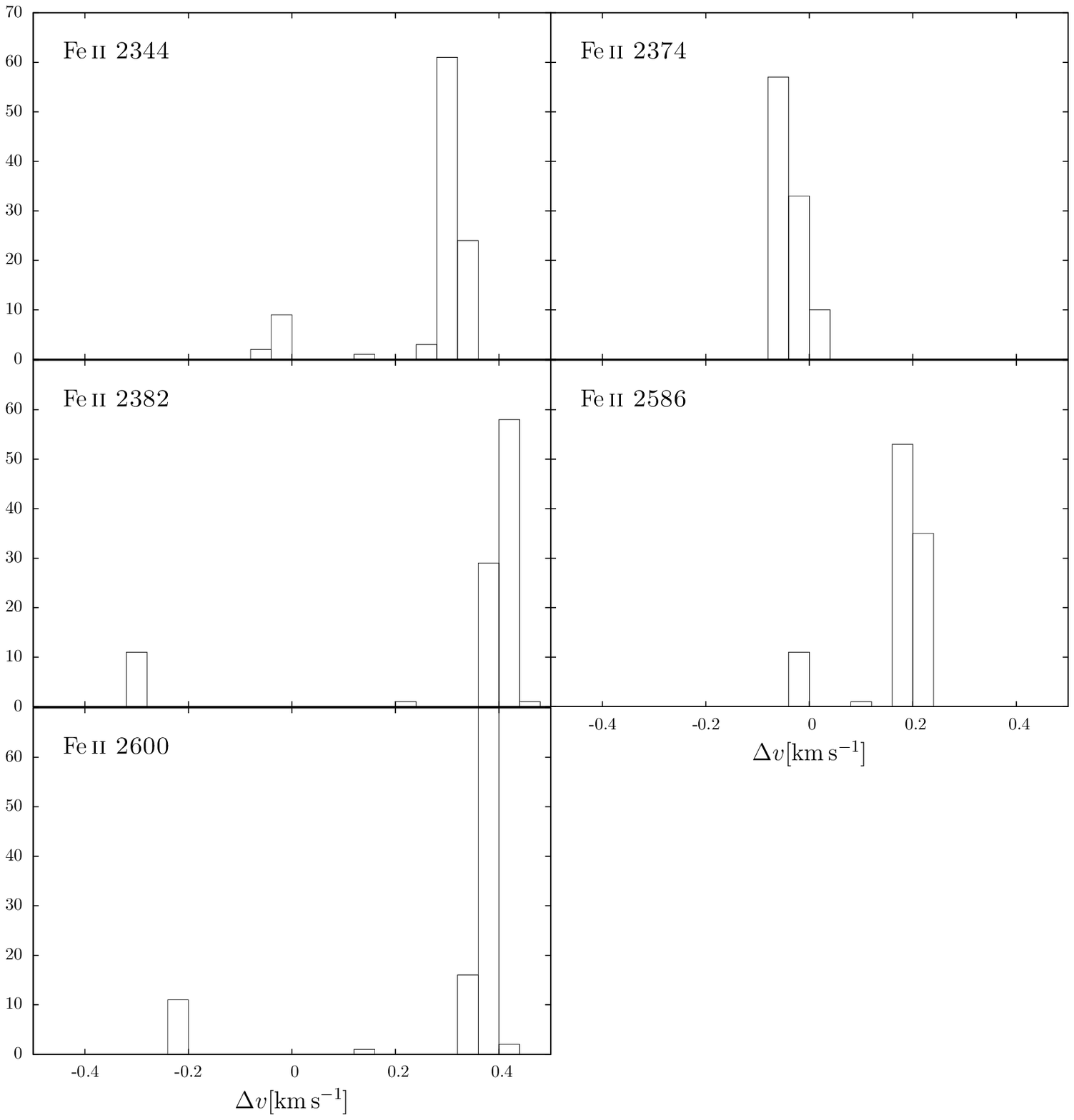}
  \caption{Histograms of the apparent velocity shifts relative to \ion{Fe}{ii} 1608 of close line blends using the first set-up. Two-component fit for 100 realizations with random noise.}
 \label{fig:hist_res_bl_130_2}
 \end{figure}
 
 \begin{figure}
  \includegraphics[width=\columnwidth, bb=40 170 518 664]{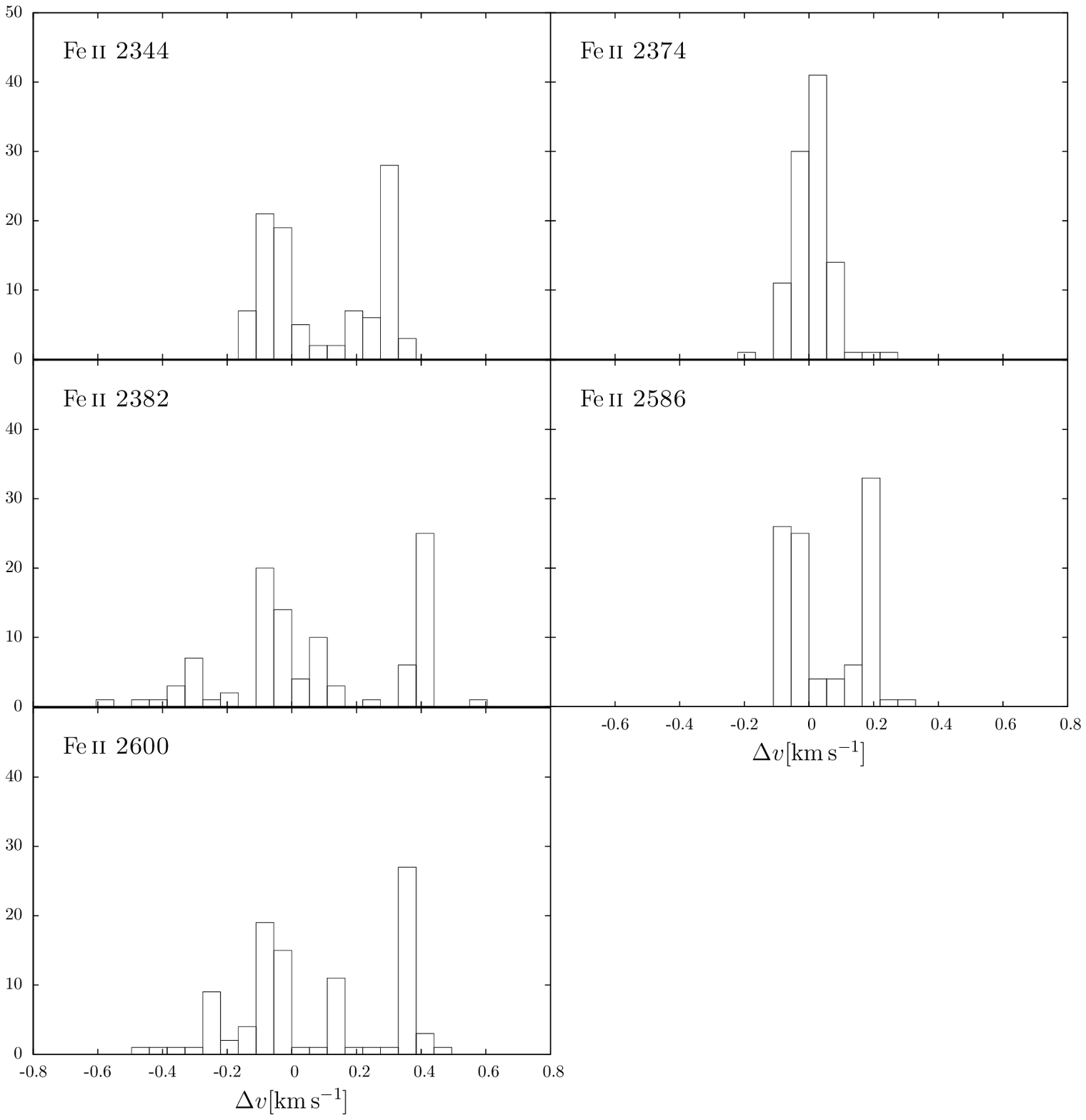}
  \caption{As in Fig.\ \ref{fig:hist_res_bl_130_2}. Histograms of close line blends using the first set-up. Three-component fit for 100 realizations with random noise.}
 \label{fig:hist_res_bl_130_3}
 \end{figure}
 
 \begin{figure}
  \includegraphics[width=\columnwidth, bb=40 170 518 664]{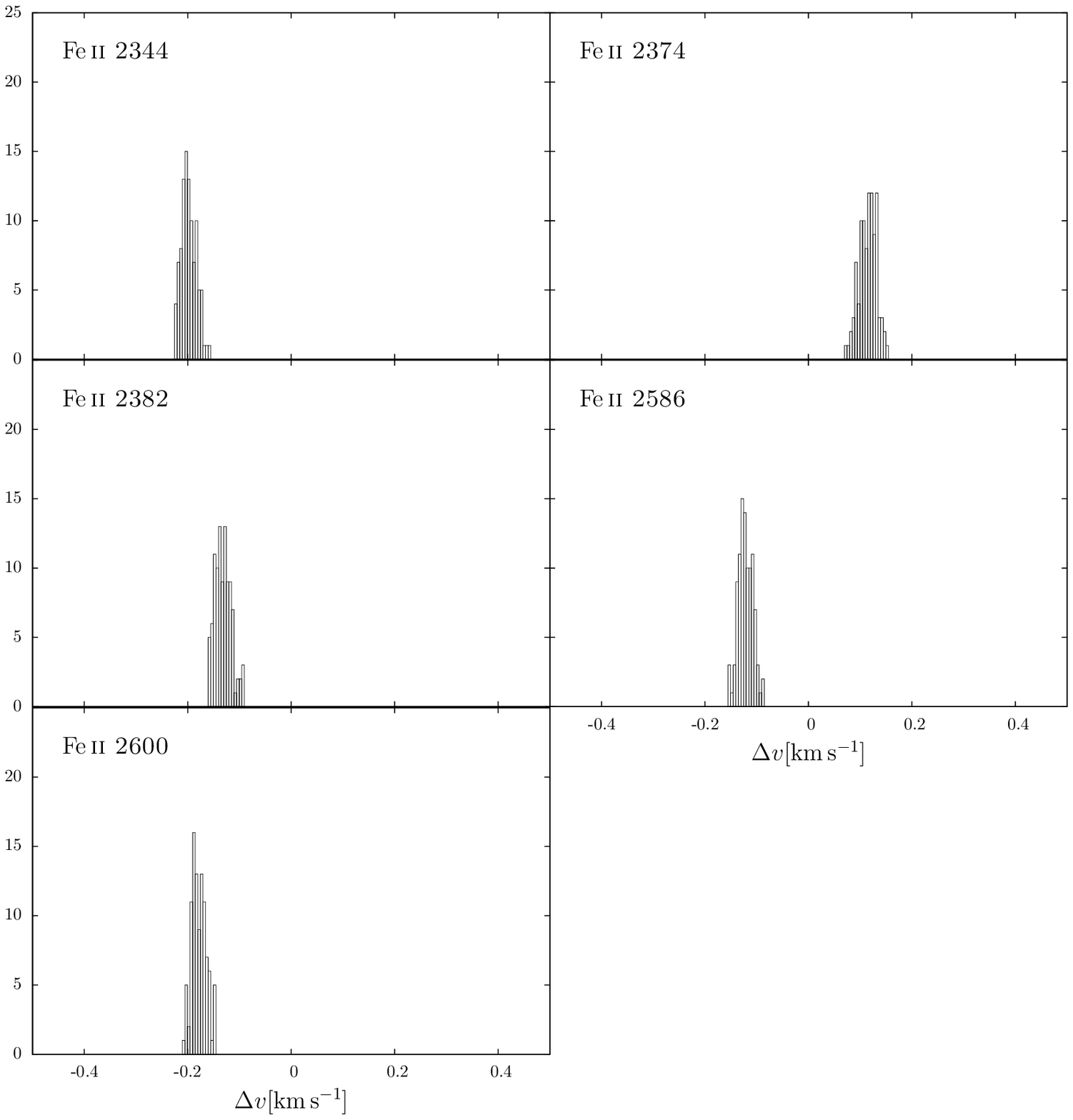}
  \caption{As in Fig.\ \ref{fig:hist_res_bl_130_2}. Histograms of close line blends using the second set-up. Two component fit for 100 realizations with random noise.}
 \label{fig:hist_res_bl_135_2}
 \end{figure}
 
 \begin{figure}
  \includegraphics[width=\columnwidth, bb=40 170 518 664]{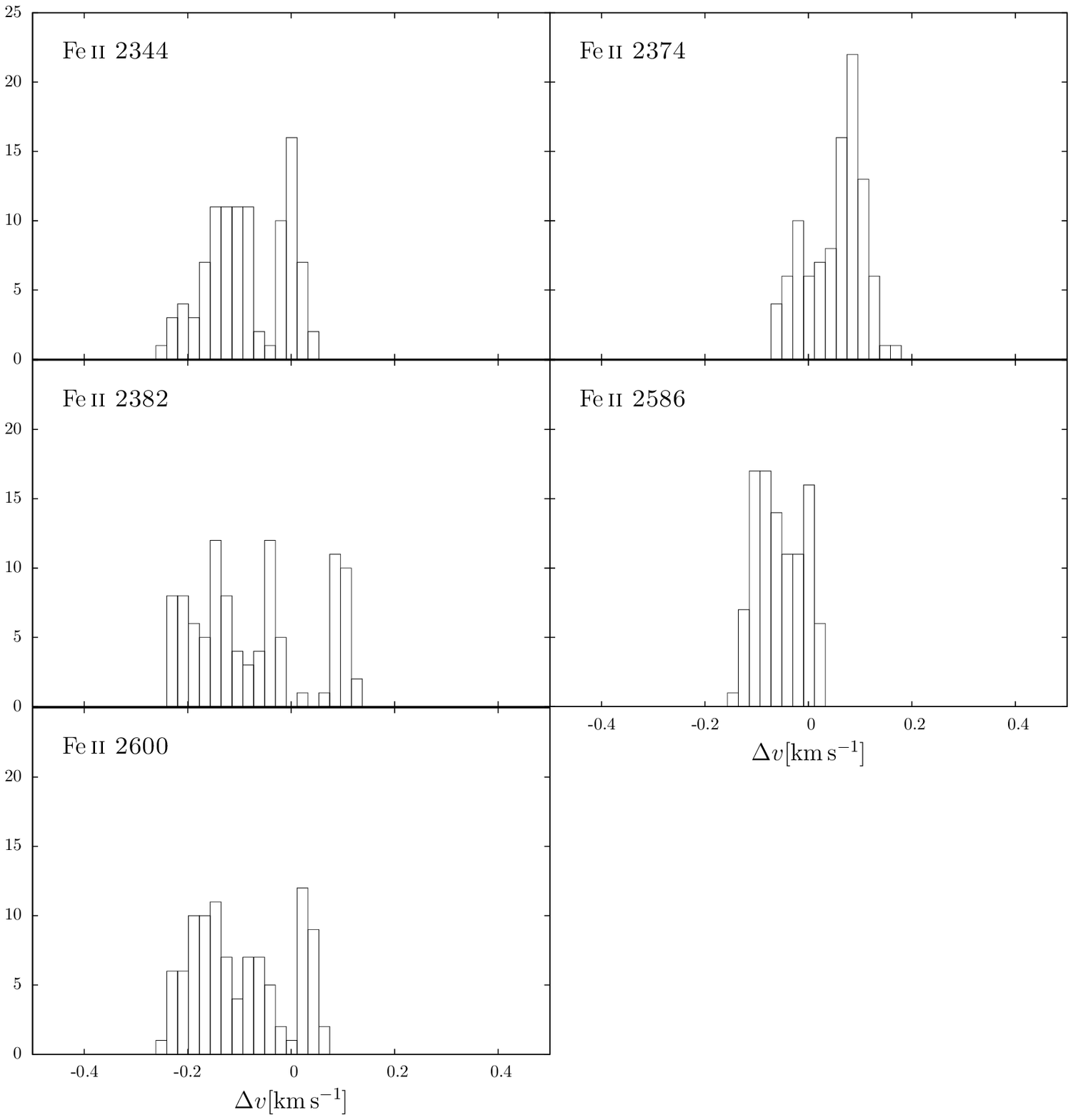}
  \caption{As in Fig.\ \ref{fig:hist_res_bl_130_2}. Histograms of close line blends using the second set-up. Three component fit for 100 realizations with random noise.}
 \label{fig:hist_res_bl_135_3}
 \end{figure}

The \ion{Fe}{ii} 1608 \AA\ transition is opposite in sensitivity to the other Fe ii transitions; therefore, it is effective to search for a varying fine-structure constant $\alpha$ by comparing only other \ion{Fe}{ii} lines with \ion{Fe}{ii} 1608 \AA\ as an anchor line (Single Ion Differential $\alpha$ measurement (SIDAM) \citep{Levshakov2005}). This has the advantage that the parameters that define the shape of the lines are the same for all lines used. This eliminates all systematic effects that can occur because of ionization substructure of the absorbing medium, when using different ions.    \\
While the SIDAM method has the disadvantage that there are basically just two different sensitivities available, \(q\sim-0.02\) for the \ion{Fe}{ii} 1608 \AA\ transition and \(q\sim0.03-0.04\) for the other \ion{Fe}{ii} transitions, the observed wavelengths of transitions with comparable sensitivities \(Q\) provide a test on the accuracy of the wavelength calibration. When the positions of transitions with similar sensitivities are not coherent, the reasons for the discrepancies have to be probed. In our simulations the only source for line position shifts are assumed to be unresolved components. These are negligible when comparing transitions with similar strength, e.g. the 1608 \AA\ and the 2374 \AA\ transition. In this case the parameter \(\kappa_{\alpha}\)  (Eq.\ \ref{eq:slope}) is simply the slope of a line through two points. In the following analysis \(\Delta\alpha/\alpha\) values are calculated using the regression analysis with all available \ion{Fe}{ii} transitions to probe the order of magnitude of this potential error source.\\
At a gas temperature of $100\,\mathrm{K} < T_\mathrm{kin} < 10^4\,\mathrm{K}$, as is expected in interstellar clouds, the thermal width of \ion{Fe}{ii} absorption features is less than \(2\,\mathrm{km\,s^{-1}}\). In the Galactic (or in the Milky Way) interstellar clouds b-parameters as low as \(b\approx0.5\,\mathrm{km\,s^{-1}}\) have been observed in \ion{Ca}{ii} and \ion{Na}{i} \citep{Welty1998}. Since many observed systems are broader, these are either broadened by turbulence, formed in galactic halos or are blends of narrow lines. \\
Two examples for simulated narrow line blends are shown in Figs.\ \ref{fig:spec_res_bl_130} and \ref{fig:spec_res_bl_135}. The first feature is composed of four components with the column densities $N_1=13.0\, N_2=13.5\, N_3=13.0\, N_4=12.5$ and the second with $N_1=13.5\, N_2=13.0\, N_3=12.5\, N_4=12.0$, respectively. The Doppler parameters for each component are $b=1\,\mathrm{km\,s^{-1}}$, the redshift of the first component \(z=1.15\), and the redshift separation between the components $\Delta z=2\cdot10^{-5}$ \((\sim2.8\,\mathrm{km\,s^{-1}})\). Each of the relevant six \ion{Fe}{ii} transitions is shown. For the strong transitions it can be seen that though the original lines are highly saturated, the resulting feature no longer shows signs of saturation. The distortion of the line shape by this effect varies with the strength of the transition and the composition of the original spectrum. \\
The resulting profiles are fitted as a sum of Doppler profiles with an increasing number of components. The Doppler parameters, column densities and separations between individual components of all \ion{Fe}{ii} transitions were fitted simultaneously, while the integral position of each transition was fitted individually. This procedure yields relative positions of all \ion{Fe}{ii} lines on a velocity or redshift scale. \\
Figures \ref{fig:hist_res_bl_130_2} and \ref{fig:hist_res_bl_130_3} show histograms of the velocity shift between the corresponding \ion{Fe}{ii} transitions to the 1608 \AA\ transition for the first set-up with two and three fitted components, respectively. The results for the second set-up is shown in Figs.\ \ref{fig:hist_res_bl_135_2} and \ref{fig:hist_res_bl_135_3}. Theses were created by fitting the simulated spectra 100 times with different random noise. \\
It is not trivial to determine the optimum number of fitted components. While an increase of the number of components decreases the $\chi^2$ value until a certain number is reached, the scatter of the results increases. 
When the shape of the feature is reasonably well approximated by a certain number of components, noise effects are primarily responsible for the location of further components. In these simulations we have chosen a two-component fit as best solution, though the three-component fits give smaller velocity shifts and have a lower $\chi^2$ value. The scatter of the fitting results is at a minimum for the two-component fit, allowing the best predictability for real data fits. Table \ref{tab:close_blend} shows the mean apparent \(\alpha\) variation for both set-ups and an increasing number of components. The error represents the spread of the results. Averaged \(\chi^2\) values are given for each number of components. Fitting more than two components increases the spread of the results in both cases, compare Fig.\ \ref{fig:hist_res_bl_135_2} and Fig.\ \ref{fig:hist_res_bl_135_3}. This shows the main danger when using too many components. Apparently the result becomes less predictable, because the position of the third component is mainly governed by noise. In many cases the fitting code could not clearly place a third component. These cases naturally have a higher statistical error in the total position fit since the location of all components are correlated. \\
\begin{table}[ht]
  \caption{Simulation results of narrow line blends.}
 \begin{tabular}{c|cc|cc}
   & Set-up 1  &   & Set-up 2 &  \\
  \hline
  \(\#_\mathrm{comp}\) & \(\Delta\alpha/\alpha [10^{-6}]\) & \(\chi^2\) &  \(\Delta\alpha/\alpha [10^{-6}]\) & \(\chi^2\) \\
  \hline
  1 &  \(-18.1\pm2.6\) & 1.9 &  \(-44.5\pm4.5\) & 4.5 \\
  2 &  \(  6.0\pm1.0\) / \(- 3.7\pm0.7\)  & 1.3 &  \(- 1.8\pm0.7\) & 1.2 \\
  3 &  \(  1.8\pm4.4\) & 1.2 &  \(- 0.8\pm1.3\) & 1.1  \\
  4 &  \(  3.0\pm4.6\) & 1.3 &  \(- 1.1\pm1.6\) & 1.1 \\
  5 &  \(  1.6\pm4.1\) & 1.3 &  \(- 2.3\pm2.2\) & 1.1  \\

 \end{tabular}
 \tablefoot{The \(\Delta\alpha/\alpha\) values are determined for two different simulation set-ups, fitted with up to five components each. They are averaged over 100 fits, each with random noise. The error represents the spread of the results. The second column for each set-up shows the averaged \(\chi^2\) values of the line profile fits.}
\label{tab:close_blend}
\end{table}

The two-component fit of the first set-up gave two separated solutions (see Fig.\ \ref{fig:hist_res_bl_130_2}). It can be seen in the two-component fit that systematic shifts of up to \(\Delta v \approx400\,\mathrm{m\,s^{-1}}\), depending on the transition strength, can occur. The effect can go in either direction even when the original line composition is the same, depending on the resulting best-fit parameters. 
In the first set-up, in 96\% of the cases a composition of $N_1=13.5, b_1=1.6\,\mathrm{km\,s^{-1}}, N_2=13.3, b_2=3.2\,\mathrm{km\,s^{-1}}$ is favoured by the $\chi^2$ analysis, while in the other cases $N_1=13.6, b_1=2.6\,\mathrm{km\,s^{-1}}, N_2=12.7, b_2=0.7\,\mathrm{km\,s^{-1}}$ has the lowest $\chi^2$. The second set-up shows a homogeneous shift to the other direction though the shape of the line is asymmetric in the same direction. The best fit gives a composition of \(N_1=13.5, b_1=1.7\,\mathrm{km\,s^{-1}}, N_2=12.8, b_2=3.2\,\mathrm{km\,s^{-1}}\). An $\alpha$ variation of $\Delta\alpha/\alpha = (6.0 \pm 1.0)\cdot10^{-6} / (-3.7\pm0.7)\cdot10^{-6}$ for the first set-up and $\Delta\alpha/\alpha = (-1.8 \pm 0.7)\cdot10^{-6}$ for the second set-up respectively, is mimicked. \\
The statistical error is quite low because of the assumed high data quality. The systematic error introduced by this effect is up to four times higher. The nature of the problem involved is the incorrect deconvolution of the original spectrum. The narrow lines of the simulated systems are often affected by unresolved saturation while the fitted, broader lines are not. The degree of saturation depends on the transition strength. Thus the effect decreases when two transitions with the same strengths are compared. For narrow lines the strong transitions will, in most cases, be saturated when the weak 1608 \AA\ transition is just strong enough to be seen.

\subsection{Velocity fields}
\label{subsec:vel}
\begin{figure}
\resizebox{0.5\columnwidth}{!}{\input{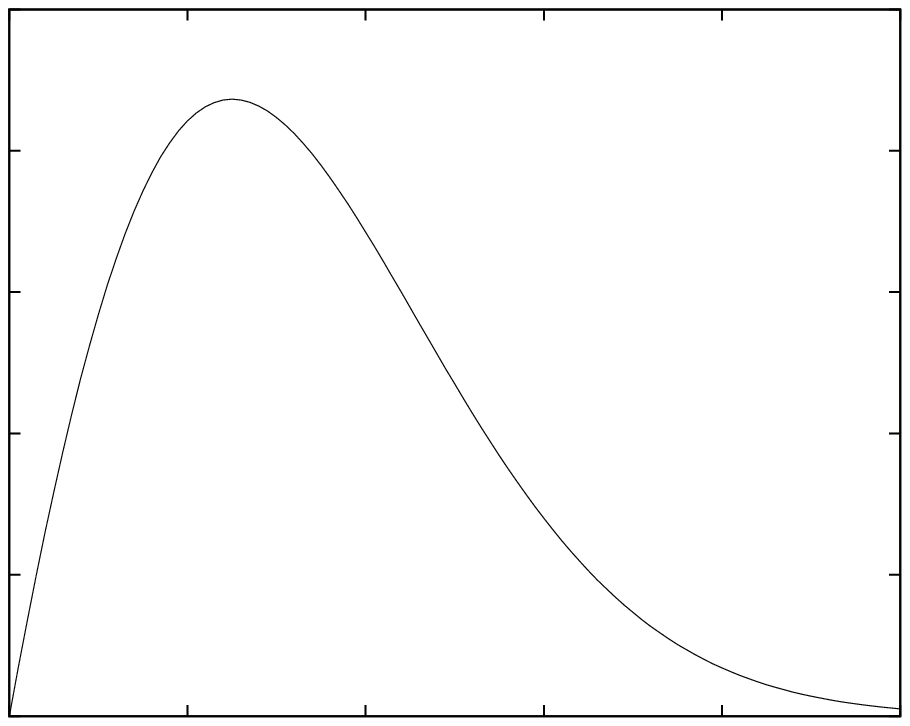}}
\hspace{-0.1cm}\resizebox{0.5\columnwidth}{!}{\input{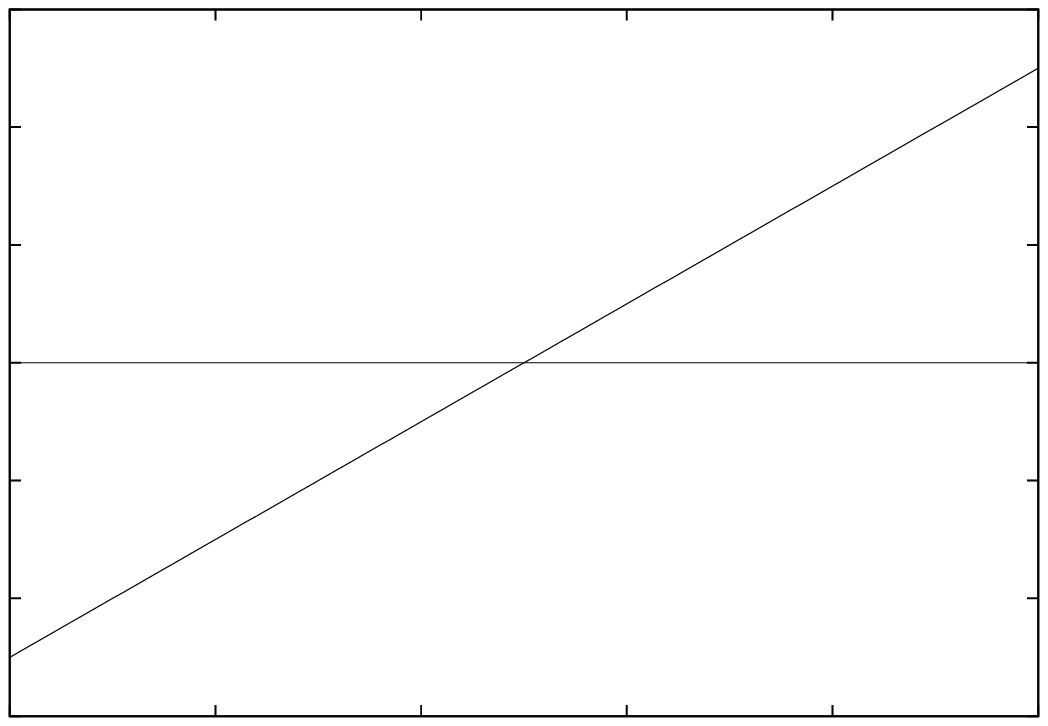}} \\
\caption{Number density (a) and velocity field (b) of absorbing medium used in the simulation of asymmetric line profiles, parametrized along the line of sight s with \(N_1=13\) and \(v_\mathrm{p}=10\mathrm{km\,s^{-1}}\). \(n(s)\) is the density distribution in \(\mathrm{cm^{-3}}\) and \(v(s)\) the velocity field in \(\mathrm{km\,s^{-1}}\).}
\label{fig:vel}
\end{figure}

\begin{figure}
\includegraphics[width=\columnwidth,bb= 45 160 506 646]{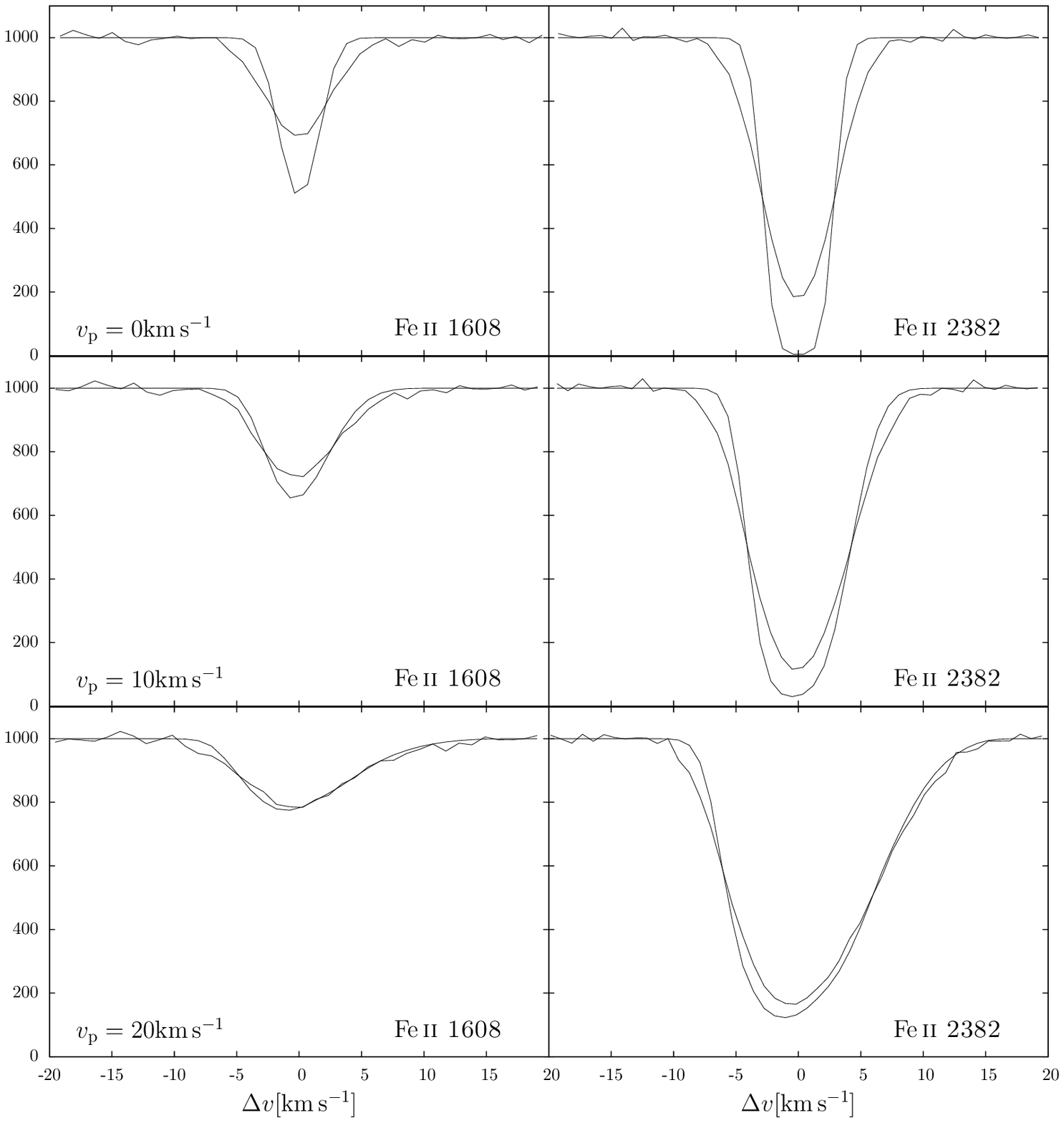}
\caption{Simulated spectra of gas with an underlying velocity field according to Fig. \ref{fig:vel} with \(N_1=13.0\). The peak velocities are \(v_\mathrm{p}=0\mathrm{km\,s^{-1}}\),  \(v_\mathrm{p}=10\mathrm{km\,s^{-1}}\) and \(v_\mathrm{p}=20\mathrm{km\,s^{-1}}\). The different curves show the flux before and after convolution with the instrument profile.}
\label{fig:spec_vel}
\end{figure}

\begin{figure}
\includegraphics[width=\columnwidth,bb= 45 160 506 646]{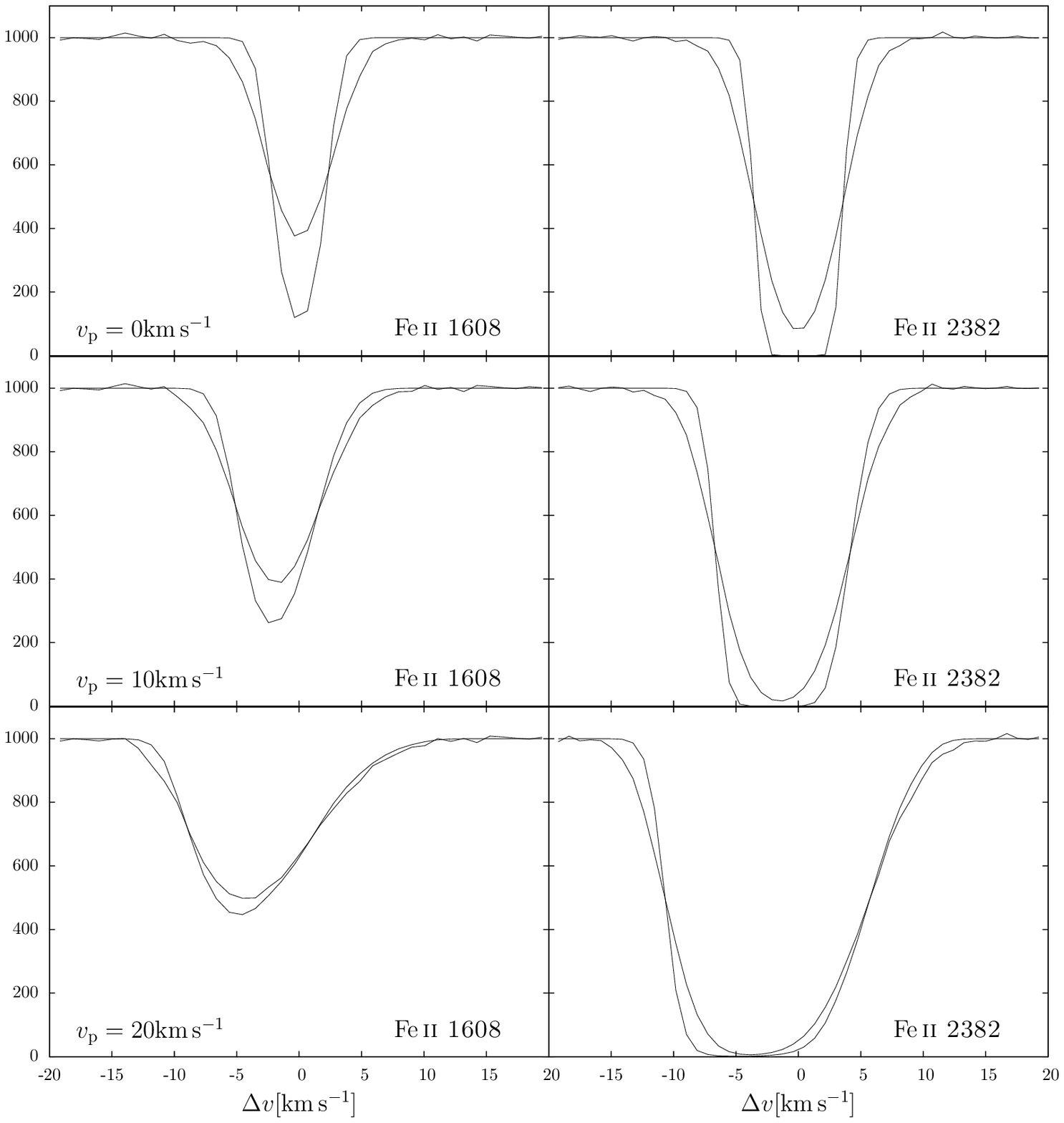}
\caption{Simulated spectra of gas with an underlying velocity field according to Fig. \ref{fig:vel} with \(N_2=13.5\). The peak velocities are \(v_\mathrm{p}=0\mathrm{km\,s^{-1}}\), \(v_\mathrm{p}=10\mathrm{km\,s^{-1}}\) and \(v_\mathrm{p}=20\mathrm{km\,s^{-1}}\). The different curves show the flux before and after convolution with the instrument profile. Saturated version.}
\label{fig:spec_vel_sat}
\end{figure}

When we abandon the assumption of a constant velocity of the absorbing medium, Eq.\ \ref{eq:opt_depth_2} has to be calculated numerically for a given density distribution $n(s)$ and velocity field $v(s)$. This is a simplified model which excludes possible mesoturbulence \citep{Levshakov1996}. It is, however, the simplest realistic model that produces asymmetric line profiles. Nevertheless, it would be impractical to use it in a fitting procedure since there are too many parameters, which would result in ambiguous solutions.  \\
By exploring several possibilities, it can be shown that a wide variety of line shapes can be produced with realistic parameters. As an example a continuous density distribution and velocity field are used. The size of the absorber is parametrized along the line of sight \(s\). A thermal broadening of \(b=2 \mathrm{km\,s^{-1}}\) is used to approximate the usual \ion{Fe}{ii} line width found in quasar spectra. The density distribution is adjusted which results in a column density of \(N_\mathrm{1}=\frac{\log}{\mathrm{cm^{-2}}}\int n_\mathrm{1}(s)ds=13.0\) or \(N_\mathrm{2}=13.5\). In the first case the lines are not saturated, in the second case the strong transitions are saturated. The mean gas velocity is $v_m=0\,\mathrm{km\,s^{-1}}$. Artificial spectra are created, differing in column density and peak velocity \(v_\mathrm{p}\), meaning that this is the highest velocity difference in the system. Figure\ \ref{fig:vel} shows the density distribution and velocity field for the parameters \(N_\mathrm{1}=13.0\) and \(v_\mathrm{p}=10\mathrm{km\,s^{-1}}\). Since the size of the absorption system has no direct influence on the shape of the absorption lines, it is parametrized from 0 to 1. We note that the high density values given in Fig.\ \ref{fig:vel} are a consequence of the parametrization of the sightline. A physically small absorber with a high number density gives the same absorption profile as an extended system with a low density. Figures \ref{fig:spec_vel} and \ref{fig:spec_vel_sat} show the resulting spectra before and after convolution with the instrument profile for \(N_1=13.0\) and \(N_2=13.5\), respectively. The peak velocities are \(v_\mathrm{p}=0\mathrm{km\,s^{-1}}\), \(v_\mathrm{p}=10\mathrm{km\,s^{-1}}\), and \(v_\mathrm{p}=20\mathrm{km\,s^{-1}}\). Naturally broader and therefore more asymmetric profiles are less influenced by the instrument profile, and the problem of unresolved saturation decreases. Small asymmetries, which are not visible by eye, are more prone to errors.  \\

\begin{figure}
\includegraphics[width=\columnwidth, bb=40 170 518 664]{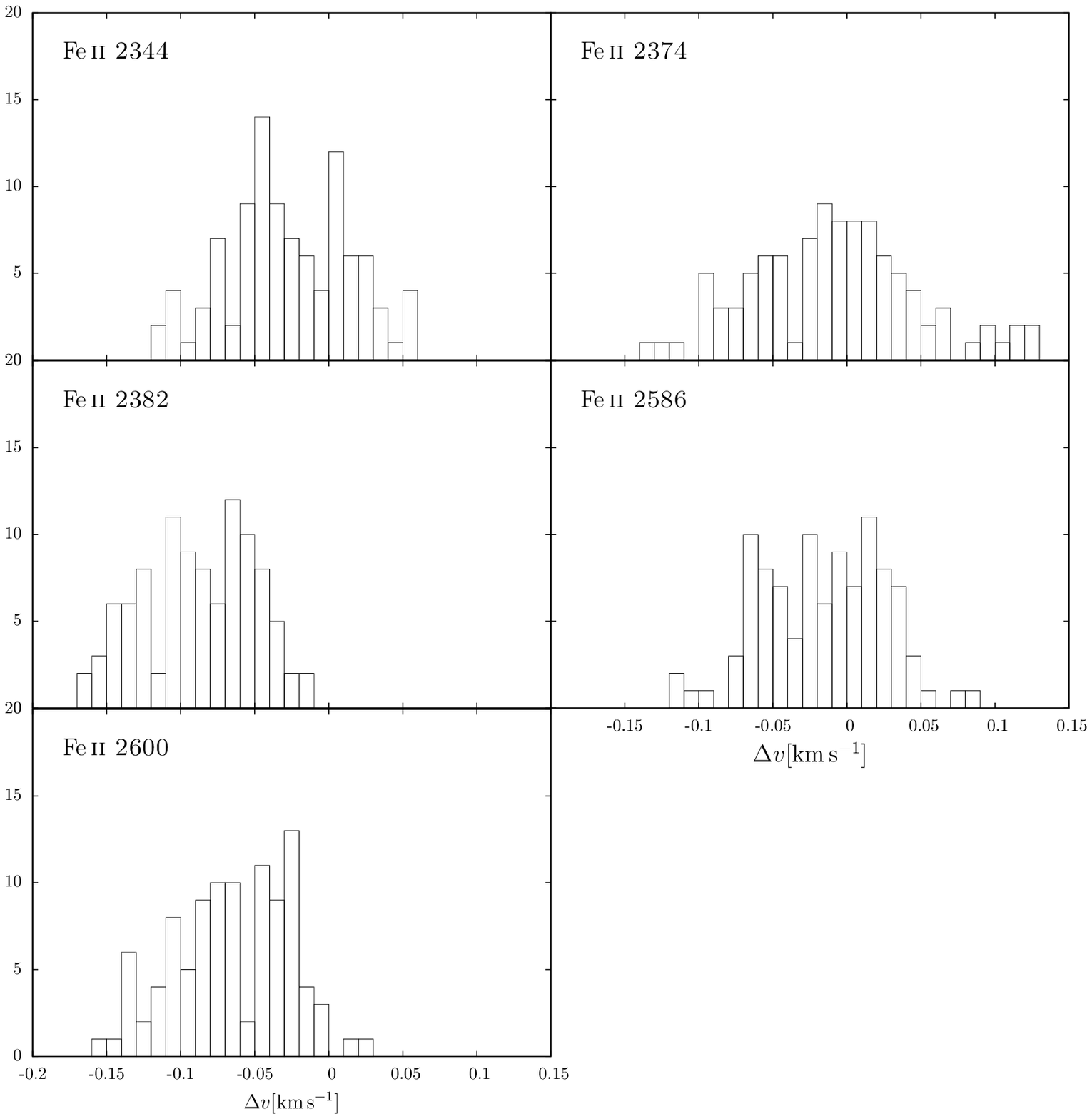}
\caption{Histograms of apparent velocity shifts relative to \ion{Fe}{ii} 1608 of simulated lines with an underlying velocity field with peak velocity \(v_\mathrm{p}=10\mathrm{km\,s^{-1}}\). One-component fit of 100 realizations with random noise.}
\label{fig:hist_vel_field_10_1}
\end{figure}

\begin{figure}
\includegraphics[width=\columnwidth, bb=40 170 518 664]{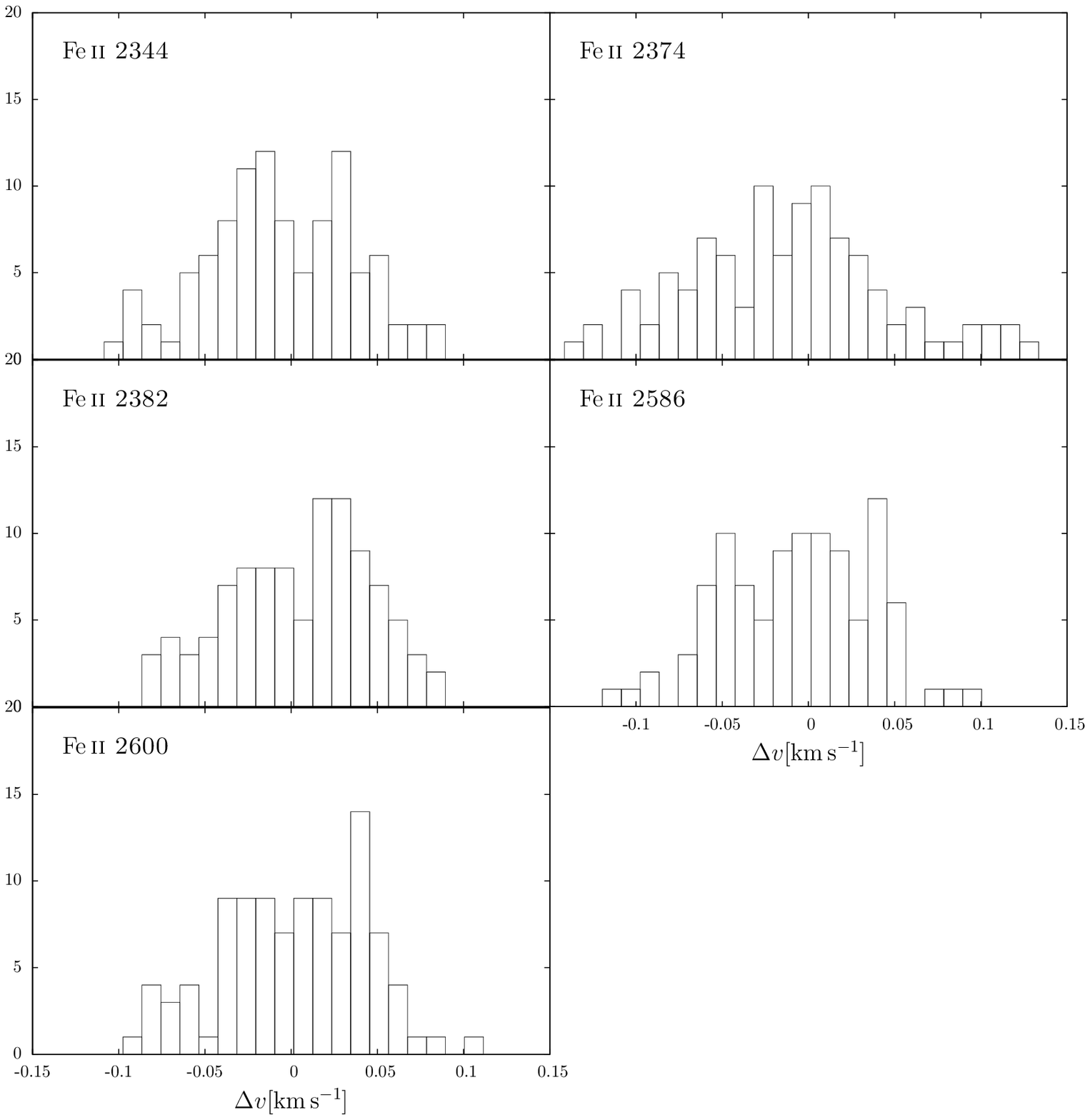}
\caption{As in Fig.\ \ref{fig:hist_vel_field_10_1}. Histograms of simulated lines with underlying velocity field with \(v_\mathrm{p}=10\mathrm{km\,s^{-1}}\). Two-component fit of 100 realizations with random noise.}
\label{fig:hist_vel_field_10_2}
\end{figure}

When fitting these profiles, the fitting code cannot recover the original velocity field, because the fit is made by assuming a finite number of Doppler profiles. The best we can hope for is a good approximation of the resulting profile. The same is true for real data, since the properties of the absorbing medium are generally unknown and supposedly complex.\\
To ascertain the best number of components, histograms are created for a wide range of gas velocities. As an example, Figs. \ref{fig:hist_vel_field_10_1} and \ref{fig:hist_vel_field_10_2} show the histograms of the not saturated version with \(v_\mathrm{p}=10\,\mathrm{km\,s^{-1}}\) for each transition, fitted with one and two components, respectively. \\
Table \ref{tab:vel_field_comp} shows the averaged mimicked \(\alpha\) variation for gas velocities from \(v_\mathrm{p}=5\,\mathrm{km\,s^{-1}}\) to \(v_\mathrm{p}=20\,\mathrm{km\,s^{-1}}\). Each is fitted with up to four components. The lowest velocity shifts are in this case achieved by using two or three component fits. There are a few cases where additional components lead to a lower precision in line positioning. This can happen when the additional components fit line distortions of the stronger lines that are created by noise. \\
The \(\chi^2\) value varies very little with the number of components. For small asymmetries, with this procedure it is not possible to determine the best number of components. Better methods are described in Sects. \ref{subsec:reg} and \ref{subsec:bis}. When the asymmetry cannot be seen by eye and adding further components does not decrease the \(\chi^2\) value, a one component fit would naturally be used. The corresponding velocity shifts between the 1608 \AA\ and the other transitions are shown in Table \ref{tab:vel_field_vel}.

For the smallest simulated peak velocity \(v_\mathrm{p}=5\mathrm{km\,s^{-1}}\) the best number of fitting components could not be determined by the \(\chi^2\) value alone. Fitting one component results in a shift of $\Delta v\approx 0.1\,\mathrm{km\,s^{-1}}$ between the stronger and the weaker transitions. Using all transitions, an $\alpha$ variation of $\Delta\alpha/\alpha=(-0.31\pm0.17)\cdot10^{-5}$ is mimicked, using just the weak transitions gives $\Delta\alpha/\alpha=(-0.03\pm0.18)\cdot10^{-5}$.

\begin{table*}[ht]
\centering
  \caption{Mimicked \(\alpha\) variation and \(\chi^2\) values for simulated spectra with an underlying velocity field.} 
 \begin{tabular}{c|cc|cc|cc|cc|cc}
   & \(v_\mathrm{p}=0\mathrm{km\,s^{-1}}\)  &  & \(v_\mathrm{p}=5\mathrm{km\,s^{-1}}\)  &  &\(v_\mathrm{p}=10\mathrm{km\,s^{-1}}\) & & \(v_\mathrm{p}=15\mathrm{km\,s^{-1}}\) & & \(v_\mathrm{p}=20\mathrm{km\,s^{-1}}\) & \\
  \hline
   \(\#_\mathrm{c}\) & \(\Delta\alpha/\alpha [10^{-6}]\) & \(\chi^2\) & \(\Delta\alpha/\alpha [10^{-6}]\) & \(\chi^2\) &  \(\Delta\alpha/\alpha [10^{-6}]\) & \(\chi^2\) & \(\Delta\alpha/\alpha [10^{-6}]\) & \(\chi^2\) & \(\Delta\alpha/\alpha [10^{-6}]\) & \(\chi^2\)\\
  \hline
  1 & \(-0.22\pm1.50\) & 1.0 & \(-3.11\pm1.93\) & 1.0 &  \(-8.62\pm2.26\) & 1.1 & \(-9.28\pm3.00\) & 1.4 & \(-8.79\pm4.42\) & 1.8 \\
  2 & \(-0.46\pm1.29\) & 1.0 & \(-1.93\pm2.33\) & 1.0  & \( 0.05\pm1.17\) & 1.0 & \( 0.20\pm2.32\) & 1.0 & \(-0.49\pm3.10\) & 1.0 \\
  3 & \( 0.14\pm1.96\) & 1.0 & \( 0.19\pm1.89\) & 1.0  & \( 0.49\pm1.74\) & 1.0 & \( 0.04\pm2.32\) & 1.0 & \(-0.56\pm2.92\) & 1.0  \\
  4 & \(1.19\pm3.08\) & 1.0 & \( 0.36\pm2.30\) & 1.0  & \( 0.52\pm2.17\) & 1.0 & \( 0.05\pm2.50\) & 1.0 & \(-0.50\pm2.95\) & 1.0 \\
\hline
 1 & \(-0.04\pm0.31\) & 1.0 & \(-2.57\pm0.42\) & 1.0 &  \(-14.26\pm1.66\) & 1.4 & \(-26.38\pm2.24\) & 5.6 & \(-26.11\pm1.98\) & 15.3 \\
  2 & \(-0.11\pm0.33\) & 1.0 & \(-0.49\pm1.14\) & 1.0  & \( 0.31\pm0.37\) & 1.0 & \( 0.54\pm0.42\) & 1.1 &  \( 0.31\pm0.63\) & 1.6 \\
  3 & \(0.02\pm0.63\) & 1.0 & \(-0.13\pm0.76\) & 1.0  & \(-0.12\pm01.01\) & 1.0 & \( 0.02\pm0.42\) & 1.0 &  \( 0.12\pm0.64\) & 1.0  \\
  4 & \(0.05\pm0.53\) & 1.0 & \( 0.06\pm0.80\) & 1.0 & \( 0.00\pm0.54\) & 1.0 & \( 0.02\pm0.42\) & 1.0 & \( 0.10\pm0.68\) & 1.0 \\
 \end{tabular}
\tablefoot{The first four rows show the results for \(N_1=13.0\) and the last four rows for \(N_2=13.5\). Simulated spectra with peak gas velocities from \(v_\mathrm{p}=05\mathrm{km\,s^{-1}}\) to \(v_\mathrm{p}=20\mathrm{km\,s^{-1}}\) were fitted with up to four components. Results are averaged over 100 fits with random noise. Errors represent the spread of the values.}
\label{tab:vel_field_comp}
\end{table*}

\begin{table*}[ht]
\centering	
  \caption{Velocity shifts between transitions for asymmetric lines with an underlying velocity field.}
 \begin{tabular}{c|ccccc}
  \(v_\mathrm{p} [\mathrm{km\,s^{-1}}]\) & \(\Delta v_{2344}[\mathrm{km\,s^{-1}}]\) & \(\Delta v_{2374}[\mathrm{km\,s^{-1}}]\) & \(\Delta v_{2382}[\mathrm{km\,s^{-1}}]\) &  \(\Delta v_{2586}[\mathrm{km\,s^{-1}}]\) & \(\Delta v_{2600}[\mathrm{km\,s^{-1}}]\) \\
  \hline
  0  & \(0.01\pm0.04\) & \(0.01\pm0.06\) & \(0.01\pm0.04\) & \(0.01\pm0.05\) & \(0.01\pm0.04\) \\
  5  & \(0.02\pm0.05\) & \(0.00\pm0.07\) & \(0.05\pm0.04\) & \(0.01\pm0.05\) & \(0.04\pm0.04\) \\
  10 & \(0.02\pm0.06\) & \(-0.02\pm0.08\) & \(0.11\pm0.06\) & \(0.01\pm0.06\) & \(0.09\pm0.06\) \\
  15 & \(0.03\pm0.08\) & \(0.01\pm0.11\) & \(0.13\pm0.08\) & \(0.02\pm0.08\) & \(0.10\pm0.08\) \\
  20 & \(0.06\pm0.10\) & \(0.02\pm0.15\) & \(0.12\pm0.10\) & \(0.03\pm0.11\) & \(0.11\pm0.10\) \\
\hline
 0  & \(0.00\pm0.01\) & \(0.00\pm0.02\) & \(0.00\pm0.01\) & \(0.00\pm0.01\) & \(0.00\pm0.01\) \\
  5  & \(0.03\pm0.01\) & \(-0.01\pm0.02\) & \(0.09\pm0.01\) & \(0.02\pm0.01\) & \(0.08\pm0.01\) \\
  10 & \(0.10\pm0.01\) & \(0.01\pm0.02\) & \(0.36\pm0.01\) & \(0.05\pm0.01\) & \(0.30\pm0.01\) \\
  15 & \(0.10\pm0.01\) & \(0.01\pm0.02\) & \(0.50\pm0.01\) & \(0.05\pm0.02\) & \(0.33\pm0.01\) \\
  20 & \(0.09\pm0.02\) & \(0.01\pm0.03\) & \(0.36\pm0.10\) & \(0.04\pm0.02\) & \(0.28\pm0.02\) \\
 \end{tabular}
\tablefoot{The first four rows show the results for \(N_1=13.0\) and the last four rows for \(N_2=13.5\). Column 1 shows the peak velocity \(v_\mathrm{p}\) of the velocity field, Cols.\ two to six the velocity shifts between the stated transition and the 1608\ \AA\ transition. A one-component fit is used. Results are averaged over 100 fits with random noise. Errors represent the spread of the values.}
\label{tab:vel_field_vel}
\end{table*}

\subsection{Line shift analysis}
\label{subsec:reg}
\begin{figure}
\resizebox{0.45\columnwidth}{!}{\input{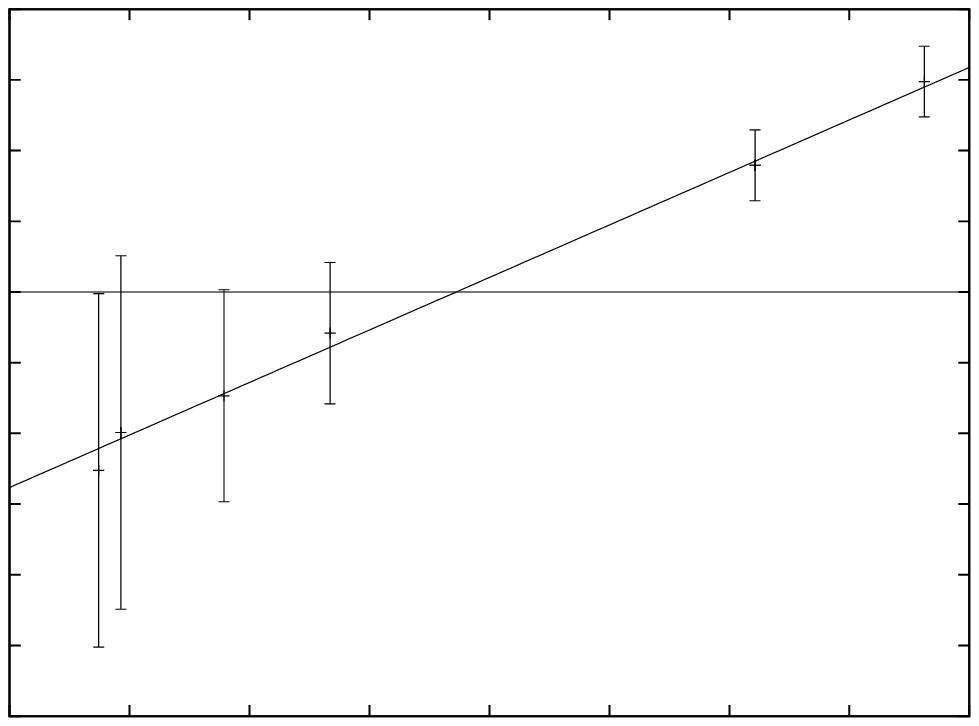}}
\resizebox{0.45\columnwidth}{!}{\input{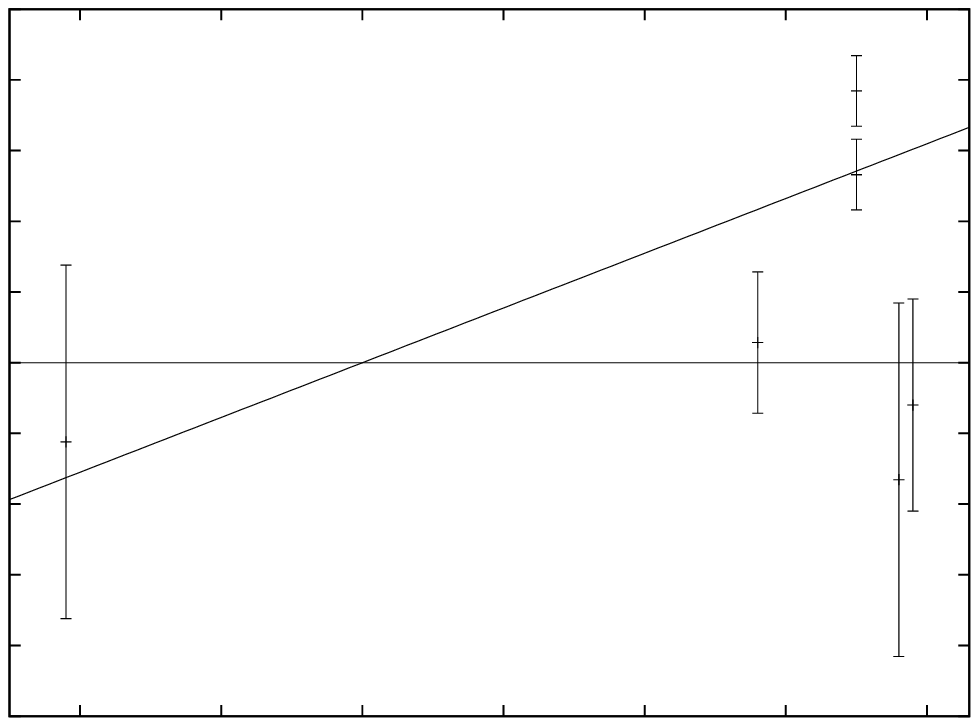}}\\
\resizebox{0.44\columnwidth}{!}{\input{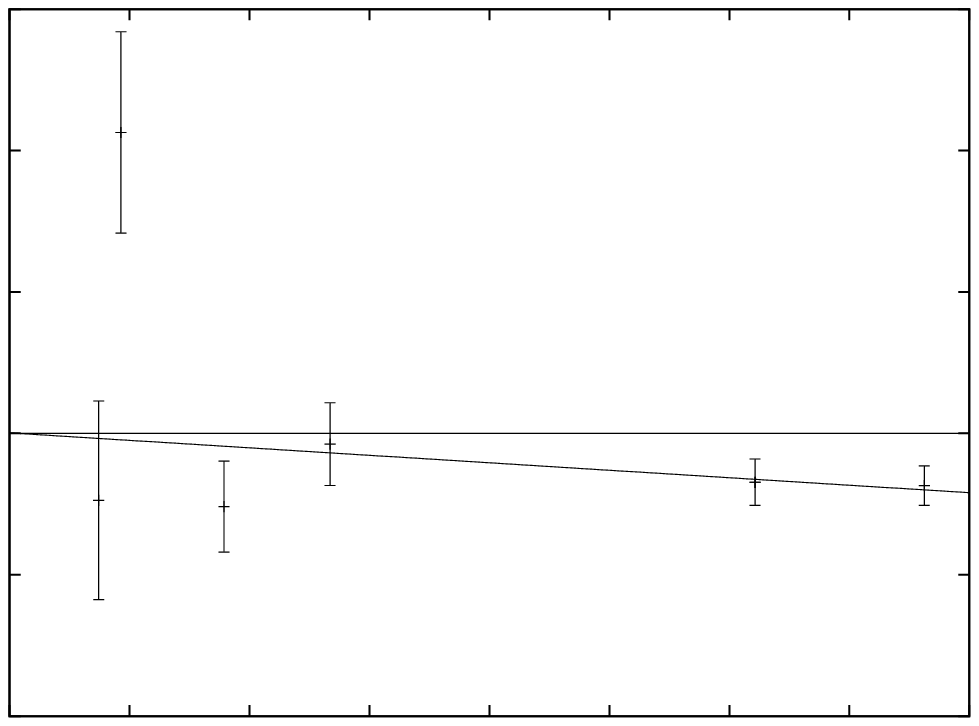}}\hspace{1mm}
\resizebox{0.44\columnwidth}{!}{\input{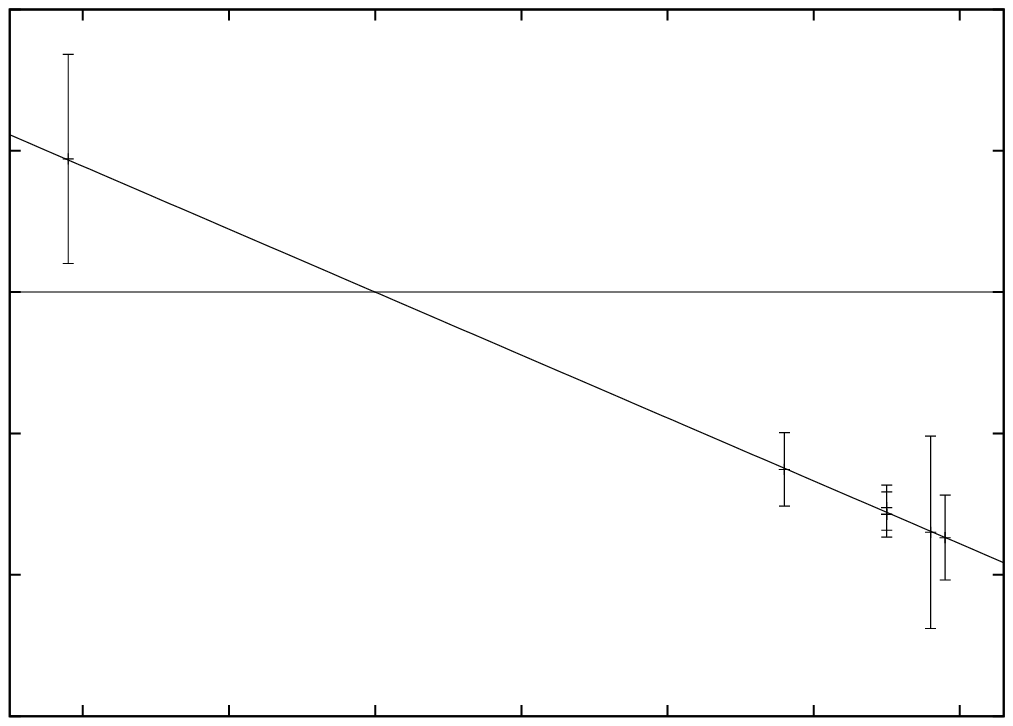}} \\
\resizebox{0.45\columnwidth}{!}{\input{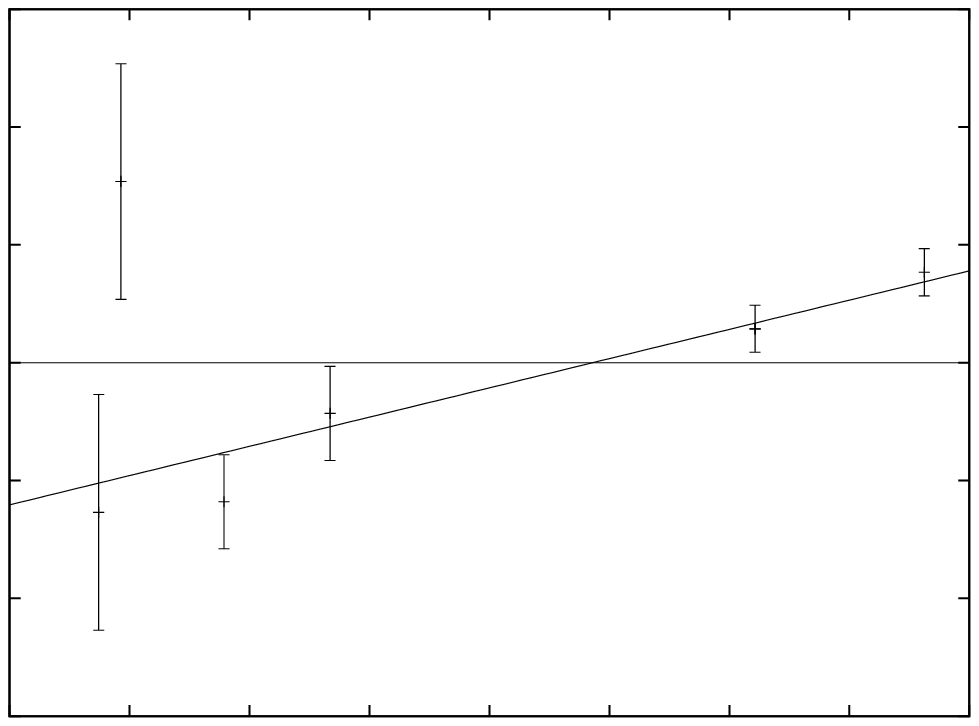}}
\resizebox{0.45\columnwidth}{!}{\input{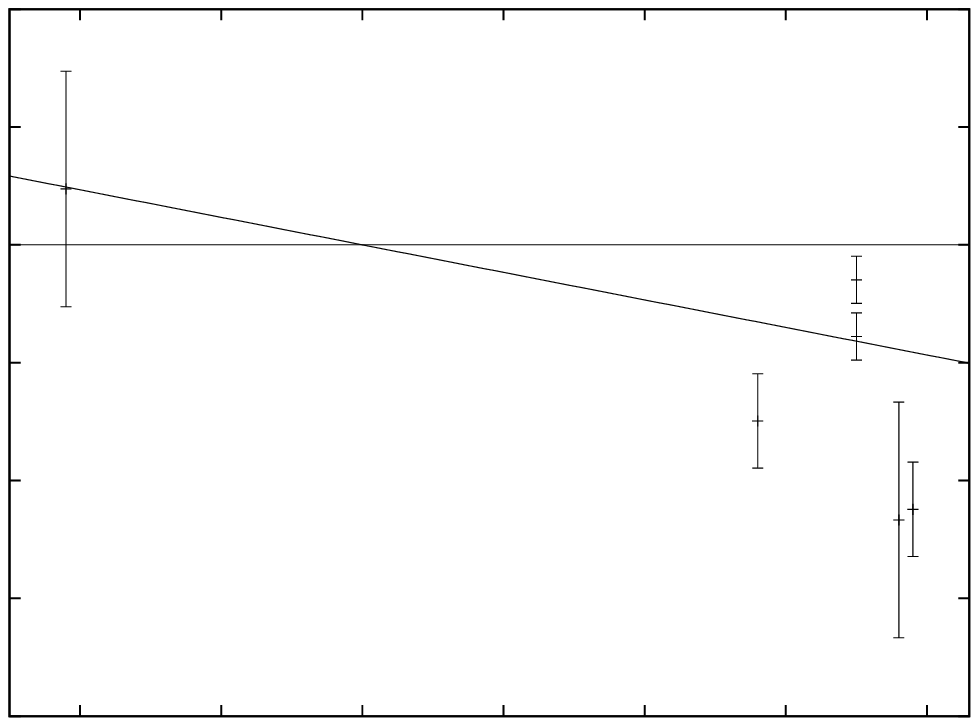}}\\
\caption{Redshift $z$ over transition strength \(f\lambda_0\) (a) and sensitivity coefficient \(Q\) (d) for an asymmetric line with underlying velocity field with peak velocity \(v_\mathrm{p}=10\,\mathrm{km\,s^{-1}}\), a symmetric line with artificial $\alpha$ variation of $\Delta\alpha/\alpha=0.5\cdot10^{-5}$ (b,e), and an asymmetric line with \(v_\mathrm{p}=10\,\mathrm{km\,s^{-1}}\) and $\Delta\alpha/\alpha=0.5\cdot10^{-5}$ (c,f).}
 \label{fig:reg}

\end{figure}

To test for possible errors of wavelength calibration, as well as for saturation effects and velocity fields, it is also helpful to look for position shifts between all the other transitions, especially the 2382\AA\ and the 2600\AA\ lines. Different sources that cause shifts between the lines will be discernible by comparing the shifts to different parameters. 
Figure \ref{fig:reg}a shows positions of lines over the transition strength \(f\lambda_0\) for simulated spectra created in \ref{subsec:vel} with a peak velocity of \(v=10\,\mathrm{km\,s^{-1}}\). The nearly linear dependence of shift and transition strength indicates a problem with saturation effects. As a comparison, in Fig.\ \ref{fig:reg}b the same information is shown for a symmetric feature with an artificial $\alpha$ variation of $\Delta\alpha/\alpha=0.5\cdot10^{-5}$. A combination of both effects is shown in Fig.\ \ref{fig:reg}c and the resulting $\alpha$ variation is $\Delta\alpha/\alpha=(0.23\pm0.17)\cdot10^{-5}$.\\
In Figs.\ \ref{fig:reg}d, \ref{fig:reg}e, and \ref{fig:reg}f the same shifts are plotted over the sensitivity coefficient \(Q\). 
Since the strong 2382 \AA\ and 2600 \AA\ transitions have the same sensitivities, all position shifts between these two lines cannot be created by $\alpha$ variations. In principle, the difference between shifts caused by an \(\alpha\) variation and those created by an incorrect line decomposition can thus be distinguished. Assuming a linear correlation between $z$ and \(f\lambda_0\), the strong lines can be used to correct the positions of the other transitions by shifting them according to a straight line fitted through the positions of the 2382 \AA\ and the 2600 \AA\ transitions (see Fig. \ref{fig:reg}a). Applying the correction would result in $\Delta\alpha/\alpha=(0.41\pm0.17)\cdot10^{-5}$ for all transitions and $\Delta\alpha/\alpha=(0.46\pm0.17)\cdot10^{-5}$ using just the weak transitions. Generally, the $z-f\lambda_0$ relation will not be exactly linear, as can be seen in Fig.\ \ref{fig:reg}a. With the data quality available, this procedure will bring no significant improvement. Simply using the weak transitions gives in this case $\Delta\alpha/\alpha=(0.45\pm0.17)\cdot10^{-5}$. If the 2374 \AA\ or the 2586 \AA\ transition are 
not available, or not usable for other reasons, the systematic error introduced by an $z-f\lambda_0$ dependence can be reduced significantly with this method.

\subsection{Bisector analysis}
\label{subsec:bis}
\begin{figure}
\resizebox{\columnwidth}{!}{\input{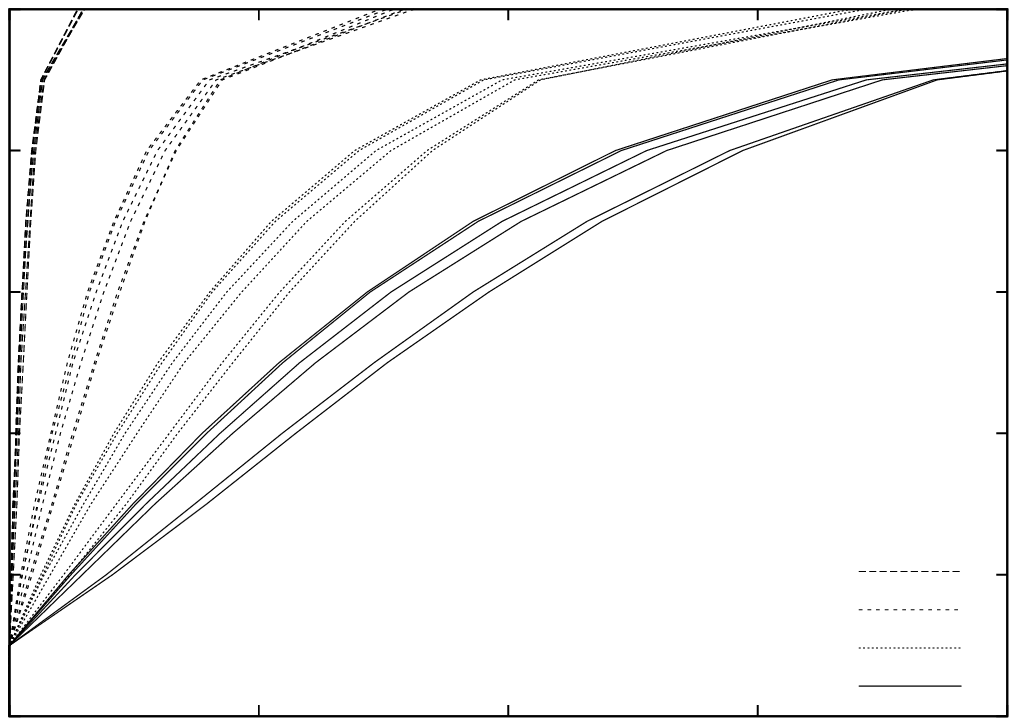}}
\caption{Bisectors of \ion{Fe}{ii} 1608 \AA, 2344 \AA, 2374 \AA, 2382 \AA, 2586 \AA, and 2600 \AA\ transitions. Macroscopic velocities of $v_\mathrm{p}=5\,\mathrm{km\,s^{-1}}$, \(10\,\mathrm{km\,s^{-1}}\), \(15\,\mathrm{km\,s^{-1}}\), and \(20\,\mathrm{km\,s^{-1}}\) are plotted.}
\label{fig:bisectors}
\end{figure}

It would be helpful to have the possibility of directly measuring the symmetry of a line as a starting point to search for unresolved line structure or velocity fields. There are several ways to create a measure of the asymmetry of an absorption line. The bisector method, originally developed in solar physics for detecting velocity fields in the atmospheres of late-type stars (e.g. \citet{Dravins1982}) has the advantage that it can identify not only to the magnitude of the asymmetry but also its general shape. For a given flux $F$, the central wavelength $\lambda_\mathrm{c}=(\lambda_2-\lambda_1)/2$ between the two flanks of the line profile at this flux is calculated. The bisector is a curve crossing the points ($\lambda_\mathrm{c,i}$,$F_\mathrm{i}$). For a perfectly symmetric line, the bisector is just a vertical straight line at the position of the line centre from the lowest flux of the line up to the continuum. \\
As an example, Fig.\ \ref{fig:bisectors} shows bisectors of artificial lines of the \ion{Fe}{ii} 1608 \AA, 2344 \AA, 2374 \AA, 2382 \AA, 2586 \AA, and 2600 \AA\ transitions and macroscopic gas velocities of $v_\mathrm{p}=5\,\mathrm{km\,s^{-1}}$, \(10\,\mathrm{km\,s^{-1}}\), \(15\,\mathrm{km\,s^{-1}}\), and \(20\,\mathrm{km\,s^{-1}}\) calculated as described in Sect. \ref{subsec:vel}. The bisectors are parametrized from the minimum of the profile \((bis=0)\) to the continuum \((bis=1)\). The lowest value is omitted because the determination of a line centre, which is used as a basis for comparing bisectors of different transitions, is strongly affected by noise at the minimum intensity especially for weak or saturated lines. The bisector value at 0.1 is thus used as the central point. The figure shows that for each line composition the bisectors of the different transitions can be distinguished. The weakest 2374 \AA\ transition is the steepest on the left side and the strong 2382 \AA\ transition on the right side for each velocity setup in Fig.\ \ref{fig:bisectors}. \\ 
Line positions are usually determined by the least-squares method. The results depend on a correct decomposition of the line profile. As was shown in Sect.\ \ref{subsec:vel}, the determination of the number of components for the best fit is often ambiguous. The bisector can be used to compare the symmetry of the involved lines and thus reveal potential decomposition problems and other error sources. \\

In principle, the bisector of each transition is slightly different when saturation effects or velocity fields are present; however, these differences are so small that they are, in nearly all cases, blurred by noise. Finding considerable differences in the bisectors between different transitions of the same ion would usually mean that some of the lines are not suitable and should not be used. \\
There is another way the bisector method can be used in this case. Even when a line looks symmetric and a one-component fit is favoured, there can be a measurable deviation from a truly symmetric line. By studying the bisector, these deviations can be detected and the potential error can be estimated. 

\begin{figure}
\resizebox{\columnwidth}{!}{\input{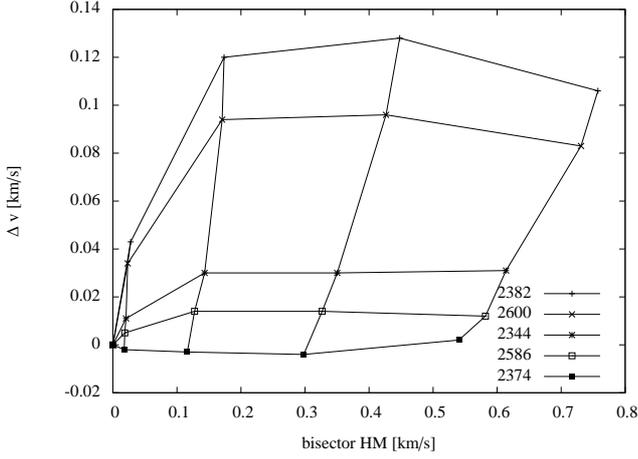}}
\caption{Velocity shifts over bisector at half maximum for different \ion{Fe}{ii} transitions. The vertical lines depict from left to right macroscopic velocities of \(v_\mathrm{p}=5\,\mathrm{km\,s^{-1}}, v_\mathrm{p}=10\,\mathrm{km\,s^{-1}}, v_\mathrm{p}=15\,\mathrm{km\,s^{-1}}\), and \(v_\mathrm{p}=20\,\mathrm{km\,s^{-1}}\).}
\label{fig:bis_diff_nonoise}
\end{figure}

\begin{figure}
\resizebox{\columnwidth}{!}{\input{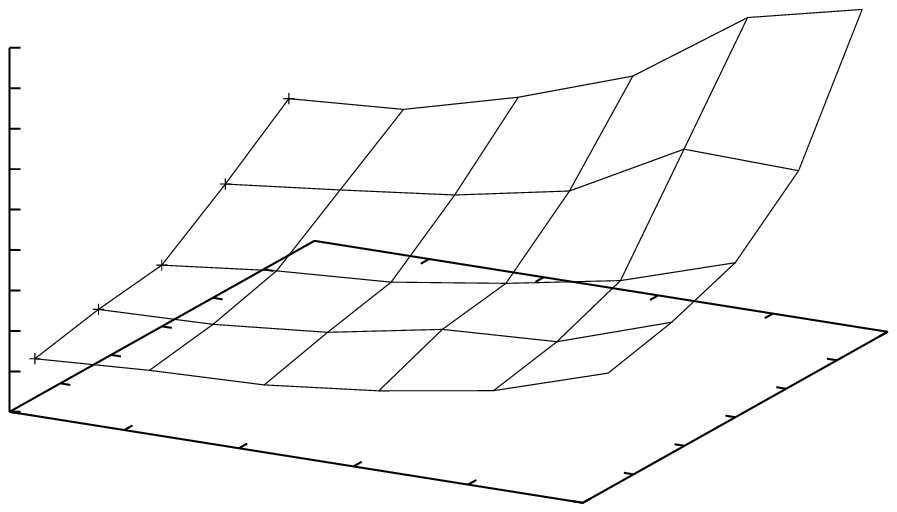}}
\caption{Standard deviation $\sigma$ of bisector at half maximum over resolution $R$ and with signal to noise ratio $S/N$.}
\label{fig:bis_error}
\end{figure}

Since the differences of the bisectors of different transitions are quite low in the most cases, for the data quality currently available they would not be detected. The asymmetry of each line can be measured, when plotting the total bisector. Figure \ref{fig:bis_diff_nonoise} shows position shifts over the total bisector at half maximum of the corresponding transition, plotted for the \ion{Fe}{ii} 2344 \AA, 2374 \AA, 2382 \AA, 2586 \AA, and 2600 \AA\ transitions with respect to the \ion{Fe}{ii} 1608 \AA\ transition. Each vertical line represents a model with macroscopic velocities of $v_p=5\,\mathrm{km\,s^{-1}}, 10\,\mathrm{km\,s^{-1}}, 15\,\mathrm{km\,s^{-1}}$, and $v_p=20\,\mathrm{km\,s^{-1}}$, seen from left to right. The main increase of position offsets comes at small asymmetries of $bis_\mathrm{HM}\lesssim0.2 \mathrm{km\,s^{-1}}$ since higher asymmetries allow more components to be fitted. The possibility of finding asymmetries on this scale depends on the data quality. Figure \ref{fig:bis_error} shows the accuracy of bisector measurements at half maximum for different resolutions $R$ and signal to noise ratios $S/N$. Though total bisectors with $bis_\mathrm{HM}\sim0.2\,\mathrm{km\,s^{-1}}$ would be detectable with the data quality currently available, an asymmetric line does not necessarily imply a position shift. Bisector differences that detect saturation effects would need very high quality data with $R\gtrsim80.000$ and $S/N\gtrsim140$. Since these will not be detectable in most cases with the data currently available, an upper limit to the total bisector can be used to estimate the shift that could be introduced by saturation effects according to Fig. \ref{fig:bis_diff_nonoise}. However, for future spectra taken with ESPRESSO at the VLT or PEPSI at the LBT for example, the bisector method might be a useful instrument.

\section{Data analysis}
\label{sect:observations}
To compare the simulations with real data, spectra of the ESO-VLT Large Program ``The Cosmic Evolution of the IGM'' from 2004 are used. The data were reduced by \citet{Aracil2004}. A set of 19 quasar spectra were taken over a period of two years. The quality of the data is lower than used in the simulations, resulting in a lower accuracy. The bisector analysis will only be possible in special cases. However, the advantage is that many different systems are available. We analysed 14 \ion{Fe}{ii} systems in the spectra of seven quasars. 
Each system is studied carefully to detect potential sources for position shifts that could mimic an $\alpha$ variation. To detect possible decomposition problems, for each system a $z-f\lambda_0$ diagram is plotted. Additionally, for single isolated lines, bisectors are plotted. The minimizing algorithm based on an evolution strategy \citep{Quast2005} used for fitting the data reduces the danger of finding only a local \(\chi^2\) minimum. To further reduce potential fitting problems, each system was fitted several times with an increasing number of components until the minimum \(\chi^2\) value change is less than 10\%. To account for the possibility that the FWHM of the instrument profile fluctuates, the fits were repeated with a change of the FWHM of up to 20\% in either direction. In all cases the changes in position measurements were well within the error limits of each other and thus had no significant influence on the results of the \(\alpha\) variation measurement. \\
When calculating the apparent \(\alpha\) variation, different methods are used. Since line shifts due to wavelength calibration errors are hard to detect, the selection of suitable lines is mainly done by studying the bisector (Sect \ref{subsec:bis}) and the \(z-f\lambda_0\) diagram (Sect.\ \ref{subsec:reg}). For comparison, the results of using on the one hand all available transitions in a regression analysis and on the other hand just two line positions of transitions with similar strengths, are given separately in each case. \\
In this chapter only statistical errors are given. The systematic errors will be discussed in Sect. \ref{sect:discussion}. 

\subsection{HE0001-2340}
\begin{figure*}[!htb]
\centering
 \includegraphics[width=0.9\textwidth,bb = 30 28 558 784]{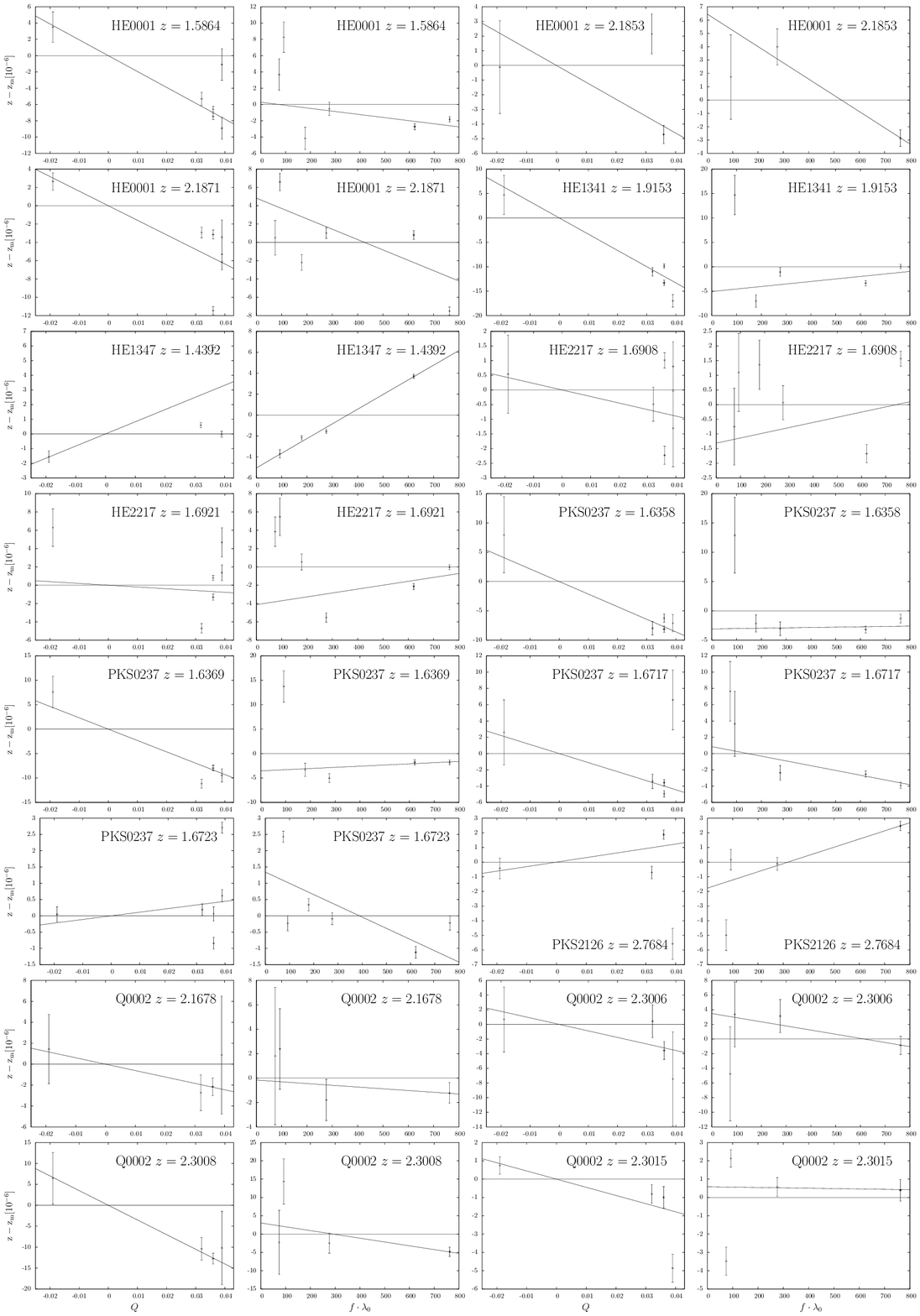}
\caption{Line shift analysis of 16 \ion{Fe}{ii} systems in eight quasar spectra. The relative position shift is plotted against the sensitivity coefficient $Q$ and the transition strength $f\lambda_0$ for each system. \(z_\mathrm{m}\) is the intercept term of the depicted regression.}
\label{fig:reg_data}
\end{figure*}

The bright quasar \object{HE0001-2340} has an emission redshift of $z_\mathrm{em}=2.28$. It has several \ion{Fe}{ii} systems, one of them a strong damped Lyman $\alpha$ (DLA) system. For one of the \ion{Fe}{ii} systems ($z=0.44$) the important 1608 \AA\ transition is outside of the range of optical telescopes. The system at $z=1.59$, composed of a single visual component, is quite weak, so the important transitions are highly influenced by noise. Assuming that the wavelength shifts are created by an $\alpha$ variation, a one-component fit would yield $\Delta\alpha/\alpha=(3.8\pm0.6)\cdot10^{-5}$. Fitting in a second component does not change the result within the error limits ($\Delta\alpha/\alpha=(3.9\pm0.6)\cdot10^{-5}$). The ratio of the minimum \(\chi^2\) values of the two-component fit to the one-component fit is \(\chi^2_{2}/\chi^2_{1}=1.0\). There is no strong correlation of position and transition strength (Fig.\ \ref{fig:reg_data}). Using just the 1608 \AA\ and the 2374 \AA\ transitions, the result would change to $\Delta\alpha/\alpha=(1.5\pm0.8)\cdot10^{-5}$. \\
The bisectors of the lines (Fig.\ \ref{fig:bis}a) show that all three of the weaker components deviate strongly from a symmetric shape while the strong transitions are symmetric. These lines do not show a velocity shift bigger than 1$\sigma$ (statistical) to each other. It thus has to be assumed that the velocity shift between the weak and the strong transitions is created by the deformation of the lines by noise or an unknown effect. Wavelength calibration errors are always possible and not really under control at UVES \citep{Whitmore2010}, although \citet{Chand2006} have shown by comparison with HARPS with help of the bright QSO HE0515-4414 that even the UVES pipeline data are fairly accurate on a relative scale.\\
The \(z=1.59\) system in HE0001-2340 has recently been analysed by \citet{Agafonova2011} with a new set of data obtained in 2009. They compared the 1608 \AA\ transition with the 2382 \AA\ transition and found $\Delta\alpha/\alpha=(-0.05\pm1.1)\cdot10^{-5}$. 

The DLA system at $z=2.19$ has two distinct \ion{Fe}{ii} features at $z_1=2.1853$ and $z_2=2.1871$. This system was previously analysed by \citet{Molaro2008}. System 1 is quite weak and only three transitions are usable (1608 \AA, 2344 \AA, and 2382 \AA). The line shift analysis is highly dominated by the strong 2382 \AA\ transition (Fig. \ref{fig:reg_data}). Using all three transitions with a one component fit gives an apparent variation of $\Delta\alpha/\alpha=(1.8\pm0.9)\cdot10^{-5}$. Fitting two components gives the same result within the error limits ($\Delta\alpha/\alpha=(2.6\pm0.9)\cdot10^{-5}$), with \(\chi^2_{2}/\chi^2_{1}=1.0\). \citet{Molaro2008} only used the 2382 \AA\ transition in comparison with the 1608 \AA\ transition and got a similar result ($\Delta\alpha/\alpha=(2.3\pm1.0)\cdot10^{-5}$).
Figure \ref{fig:reg_data} shows a stronger correlation of the line shifts with transition strength \(f\lambda_0\) than with sensitivity coefficient $Q$, indicating asymmetry effects. Without the 2600 \AA\ transition, the influence of this effect on the total position shifts cannot be quantified. Since the weak transitions are not available in this system, the way to proceed would be to use just the 2344 \AA\ transition. The bisector of the 1608 \AA\ transition shows a slight slope, possibly created by noise, which can account for some unwanted shift. Disregarding the 2382 \AA\ transition would give $\Delta\alpha/\alpha=(-0.7\pm1.0)\cdot10^{-5}$. \citet{Molaro2008} concluded that the shift was created by wavelength calibration problems. We propose that the effect is mainly based on an unresolved substructure of the lines. \\
The second \ion{Fe}{ii} feature in this subDLA system is stronger and quite promising. The 2382 \AA\ transition has a strong shift which cannot be accounted for. The bisector looks identical to that of the 2600 \AA\ transition. It is possible that some unresolved blend with another line shifts this transition or that there is some local error in the wavelength calibration. Since the shift is definitely not created by an $\alpha$ variation, the transition is left out of the analysis. The line shift analysis (Fig. \ref{fig:reg_data}) shows similarities with the artificial spectrum in Fig.\ \ref{fig:reg}e. The absence of the 2382 \AA\ transition makes it difficult to disentangle the different effects. All remaining transitions would give $\Delta\alpha/\alpha=(1.8\pm0.3)\cdot10^{-5}$, using a two component fit. Using only the 2374\AA\ transition in comparison with the 1608\AA\ transition, and thus excluding possible asymmetry effects, gives $\Delta\alpha/\alpha=(1.6\pm0.4)\cdot10^{-5}$. ($\Delta\alpha/\alpha=(1.4\pm0.5)\cdot10^{-5}$ for a three component fit with \(\chi^2_{3}/\chi^2_{2}=1.0\). The result is similar to that obtained by \citet{Molaro2008}. \citet{Agafonova2011} compared the position of the 1608 \AA\ transition with that of the 2344 \AA\ transition. They found a slightly lower value of \(\Delta\alpha/\alpha=(0.96\pm0.45)\cdot10^{-5}\).

\begin{table*}
 \caption{Position shifts of \ion{Fe}{ii} 2344 \AA, 2374 \AA, 2382 \AA, 2586 \AA\ and 2600 \AA\ transition with respect to the 1608 \AA\ transition for each analysed system.}
\centering
\begin{tabular}{lcrrrrr}
 QSO & \(z\) & $\Delta v_{2344} [\mathrm{km\,s^{-1}}]$ & $\Delta v_{2374} [\mathrm{km\,s^{-1}}]$ & $\Delta v_{2382} [\mathrm{km\,s^{-1}}]$ & $\Delta v_{2586} [\mathrm{km\,s^{-1}}]$ & $\Delta v_{2600} [\mathrm{km\,s^{-1}}]$\\ 
\hline
HE0001-2340 & 1.5864  & $-1.02\pm0.62$ & $-0.54\pm0.81$ & $-1.17\pm0.58$ & $-1.44\pm0.71$ & $-1.28\pm0.58$\\ 
HE0001-2340 & 2.1853  & $0.21\pm1.05$ &  & $-0.44\pm0.98$ &  &  \\ 
HE0001-2340 & 2.1871  & $-0.53\pm0.34$ & $-0.58\pm0.64$ & $-1.33\pm0.32$ & $-0.83\pm0.39$ & $-0.55\pm0.32$ \\
HE1341-1020 & 1.9153  & $-1.63\pm1.24$ &  & $-1.51\pm1.21$ & $-2.24\pm1.27$ & $-1.86\pm1.22$ \\
HE1347-2457 & 1.4392  & $0.26\pm0.13$ & & & $0.18\pm0.12$ & $0.91\pm0.12$\\
HE2217-2818 & 1.6908  & $-0.12\pm0.43$ & $-0.20\pm0.56$ & $0.06\pm0.41$ & $0.02\pm0.47$ & $-0.31\pm0.41$ \\
HE2217-2818 & 1.6921 & $-1.22\pm0.63$ & $-0.03\pm0.78$ & $-0.52\pm0.62$ & $-0.45\pm0.67$ & $-0.73\pm0.62$ \\
PKS0237-23 & 1.6358  & $-1.82\pm1.97$ &  & $-1.62\pm1.95$ & $-1.72\pm1.99$ & $-1.83\pm1.94$\\
PKS0237-23 & 1.6369  & $-2.13\pm1.00$ &  & $-1.77\pm0.97$ & $-1.94\pm1.04$ & $-1.77\pm0.97$ \\
PKS0237-23 & 1.6717  & $-0.68\pm1.23$ & $0.45\pm1.63$ & $-0.85\pm1.20$ &  & $-0.70\pm1.20$\\
PKS0237-23 & 1.6723 & $0.02\pm0.09$ & $0.30\pm0.09$ & $0.00\pm0.10$ & $0.07\pm0.09$ & $-0.10\pm0.09$\\
PKS2126-158 & 2.7684 & $-0.03\pm0.25$ & $-0.41\pm0.38$ & $0.18\pm0.23$ &  & \\
Q0002-422 & 2.1678 & $-0.40\pm1.49$ & $-0.06\pm2.34$ & $-0.34\pm1.38$ &  & \\
Q0002-422 & 2.3006 & $-0.03\pm1.47$ & $-0.74\pm1.47$ & $-0.39\pm1.37$ &  & \\
Q0002-422 & 2.3008 & $-1.53\pm2.02$ & $-1.51\pm3.21$ & $-1.74\pm1.89$ &  & \\
Q0002-422 & 2.3015 & $-0.15\pm0.21$ & $-0.52\pm0.27$ & $-0.17\pm0.23$ &  & \\

\end{tabular}
\end{table*}

\subsection{HE1341-1020}
In the spectrum of \object{HE1341-1020} ($z_\mathrm{em}=2.14$) there are two systems that show the \ion{Fe}{ii} 1608 \AA\ line. The first at $z_1=1.28$ is located in the Lyman $\alpha$ forest and will thus not be regarded. The $z_2=1.92$ system seems to show a strong signal. The correlation of position and sensitivity $Q$ is stronger than position and transition strength $f\lambda_0$; however, the offset between the strong transitions indicates some small asymmetry effect (Fig. \ref{fig:reg_data}). Using a two-component fit with just the 2586 \AA\ transition (the 2374 \AA\ transition is not available) gives a strong signal of $\Delta\alpha/\alpha=(6.4\pm1.2)\cdot10^{-5}$ (three components:  $\Delta\alpha/\alpha=(6.7\pm1.2)\cdot10^{-5}$, \(\chi^2_{3}/\chi^2_{2}=0.9\)), while all available transitions give $\Delta\alpha/\alpha=(5.7\pm1.2)\cdot10^{-5}$. The bisector of the 1608 \AA\ feature differs slightly from that of the other transitions. We note that the also weak 2568 \AA\ feature shows a line profile similar to the strong transitions (Fig. \ref{fig:bis}d). We thus have to assume that the velocity shift here is created by some unknown mechanism, e.g. an unrecognised blend, which distorts the line shape of the important 1608 \AA\ transition.

\subsection{HE1347-2457}

There is a strong and heavily blended \ion{Fe}{ii} system at $z=1.44$ in the spectrum of \object{HE1347-2457} ($z_\mathrm{em}=2.6$), which was also analysed by \citet{Molaro2008}. The lines of all transitions are saturated which would make the influence of an incorrect decomposition very strong. The 1608 \AA\ line is located in the Lyman $\alpha$ forest, so all results should be regarded with care since an undiscovered blend with a Lyman $\alpha$ line could produce a significant position shift. It is only included here to allow a comparison with \citet{Molaro2008}. The 2374 \AA\ and the 2382 \AA\ transitions fall into a data gap and are not available.
The high offset of the 2600 \AA\ line should not be too surprising, since the combination of line blends and saturation make a correct decomposition unlikely. In Fig.\ \ref{fig:reg_data} it can be seen that the correlation of position and transition strength is very strong, so the position shifts of the 2344 \AA, 2586 \AA, and 2600 \AA\ are obviously caused by an incorrect line decomposition. The offset of the 1608 \AA\ can easily be explained by an unrecognised blend with a Lyman $\alpha$ feature. \\
Assuming that this is not the case, the best approach would be just to use the 1608 \AA\ and the 2586 \AA\ feature to avoid problems with incorrect line decomposition; however, the 2586 \AA\ feature is blended with some telluric lines. Comparing the position of these two lines would, nevertheless, give an $\alpha$ variation of $\Delta\alpha/\alpha=(-0.5\pm0.1)\cdot10^{-5}$ (four components: $\Delta\alpha/\alpha=(-0.5\pm0.1)\cdot10^{-5}$, \(\chi^2_{4}/\chi^2_{3}=1.0\)), while all transitions give $\Delta\alpha/\alpha=(-1.7\pm0.1)\cdot10^{-5}$ ($\Delta\alpha/\alpha=(-1.8\pm0.1)\cdot10^{-5}$). Since the system is quite strong, the statistical error is low. \citet{Molaro2008} state a similar result of  $\Delta\alpha/\alpha=(-2.1\pm0.2_\mathrm{stat}\pm1.1_\mathrm{sys})\cdot10^{-5}$ using the 1608 \AA, 2344 \AA, and the 2586 \AA\ transitions. They have included a systematical error for the wavelength calibration (see Sect. \ref{sect:discussion}). Since the magnitude of position shifts due to unrecognised line blends can be very high, systems such as this should not be used for this analysis.

\subsection{HE2217-2818}

The quasar \object{HE2217-2818} ($z_\mathrm{em}=2.41$) has several \ion{Fe}{ii} systems, one of which has a visible 1608 \AA\ line at $z=1.69$. The system consists of two parts, at $z_1=1.6908$ and $z_2=1.6921$, which will be dealt with separately. They consist of at least five and seven blended components, respectively. It will be assumed that systems of blended lines which are not separated by a clear continuum are related and a total shift of these systems is determined. \\
The first part has a strong shift between the two strong transitions, which is a good indication of saturation effects or unresolved line blends. There is, however, no strong indication of a correlation between position shift and transition strength (Fig. \ref{fig:reg_data}). The 2344 \AA\ transition has a strong blend with a telluric line and so is not used. The 2586 \AA\ transition also shows a slight blend with a telluric feature and so is neglected. The reason for the shift of the 2600 \AA\ transition is unknown, probably an unrecognised blend. Since it cannot be caused by an $\alpha$ variation, this transition will not be used either. Using just the 2374 \AA\ transition with a five-component fit yields $\Delta\alpha/\alpha=(0.6\pm0.5)\cdot10^{-5}$ (six components: $\Delta\alpha/\alpha=(0.6\pm0.5)\cdot10^{-5}$, \(\chi^2_{6}/\chi^2_{5}=1.0\)). All components would have given $\Delta\alpha/\alpha=(0.4\pm0.4)\cdot10^{-5}$ ($\Delta\alpha/\alpha=(0.4\pm0.4)\cdot10^{-5}$).\\
The 2344 \AA\ transition of the second system is even more strongly influenced by the blend than the first component, explaining the strong position offset. The offset between the strong transitions indicates a slight decomposition problem (Fig. \ref{fig:reg_data}). The blend of the 2586 \AA\ transition that affects the first part of the system has no visible impact on the second part. There is however the possibility that the feature causing the blend has more components that also affect the second part of the system. Again using only the 1608 \AA\ and the 2374 \AA\ transitions gives $\Delta\alpha/\alpha=(0.5\pm0.7)\cdot10^{-5}$ ($\Delta\alpha/\alpha=(0.1\pm0.5)\cdot10^{-5}$, \(\chi^2_{8}/\chi^2_{7}=0.9\)). Using the 2586 \AA\ transition as well would give $\Delta\alpha/\alpha=(1.4\pm0.7)\cdot10^{-5}$ ($\Delta\alpha/\alpha=(0.9\pm0.4)\cdot10^{-5}$). All transitions, except the obviously blended 2344 \AA\ feature would give an even higher variation of $\Delta\alpha/\alpha=(1.9\pm0.7)\cdot10^{-5}$ ($\Delta\alpha/\alpha=(1.4\pm0.4)\cdot10^{-5}$).\\

\subsection{PKS0237-23}

The quasar \object{PKS0237-23}, at an emission redshift of $z_\mathrm{em}=2.22$, has several metal systems. Apart from a strong \ion{Fe}{ii} system at $z=1.36$, whose 1608 \AA\ line lies in the Lyman $\alpha$ forest and so will not be used, this quasar has three close \ion{Fe}{ii} systems at $z_1=1.64$, $z_2=1.66$ and $z_3=1.67$. The 1608 \AA\ feature of the $z=1.66$ system is heavily blended, so that a reliable position estimation is not possible. The remaining two systems are separated into two subsystems each. \\
In the first part of the $z=1.64$ system, at \(z_1=1.6358\), the 1608 \AA\ line shows a strong shift (Fig. \ref{fig:reg_data}). It is, however, slightly blended with an unidentified feature. A blend will, in most cases, create a shift since the unknown line will probably not be symmetric itself and cannot be subtracted correctly. The bisector (Fig.\ \ref{fig:bis}e) shows that only the 1608 \AA\ feature deviates obviously from a symmetrical shape. The 2374 \AA\ line is not available. The $z-f\lambda_0$ diagram shows no strong signs of correlation. Using all lines with a two-component fit, an $\alpha$ variation of $\Delta\alpha/\alpha=(4.1\pm2.0)\cdot10^{-5}$ (Three components: $\Delta\alpha/\alpha=(4.0\pm2.0)\cdot10^{-5}$, \(\chi^2_{3}/\chi^2_{2}=1.0\)) would be measured. Using just the 2586 \AA\ transition would give $\Delta\alpha/\alpha=(4.9\pm2.1)\cdot10^{-5}$ ($\Delta\alpha/\alpha=(4.9\pm2.1)\cdot10^{-5}$). \\
The second part of the system, at \(z_2=1.6369\), is an asymmetric feature consisting of at least five heavily blended components. It also shows a strong shift of the 1608 \AA\ transition (Fig. \ref{fig:reg_data}). Again, the 2374 \AA\ line is not available. Using all the remaining transitions would give $\Delta\alpha/\alpha=(4.4\pm1.0)\cdot10^{-5}$ ($\Delta\alpha/\alpha=(5.9\pm0.5)\cdot10^{-5}$, \(\chi^2_{6}/\chi^2_{5}=1.0\)), using just the 2586 \AA\ transition $\Delta\alpha/\alpha=(5.6\pm1.1)\cdot10^{-5}$ ($\Delta\alpha/\alpha=(5.9\pm0.5)\cdot10^{-5}$). Although no obvious blend is seen in this case, the position of the strong 2344 \AA\ transition is at a \(3\sigma\) distance from the regression line. No correlation of position shift and transition strength can be seen (Fig. \ref{fig:reg_data}). The bisector again shows a difference in line shape between the weak transitions and the strong (Fig.\ \ref{fig:bis}f). The data quality is too low to decide whether this is the cause for the shift. \\
The $z=1.67$ system consists of two parts with at least three and five components, respectively. The first part, at \(z_1=1.6717\), shows a position offset between the weak and the strong transitions, correlated with transition strength (Fig. \ref{fig:reg_data}). The 2586 \AA\ feature is not available. Using only the other two weak transitions gives $\Delta\alpha/\alpha=(-1.3\pm1.5)\cdot10^{-5}$ ($\Delta\alpha/\alpha=(0.1\pm1.3)\cdot10^{-5}$, \(\chi^2_{4}/\chi^2_{3}=0.9\)), while all transitions would result in  $\Delta\alpha/\alpha=(2.1\pm1.3)\cdot10^{-5}$ ($\Delta\alpha/\alpha=(1.8\pm1.1)\cdot10^{-5}$). \\
The second part of this system, at \(z_2=1.6723\), shows a shift of the 2374 \AA\ transition with respect to the other lines. The reason for this shift is unknown. Although the stronger transitions are saturated, no obvious correlation between position shift and transition strength can be seen (Fig. \ref{fig:reg_data}). A single line can always be shifted because of an unrecognised blend. Since even the 2586 \AA\ transition is incompatible with the position of the 2374 \AA\ feature, it is neglected. Using all remaining transitions gives $\Delta\alpha/\alpha=(0.1\pm0.1)\cdot10^{-5}$ ($\Delta\alpha/\alpha=(-0.1\pm0.1)\cdot10^{-5}$, \(\chi^2_{6}/\chi^2_{5}=1.0\)), while only the 2586 \AA\ transition would give $\Delta\alpha/\alpha=(-0.2\pm0.1)\cdot10^{-5}$ ($\Delta\alpha/\alpha=(-0.3\pm0.1)\cdot10^{-5}$). 

\subsection{PKS2126-158}
The quasar \object{PKS2126-158} at $z_\mathrm{em}=3.28$ has a strong system at $z=2.77$ composed of two separate parts, at \(z_1=2.7674\) and \(z_2=2.7684\), that can be used for this analysis. Because of the high redshift of the system, the 2586 \AA\ and the 2600 \AA\ transitions are not available. The 1608 \AA\ system is blended with the 1550 \AA\ transition of a \ion{C}{iv} feature. To avoid shifts due to the blend and because of a strong noise peak in the same part of the absorber in the 2374 \AA\ transition, only the second part of the system, which is apparently unaffected, is used. It consists of at least eight components. The line shift analysis shows a strong shift of the 2374 \AA\ transition (Fig. \ref{fig:reg_data}), which cannot be accounted for. Using all four of the remaining lines, we get $\Delta\alpha/\alpha=(1.0\pm0.3)\cdot10^{-5}$ ($\Delta\alpha/\alpha=(0.4\pm0.3)\cdot10^{-5}$,  \(\chi^2_{9}/\chi^2_{8}=1.0\)). To avoid effects by the heavy saturation of the 2382 \AA\ feature, the best result is given by a comparison of the 2344 \AA\ with the 1608 \AA\ lines, namely $\Delta\alpha/\alpha=(-0.2\pm0.3)\cdot10^{-5}$ ($\Delta\alpha/\alpha=(-0.2\pm0.3)\cdot10^{-5}$).

\subsection{Q0002-422}
\begin{figure}
\resizebox{0.45\columnwidth}{!}{\input{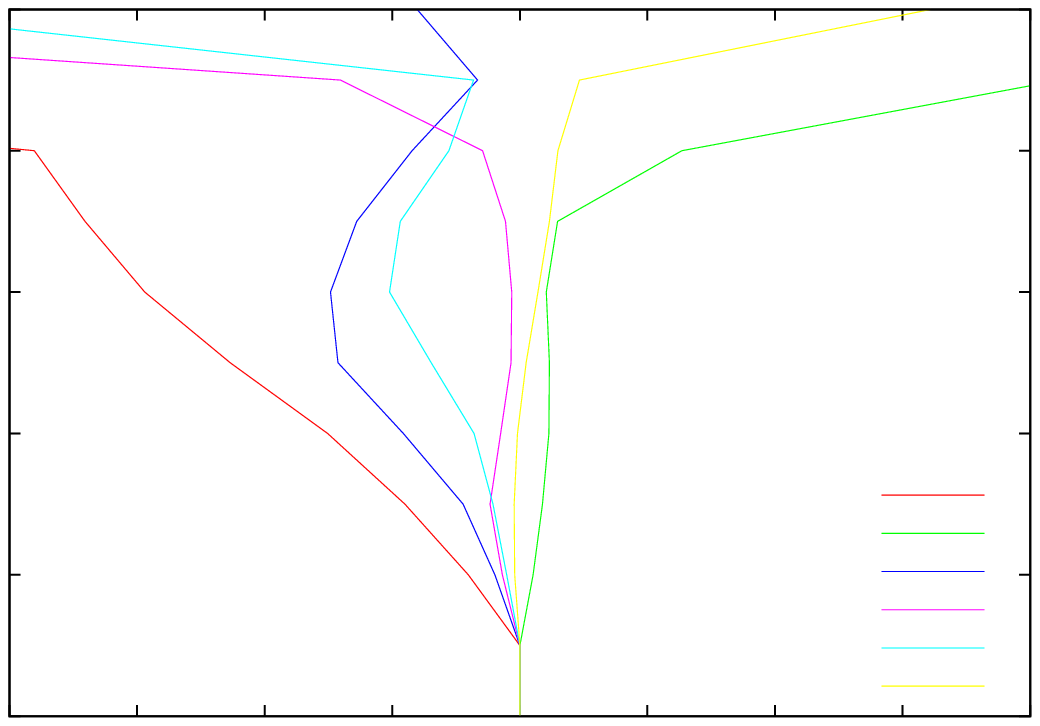}}
\resizebox{0.45\columnwidth}{!}{\input{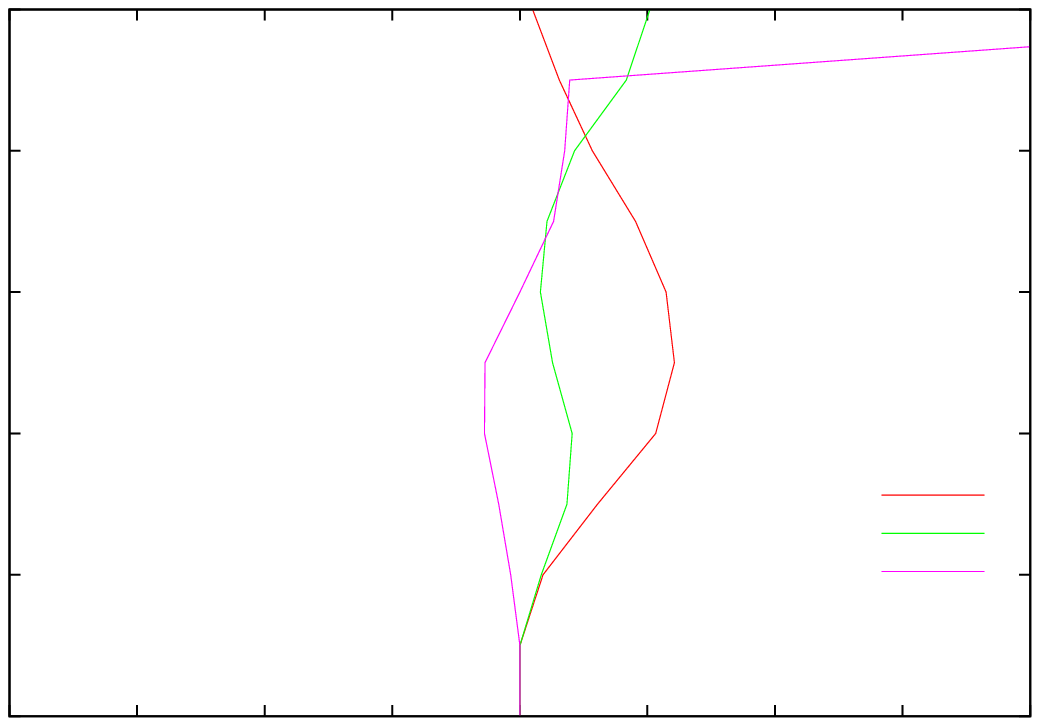}} \\
\resizebox{0.45\columnwidth}{!}{\input{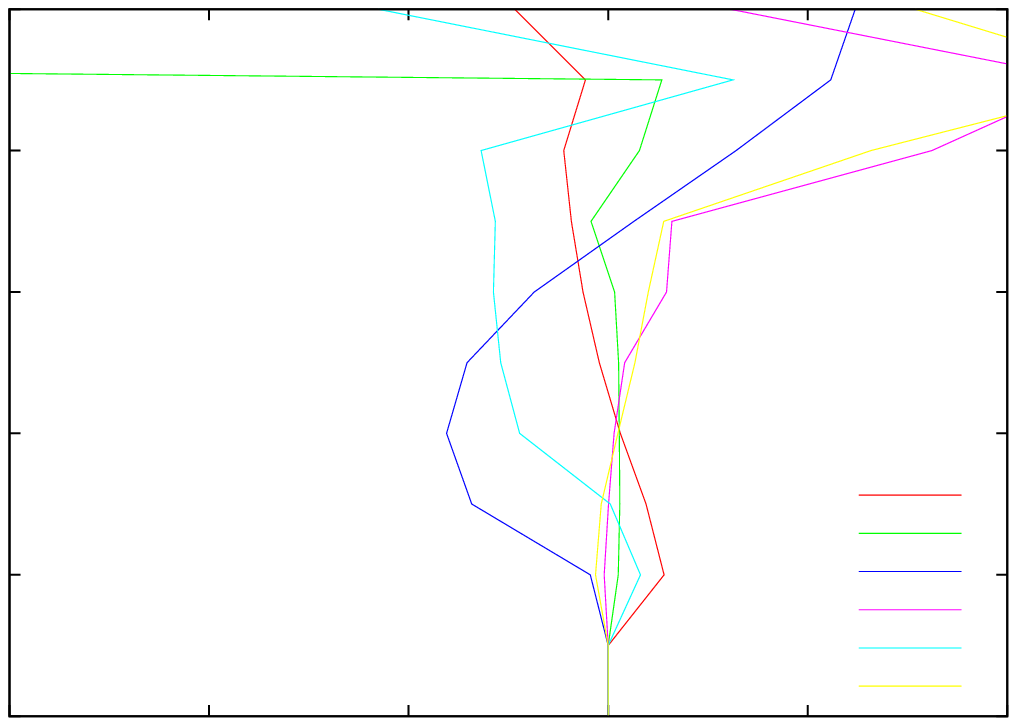}} 
\resizebox{0.45\columnwidth}{!}{\input{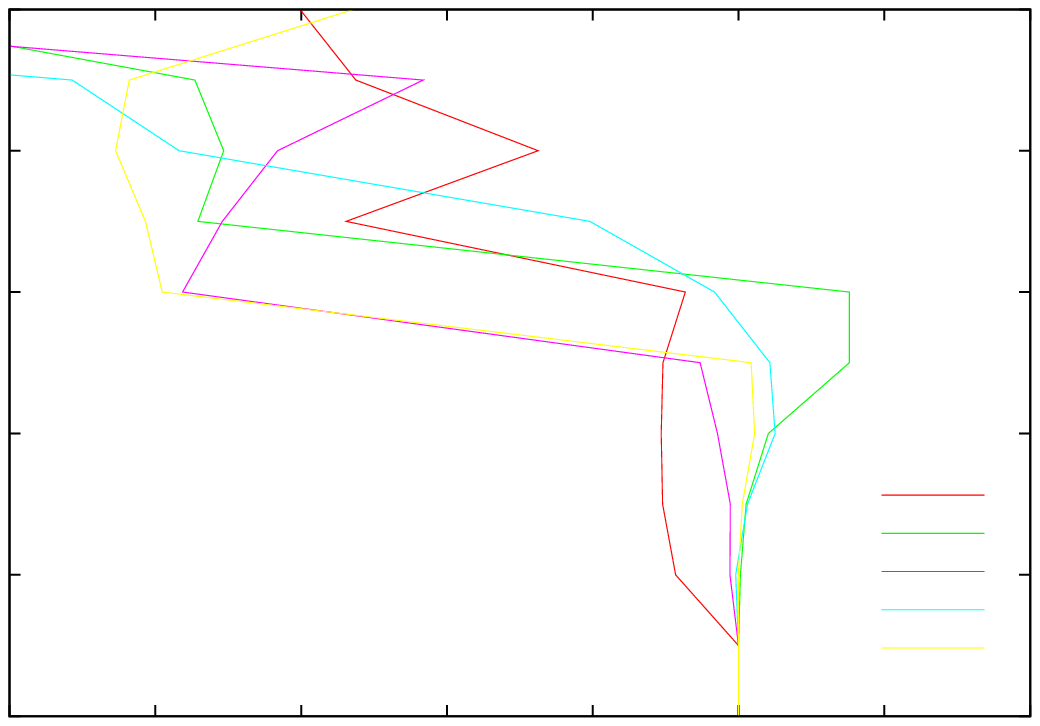}} \\
\resizebox{0.45\columnwidth}{!}{\input{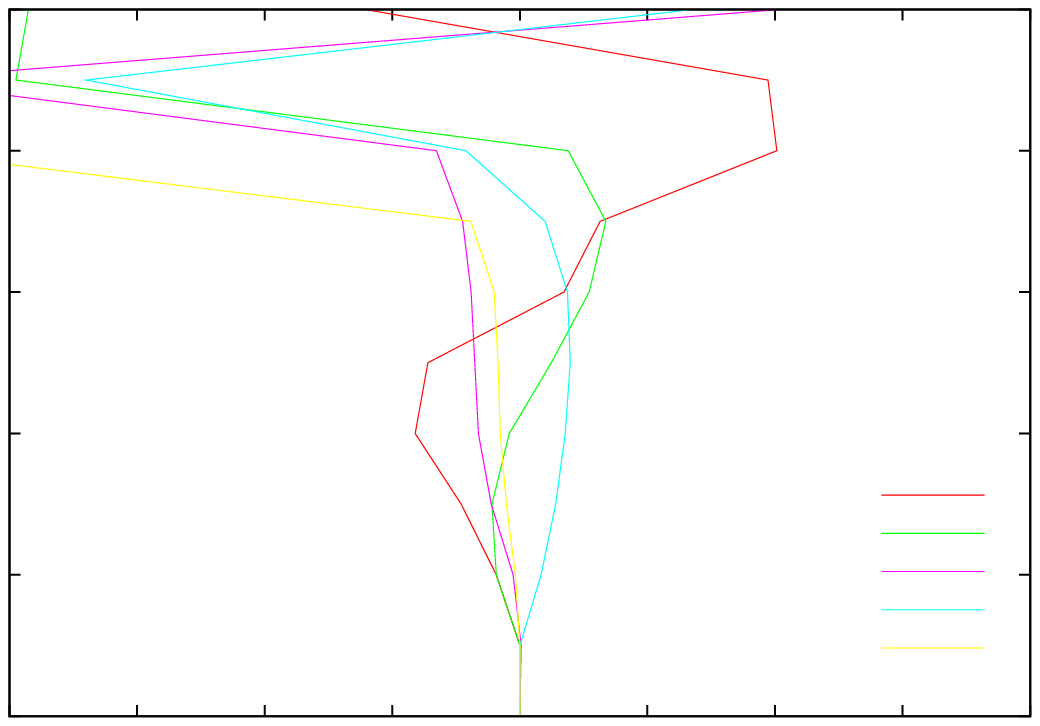}} 
\resizebox{0.45\columnwidth}{!}{\input{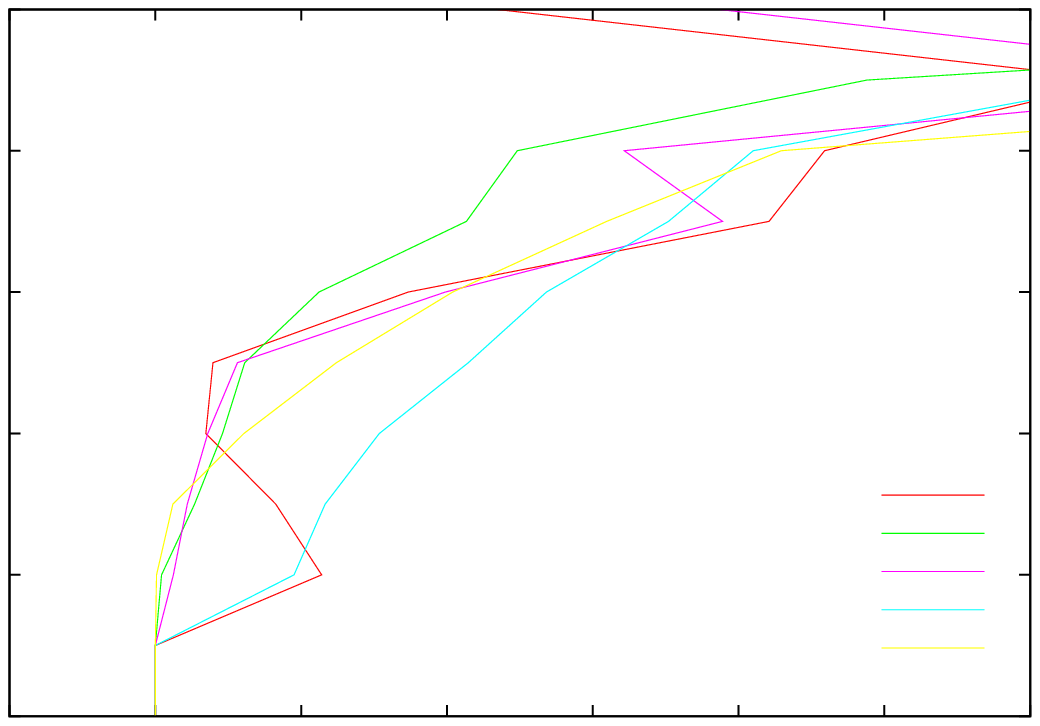}} \\
\resizebox{0.45\columnwidth}{!}{\input{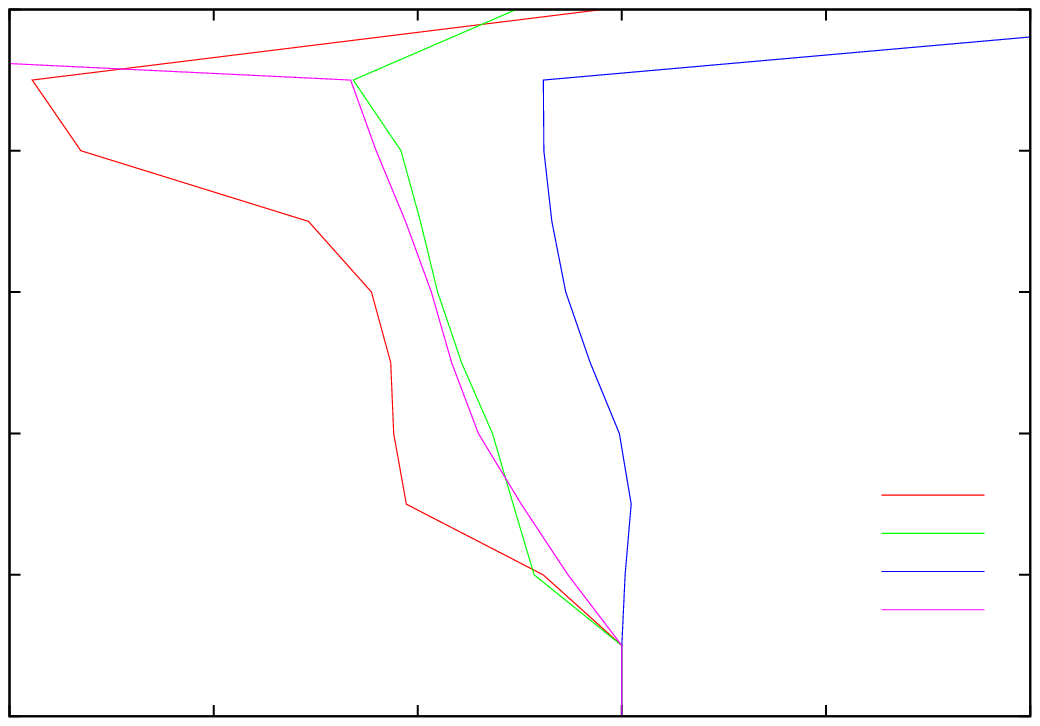}} 
\resizebox{0.45\columnwidth}{!}{\input{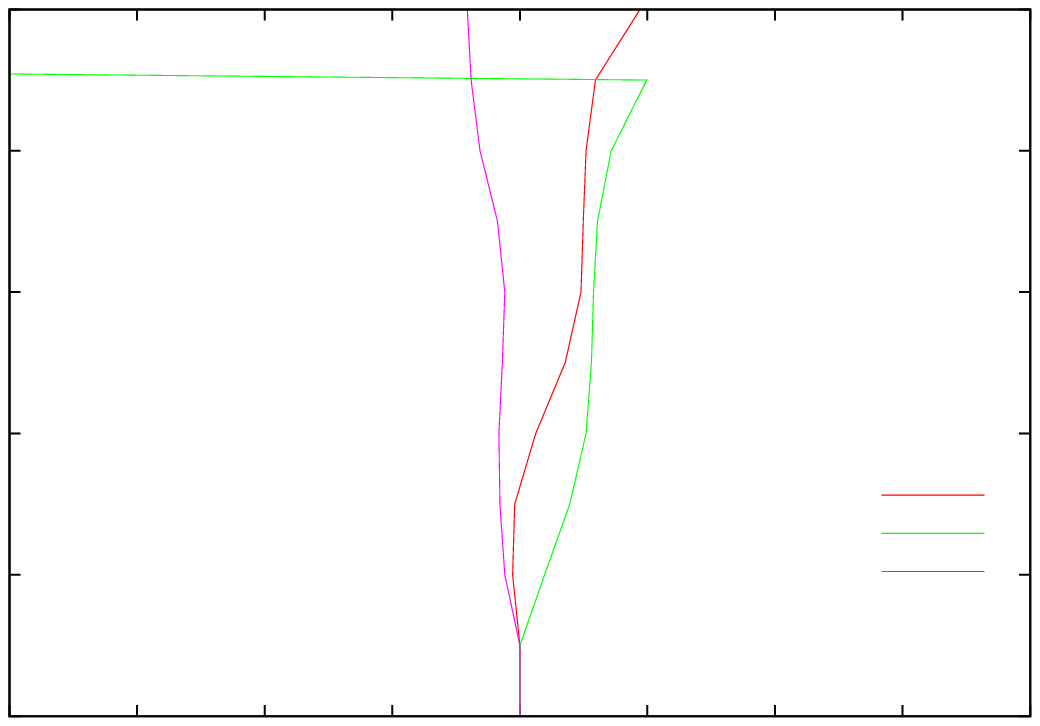}} \\
\caption{Bisectors of isolated \ion{Fe}{ii} lines. The lines are parametrized from their centres (0) up to the continuum (1) to allow a comparison between different transitions. Red: \ion{Fe}{ii} 1608\AA, green: \ion{Fe}{ii} 2344\AA, blue: \ion{Fe}{ii} 2374\AA, purple: \ion{Fe}{ii} 2382\AA, cyan: \ion{Fe}{ii} 2586\AA, yellow: \ion{Fe}{ii} 2600\AA.}
\label{fig:bis}
\end{figure}

The quasar \object{Q0002-422} at an emission redshift of $z_\mathrm{em}=2.77$ has two high redshift systems with a visible 1608 \AA\ line. The first system ($z=2.17$) seems to be a simple blend of two lines. The $z-f\lambda_0$ diagram suggests a slight correlation of position and transition strength (Fig. \ref{fig:reg_data}). The bisector shows a difference in line shape that might be created by noise, since the general slope is similar for all lines (Fig. \ref{fig:bis}g). The 2586 \AA\ and 2600 \AA\ transitions are not available. Using just the 2374 \AA\ transition gives $\Delta\alpha/\alpha=(0.1\pm1.4)\cdot10^{-5}$ ($\Delta\alpha/\alpha=(0.1\pm1.4)\cdot10^{-5}$,  \(\chi^2_{3}/\chi^2_{2}=1.0\)), while all available transitions would give $\Delta\alpha/\alpha=(1.0\pm1.0)\cdot10^{-5}$ ($\Delta\alpha/\alpha=(0.9\pm1.0)\cdot10^{-5}$).\\

The second system at z=$2.30$ is divided into three parts at \(z_1=2.3006\), \(z_2=2.3008\), and \(z_3=2.3015\). Only three transitions (1608 \AA, 2344 \AA, and 2382 \AA) are available for the whole system. A comparison with these stronger transitions always holds the danger of shifts due to saturation effects. The first part of the system, at \(z_1=2.3006\),\ consists of a single weak line. The bisector of this feature shows no strong asymmetry for all three transitions (Fig. \ref{fig:bis}h). The 2382 \AA\ transition shows a position shift in comparison with the other two transitions. The $z-f\lambda_0$ diagram shows a strong correlation of position and transition strength (Fig.\ \ref{fig:reg_data}). Using just the 2344 \AA\ transition gives $\Delta\alpha/\alpha=(0.1\pm1.4)\cdot10^{-5}$ ( $\Delta\alpha/\alpha=(0.1\pm1.4)\cdot10^{-5}$, \(\chi^2_{2}/\chi^2_{1}=1.0\), while all available transitions would yield $\Delta\alpha/\alpha=(1.3\pm1.2)\cdot10^{-5}$ ($\Delta\alpha/\alpha=(1.3\pm1.2)\cdot10^{-5}$). \\
The second part of the system, at \(z_1=2.3008\),\ is a weak and close blend of at least two components. The 1608 \AA\ feature barely exceeds the noise, making a reliable position estimation difficult; nevertheless, trying it gives a nearly perfect correlation of position shift and sensitivity coefficient (Fig. \ref{fig:reg_data}), suggesting a variation of $\Delta\alpha/\alpha=(5.0\pm1.9)\cdot10^{-5}$ ($\Delta\alpha/\alpha=(5.1\pm1.9)\cdot10^{-5}$, \(\chi^2_{3}/\chi^2_{2}=1.0\)) for the 2344 \AA\ transition. Using all three transitions gives $\Delta\alpha/\alpha=(5.3\pm1.7)\cdot10^{-5}$ ($\Delta\alpha/\alpha=(5.4\pm1.7)\cdot10^{-5}$).  \\
The third part of the system, at \(z_1=2.3015\),\ consists of a blend of at least ten partly saturated components. As in the second part of the system, there is a strong correlation of the $z-q$ diagram, however with a lower magnitude. An $\alpha$ variation of $\Delta\alpha/\alpha=(0.5\pm0.2)\cdot10^{-5}$ ($\Delta\alpha/\alpha=(0.7\pm0.2)\cdot10^{-5}$, \(\chi^2_{11}/\chi^2_{10}=1.0\)) would be measured using all three available transitions. The lack of available transitions makes a determination of possible systematic effects difficult. Because of the saturation of several components in the stronger transitions, some position shift would be expected and is supported by the $z-f\lambda_0$ correlation (Fig. \ref{fig:reg_data}). Using only the 2344 \AA\ transition also gives $\Delta\alpha/\alpha=(0.5\pm0.2)\cdot10^{-5}$ ($\Delta\alpha/\alpha=(0.5\pm0.2)\cdot10^{-5}$).\\

To summarize the results of Sect.\ \ref{sect:observations}, Table \ref{tab:results} shows the apparent \(\alpha\) variation of all studied systems. For six of them, marked bad, the \ion{Fe}{ii} 1608 \AA\ line is not usable, as shown above. For the remaining ten systems, five of which have a usable \ion{Fe}{ii} 2374 \AA\ line, we find a mean apparent variation of \(\Delta\alpha/\alpha = (0.1\pm0.8)\cdot10^{-5}\). The average is 
found without weights because the main errors are expected to be systematic with an unknown distribution. Using all available transitions in all systems including those labelled bad with a regression analysis would result in \(\Delta\alpha/\alpha = (2.1\pm2.0)\cdot10^{-5}\).

\begin{table}
\caption{Results of the \(\alpha\) variation analyses.}
 \begin{tabular}{lcccl}
  QSO & z & $\Delta\alpha/\alpha_\mathrm{all}$ [$10^{-5}$] & $\Delta\alpha/\alpha_\mathrm{weak}$ [$10^{-5}$] & Status \\
  \hline
  HE0001-2340 & 1.5864 & $3.8\pm0.6$ & $1.5\pm0.8$~ & bad \\
  HE0001-2340 & 2.1853 & $1.8\pm0.9$ & ~~$-0.7\pm1.0_{2344}$ & good \\
  HE0001-2340 & 2.1871 & $1.8\pm0.3$ & $1.6\pm0.4$~ & good \\
  HE1341-1020 & 1.9153 & $5.7\pm1.2$ & \quad$6.4\pm1.2_{2586}$ & bad \\
  HE1347-2457 & 1.4392 & $-1.7\pm0.1$~~ & ~~$-0.5\pm0.1_{2586}$ & bad\\
  HE2217-2812 & 1.6908 & $0.4\pm0.4$ & $0.6\pm0.5$~ & good \\
  HE2217-2812 & 1.6921 & $1.9\pm0.7$ & $0.5\pm0.7$~ & good \\
  PKS0237-23  & 1.6358 & $4.1\pm2.0$ & \quad$4.9\pm2.1_{2586}$ & bad \\
  PKS0237-23  & 1.6369 & $4.4\pm1.0$ & \quad$5.9\pm1.1_{2585}$ & bad \\
  PKS0237-23  & 1.6717 & $2.1\pm1.3$ & $-1.3\pm1.5$~~~~ & good \\
  PKS0237-23  & 1.6723 & $-0.1\pm0.1$~~ & ~~$-0.2\pm0.1_{2586}$ & good \\
  PKS2126-158 & 2.7684 & $1.0\pm0.3$ & ~~$-0.2\pm0.3_{2344}$ & good \\
  Q0002-422   & 2.1678 & $1.0\pm1.0$ & $0.1\pm1.4$~ & good \\
  Q0002-422   & 2.3006 & $1.3\pm1.2$ & \quad$0.1\pm1.4_{2344}$ & good  \\
  Q0002-422   & 2.3008 & $5.1\pm1.9$ & \quad$5.3\pm1.7_{2344}$ & bad \\
  Q0002-422   & 2.3015 & $0.5\pm0.2$ & \quad$0.5\pm0.2_{2344}$ & good\\
 \end{tabular}
\tablefoot{The third column shows the apparent \(\alpha\) variation when all available transitions were used, the fourth column when just the 2374 \AA\ transition is used (if not available the transition used is given as subscript).}
\label{tab:results}
\end{table}

\section{Results and Discussion}
\label{sect:discussion}

Our simulations in Sect.\ \ref{sect:simulations} and the application of our methods for detecting line asymmetries and shifts have shown that apart from wavelength calibration errors and blends, e.g. with sky lines, unresolved substructure can lead to significant errors in the \(\alpha\) variation measurements. Obviously one has to confine oneself to lines of equal strengths and sufficiently different \(Q\) values, i.e. use only \ion{Fe}{ii} 1608 \AA\ in combination with \ion{Fe}{ii} 2374 \AA.
However, even then unresolved substructure combined with noise, can lead to apparent shifts of up to \(\pm100\,\mathrm{m\,s^{-1}}\) even in the case of optically thin systems (cf. Fig.\ \ref{fig:hist_res_bl_135_3}).\\
In the systems analysed here, there was no case where an increase of the number of fitted components would change the results significantly. In the few cases where differences did occur, there was no way of judging which value was to be preferred. Simulations have shown that an increase in the number of fitted components does not necessarily give better results.
The presence of continuous velocity fields in the absorbing medium, creating a distortion of the line shapes, can cause velocity shifts of comparable amounts. In the data analysed, about 50\% of the observed systems showed signs of wavelength shifts possibly due to one of these mechanisms. While substructure could in principle be resolved with spectrographs of sufficiently high resolution, this is not the case for continuous velocity fields.\\
In some cases the bisector method, described in Sect.\ \ref{subsec:bis}, could be used to detect hidden line blends. The \(S/N\) of the available spectra was, however, too low for an efficient use of this method. With the next generation of data, e.g. ´´The UVES Large program for testing fundamental physics'', the bisector method can possibly be used to detect asymmetries that are caused by velocity substructure and hidden saturation effects. As several outliers in Table \ref{tab:results} show (e.g. the \(z=2.1871\) system in HE0001-2340), the main source of errors appears to be the wavelength calibration. This has already been shown by \citet{Molaro2008}, \citet{Griest2010}, \citet{Wendt2011}, and \citet{Agafonova2011}. Only a new spectrograph, optimized for higher wavelength accuracy, e.g. by using a frequency comb for wavelength calibration, will lead to real progress in the field.

\begin{acknowledgements}
Part of this work has been supported by the DFG Sonderforschungsbereich 676 Teilprojekt C4. \\
We wish to thank the referee for his detailed comments which helped to improve this paper.\\
Sergei Levshakov and Sebasti\'{a}n L\'{o}pez provided helpful comments.
\end{acknowledgements}

\bibliographystyle{aa}
\bibliography{meine}
\listofobjects
\appendix
\section{Plots of line fits}
\begin{figure}[h]
\includegraphics[bb=32 174 533 659,width=\columnwidth]{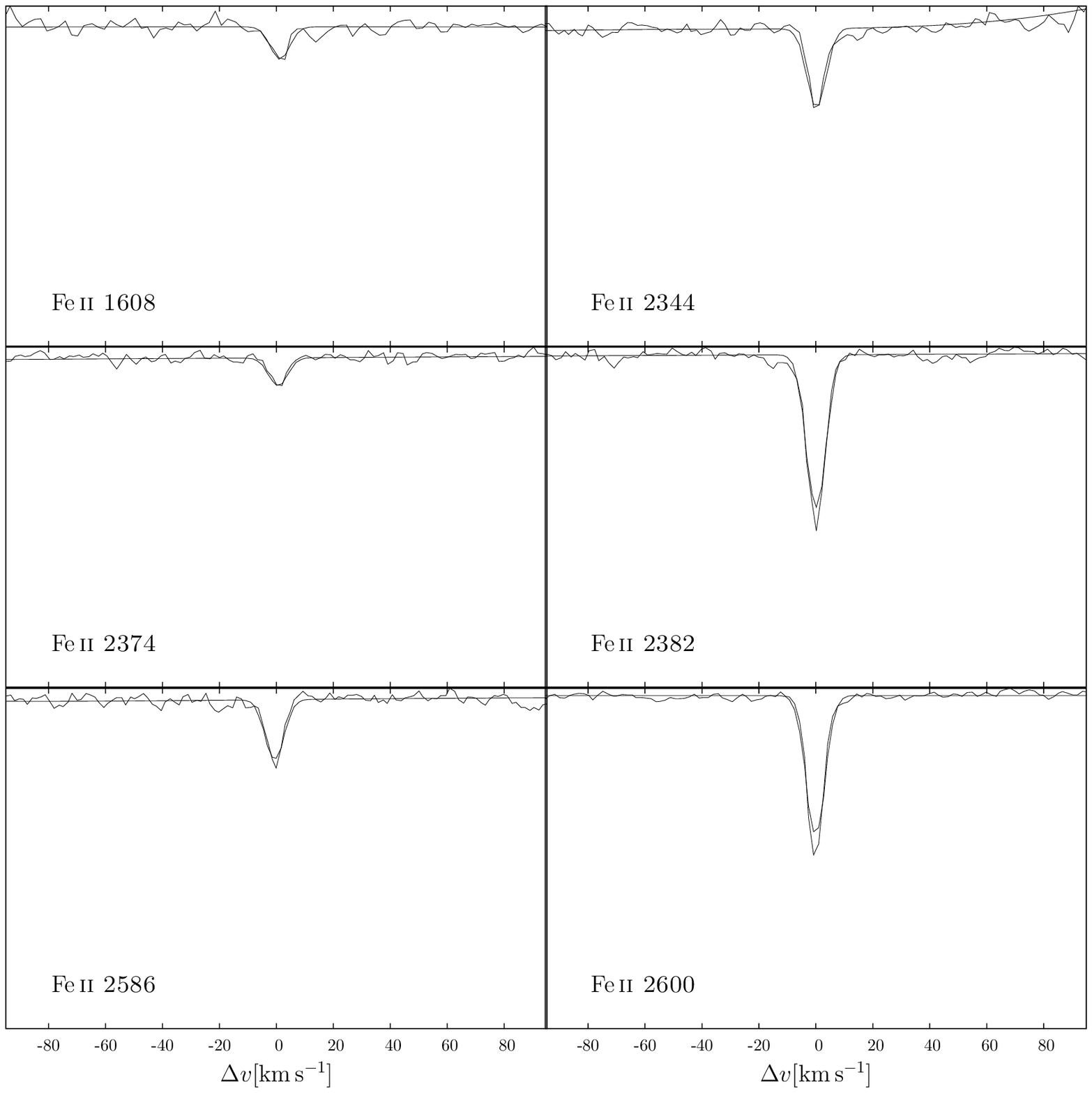}
\caption{HE0001-2340, \(z=1.5864\)}
\label{fig:fits_he0001_159}
\end{figure}
\begin{figure}[h]
\includegraphics[bb=32 238 533 570,width=\columnwidth]{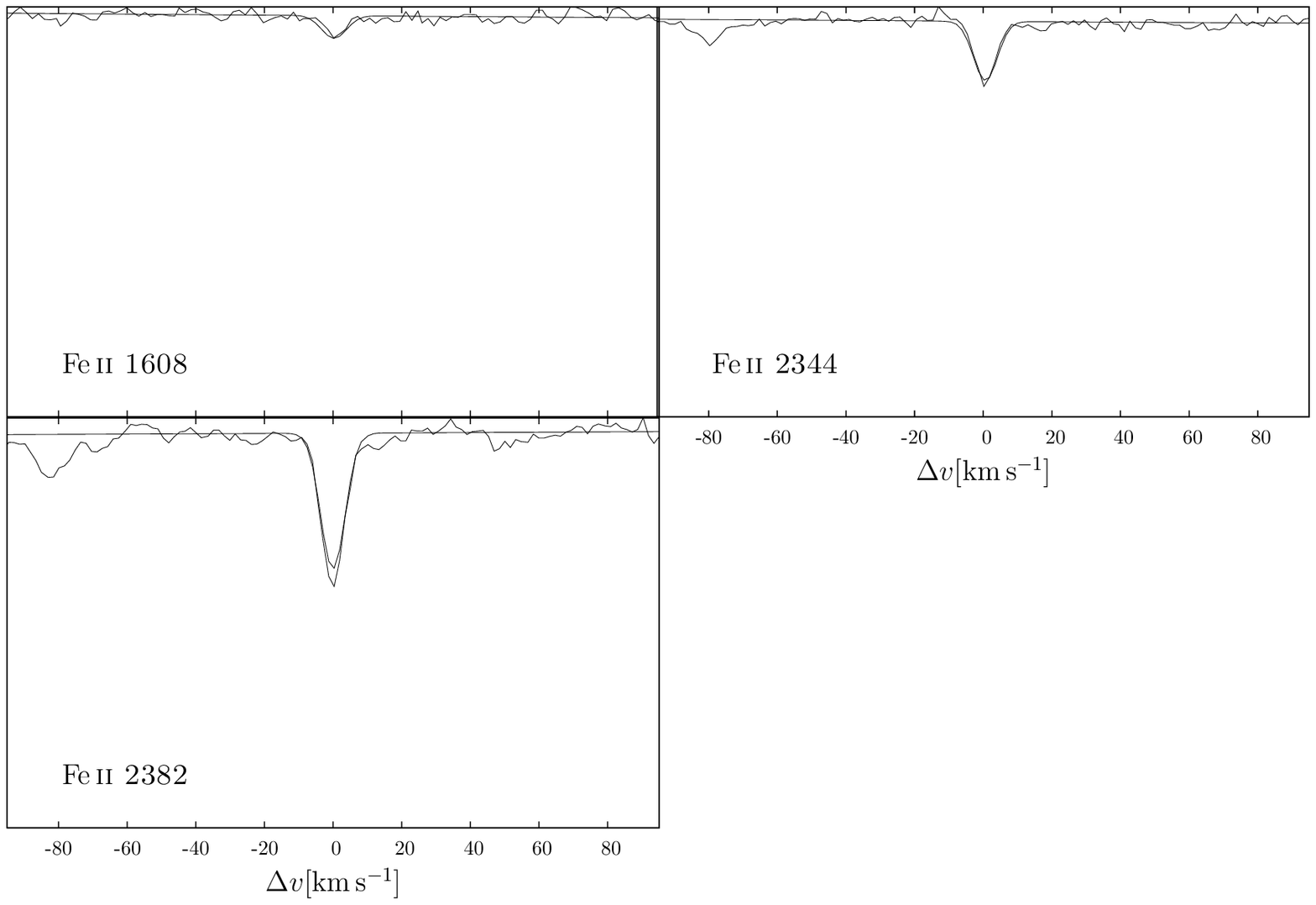}
\caption{HE0001-2340, \(z=2.1853\)}
\label{fig:fits_he0001_219}
\end{figure}
\begin{figure}
\includegraphics[bb=32 174 533 659,width=\columnwidth]{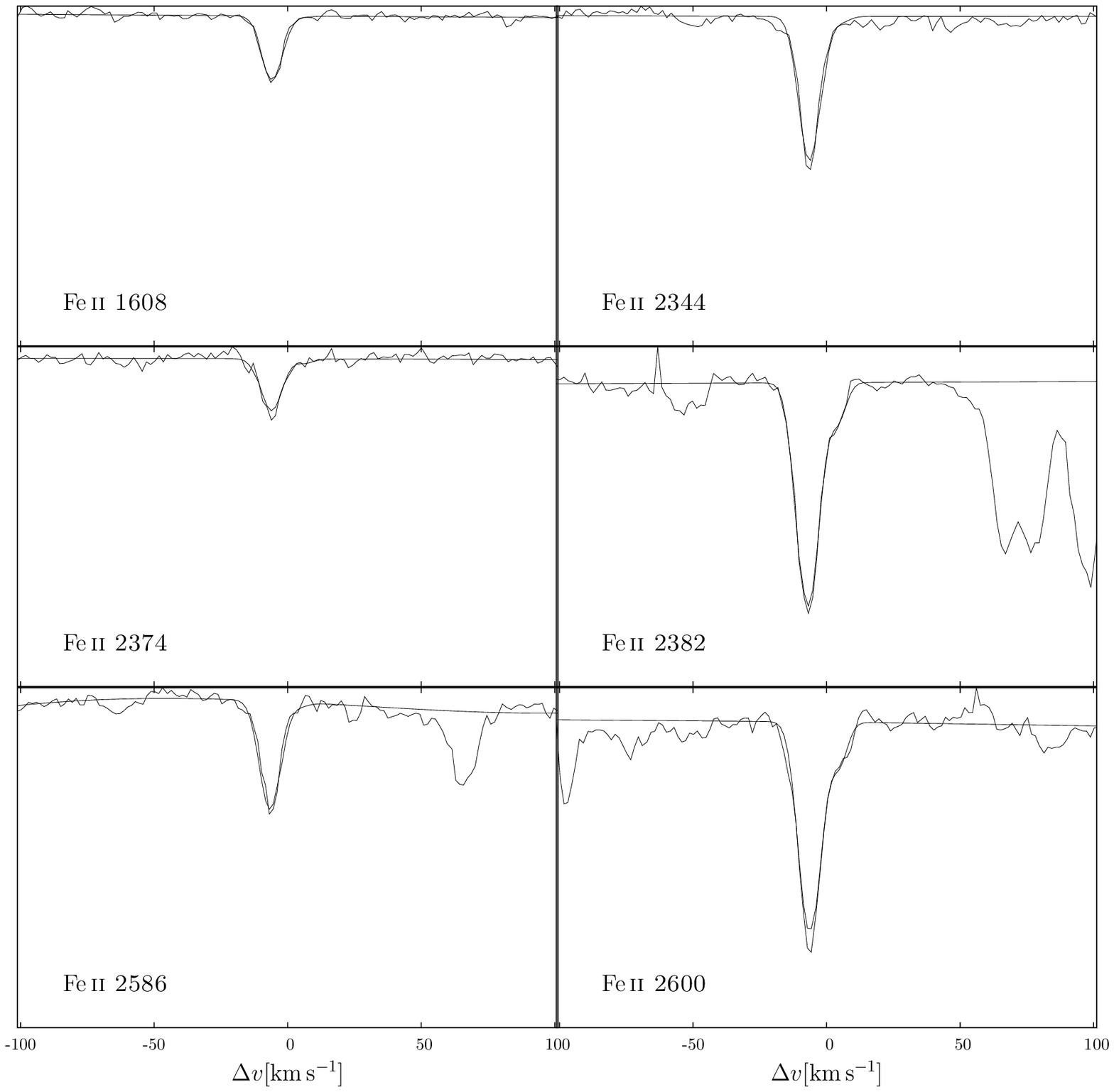}
\caption{HE0001-2340, \(z=2.1871\)}
\label{fig:fits_he0001_219_2}
\end{figure}
\begin{figure}
\includegraphics[bb=32 174 533 659,width=\columnwidth]{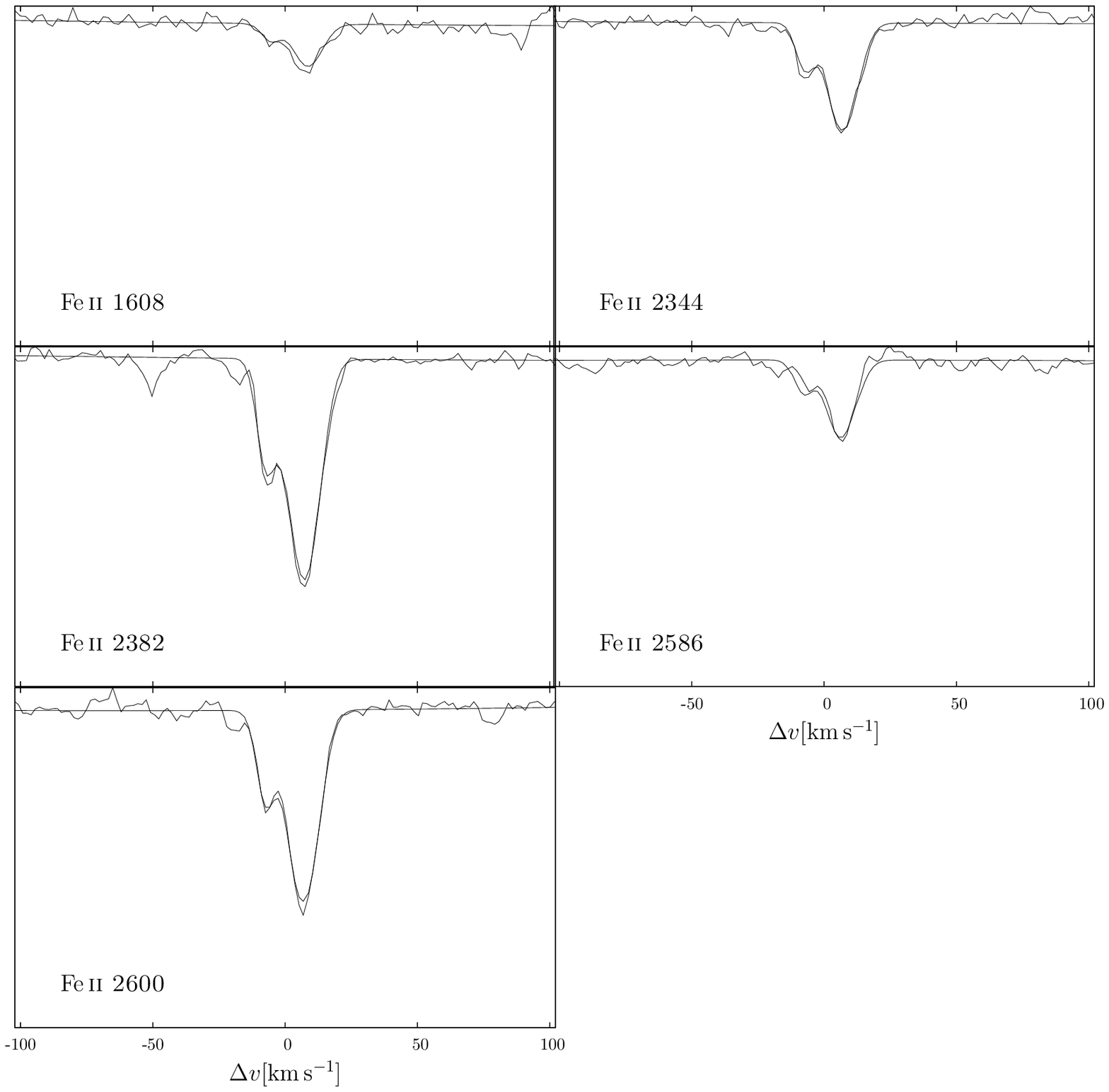}
\caption{HE1341-1020, \(z=1.9153\)}
\label{fig:fits_he1341_192}
\end{figure}
\begin{figure}
\includegraphics[bb=32 238 533 570,width=\columnwidth]{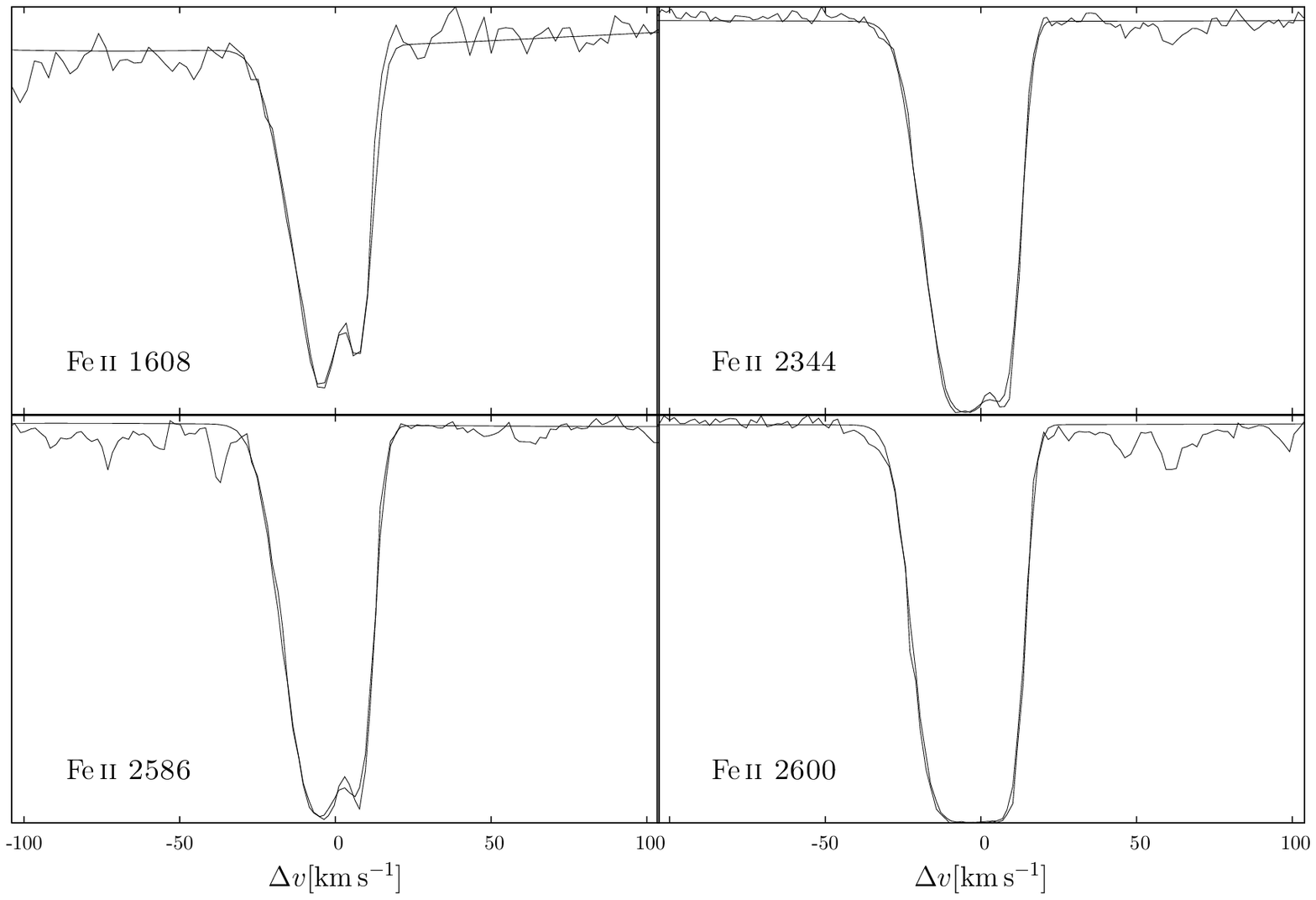}
\caption{HE1347-2457, \(z=1.4392\)}
\label{fig:fits_he1347_144}
\end{figure}
\begin{figure}
\includegraphics[bb=32 174 533 659,width=\columnwidth]{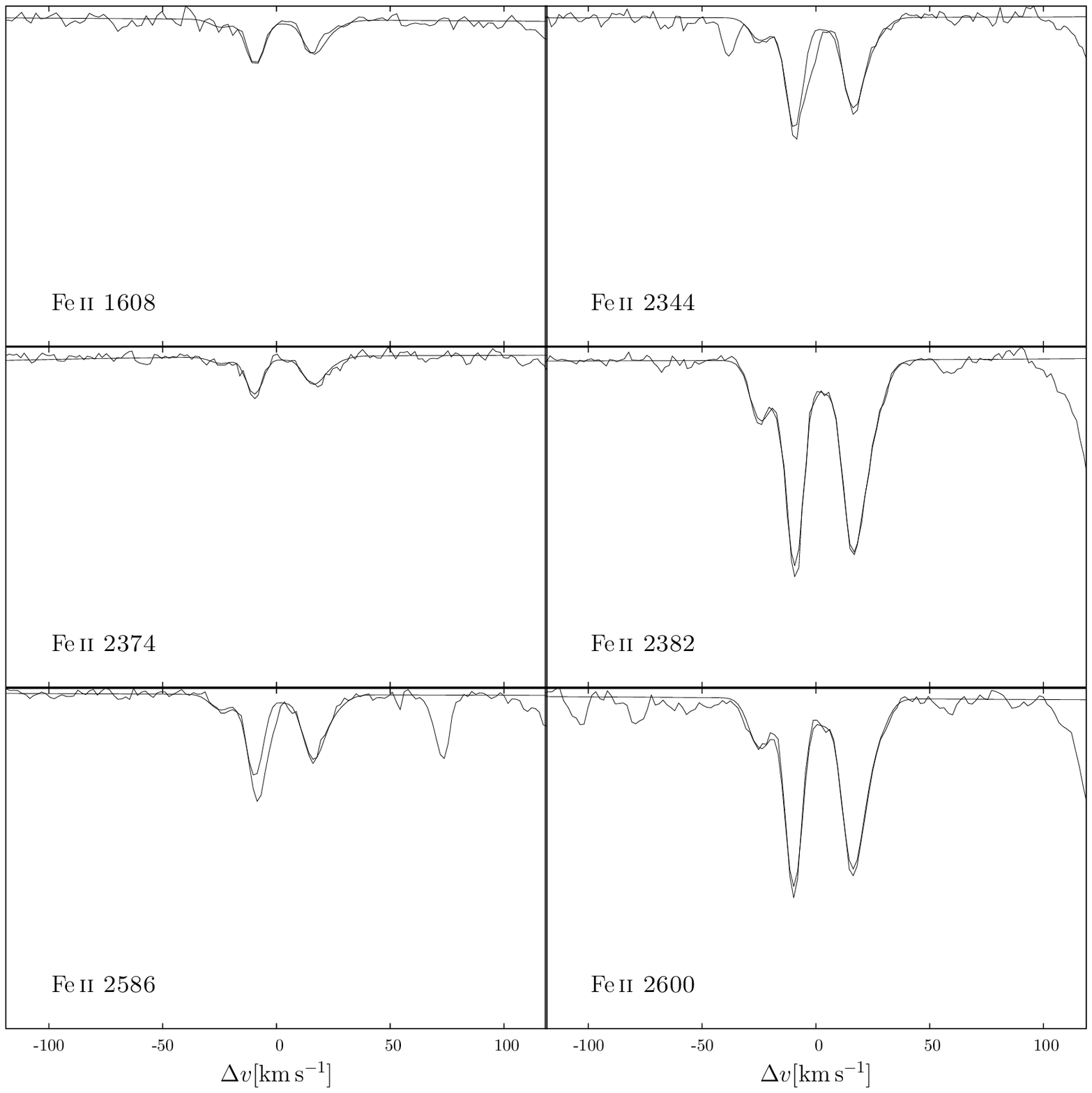}
\caption{HE2217-2818, \(z=1.6908\)}
\label{fig:fits_he2217_169}
\end{figure}
\begin{figure}
\includegraphics[bb=32 174 533 659,width=\columnwidth]{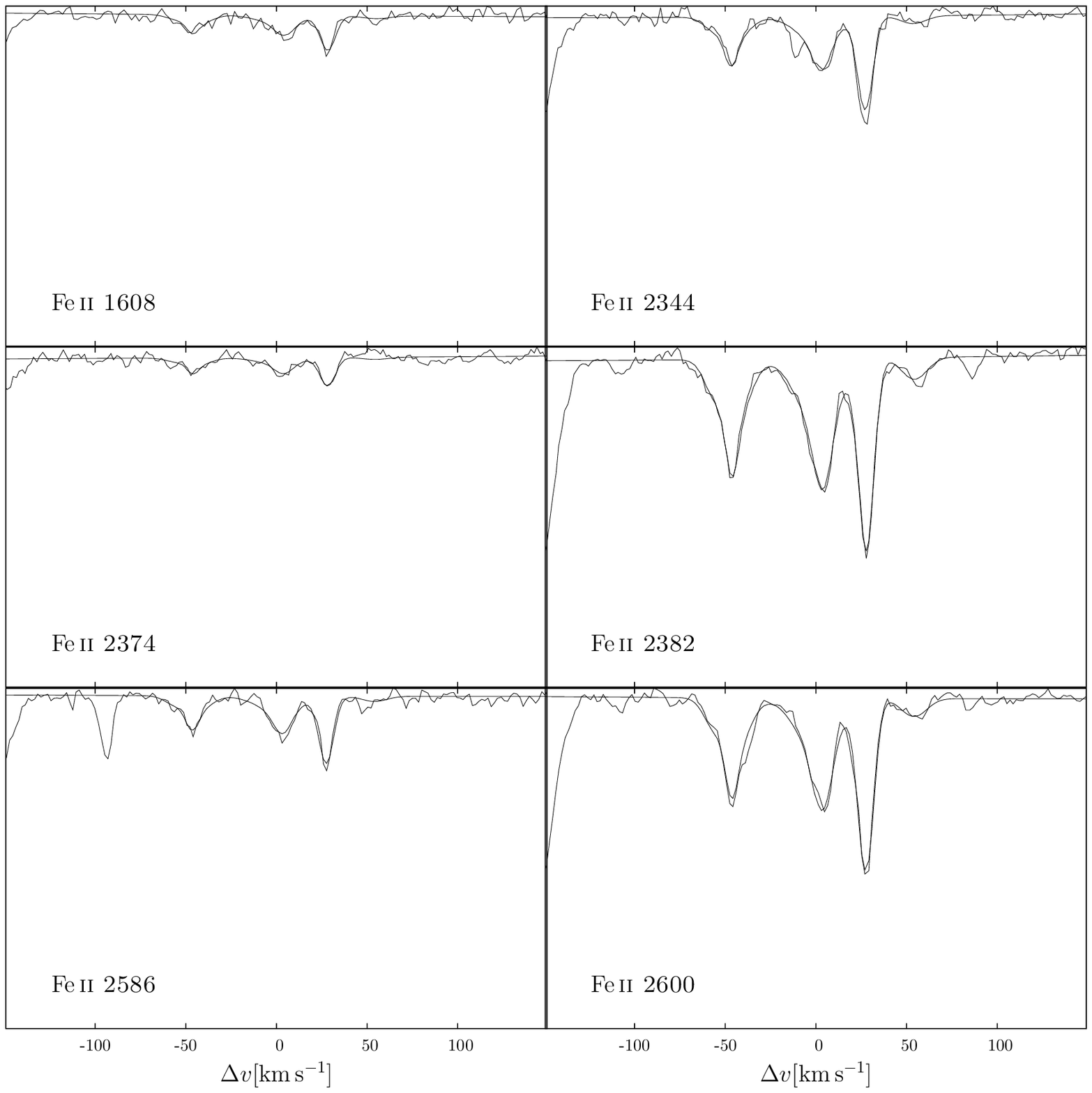}
\caption{HE2217-2818, \(z=1.6921\)}
\label{fig:fits_he2217_169_2}
\end{figure}
\begin{figure}
\includegraphics[bb=32 174 533 659,width=\columnwidth]{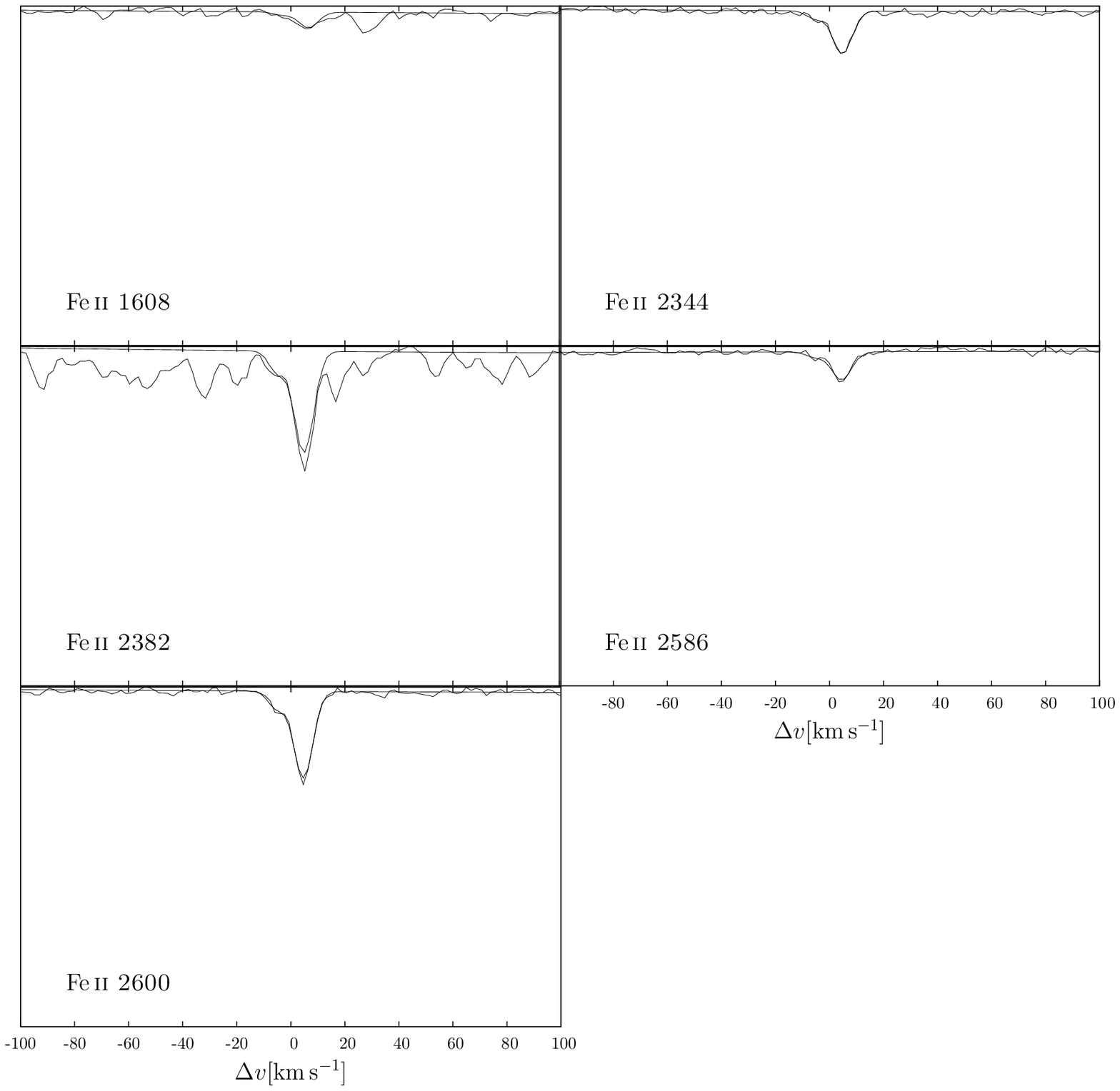}
\caption{PKS0237-23, \(z=1.6358\)}
\label{fig:fits_pks0237_164}
\end{figure}
\begin{figure}
\includegraphics[bb=32 174 533 659,width=\columnwidth]{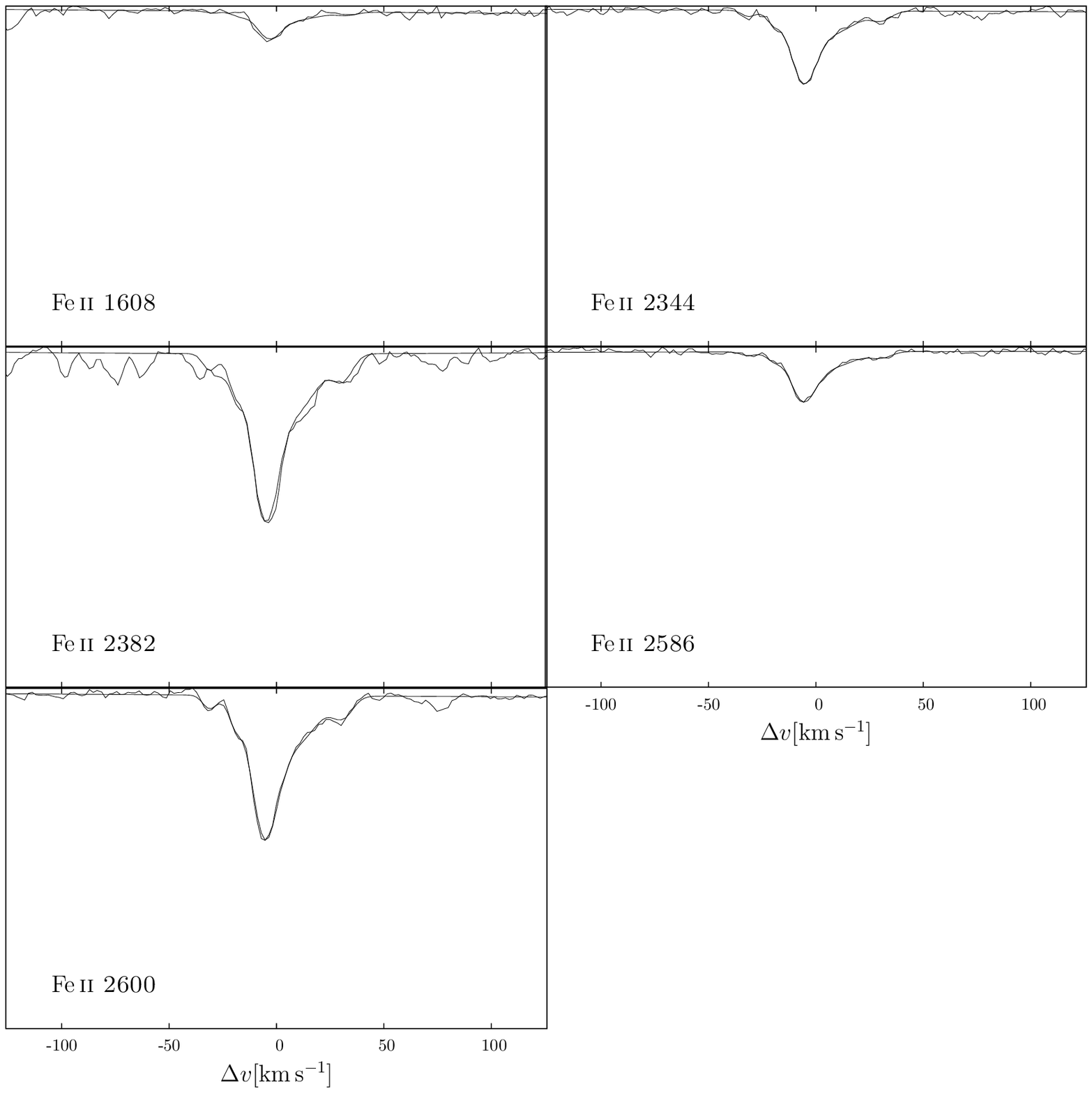}
\caption{PKS0237-23, \(z=1.6369\)}
\label{fig:fits_pks0237_164_2}
\end{figure}
\begin{figure}
\includegraphics[bb=32 174 533 659,width=\columnwidth]{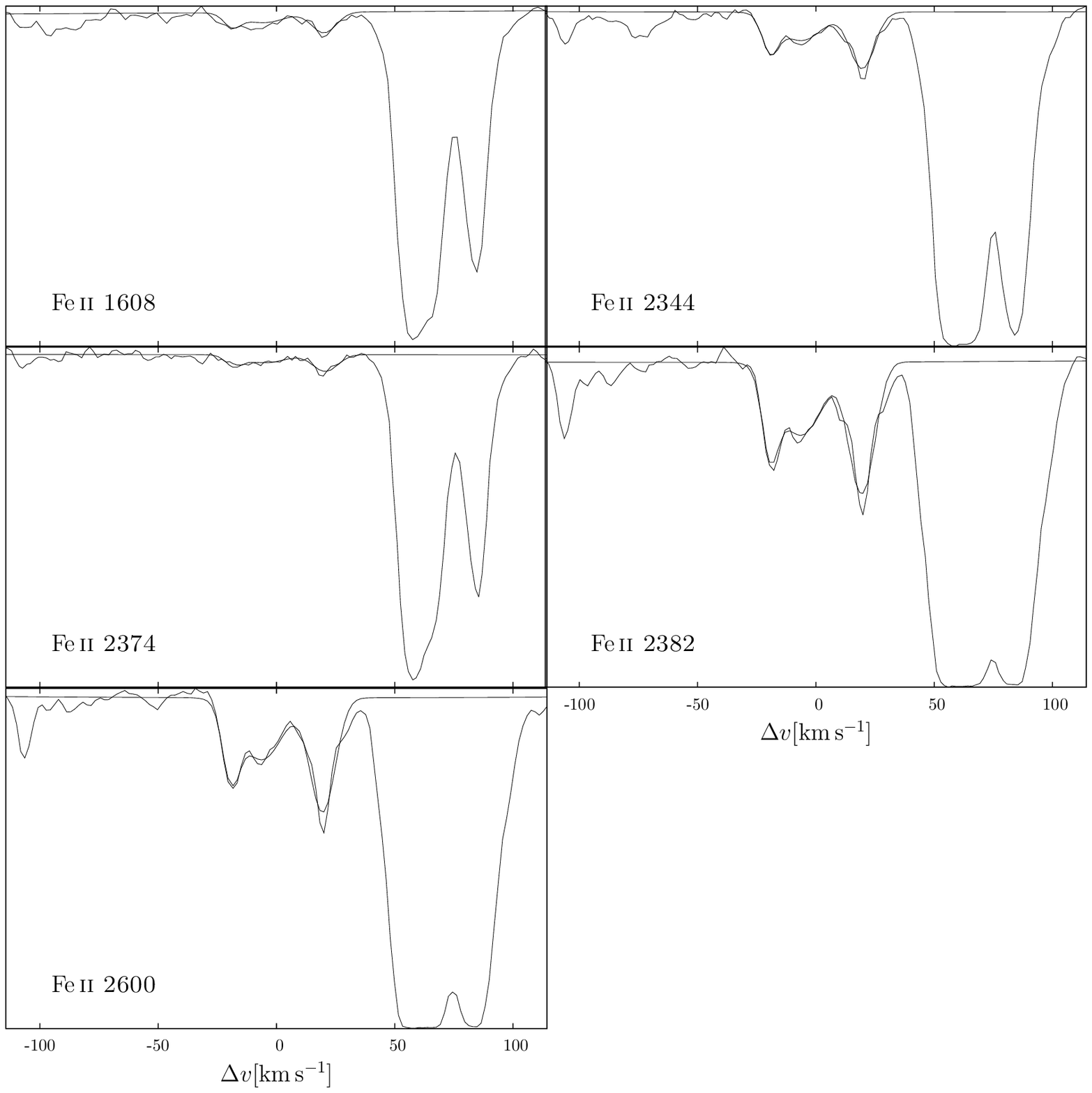}
\caption{PKS0237-23, \(z=1.6717\)}
\label{fig:fits_pks0237_167}
\end{figure}
\begin{figure}
\includegraphics[bb=32 174 533 659,width=\columnwidth]{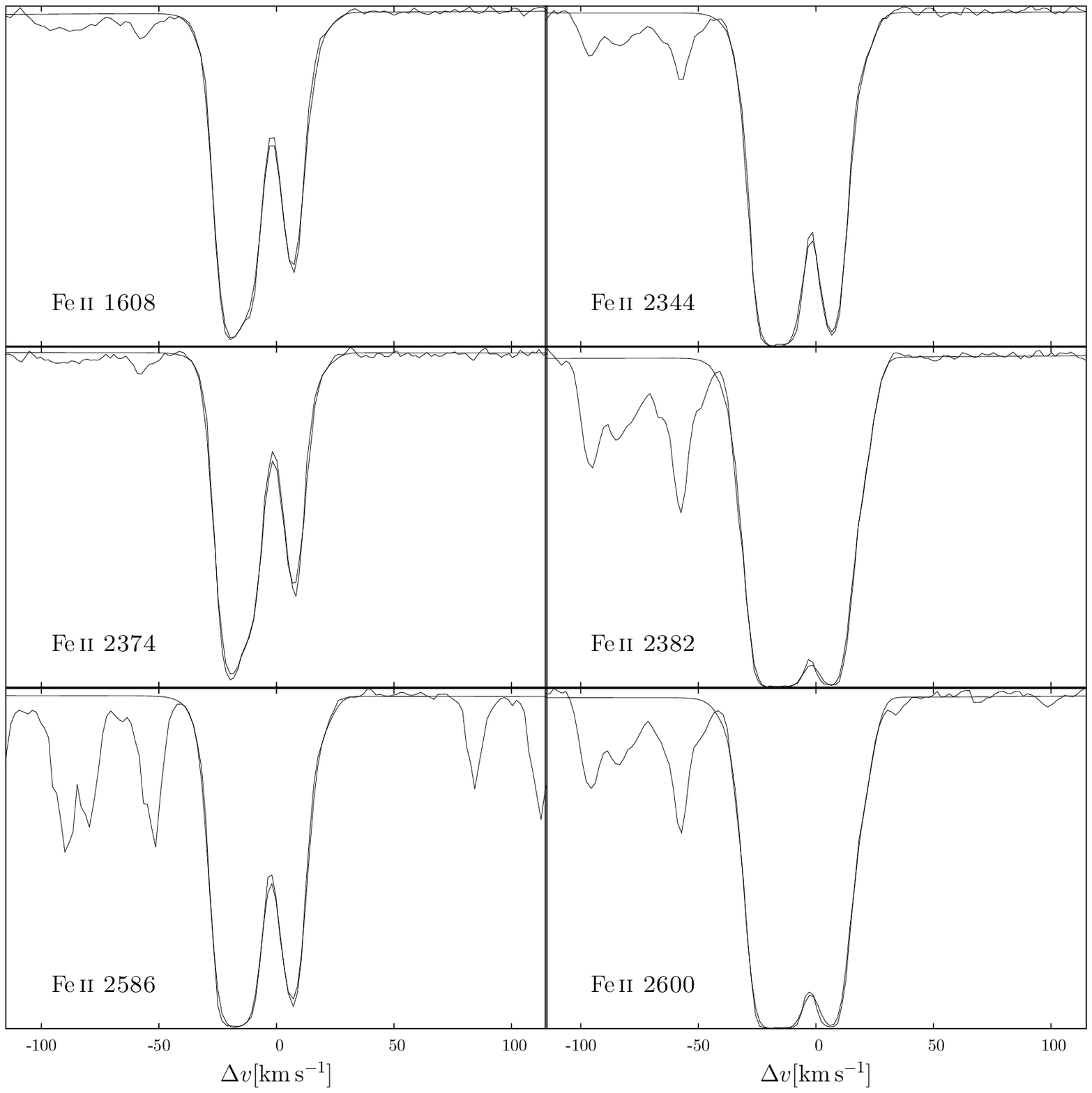}
\caption{PKS0237-23, \(z=1.6723\)}
\label{fig:fits_pks0237_167_2}
\end{figure}
\begin{figure}
\includegraphics[bb=32 238 533 570,width=\columnwidth]{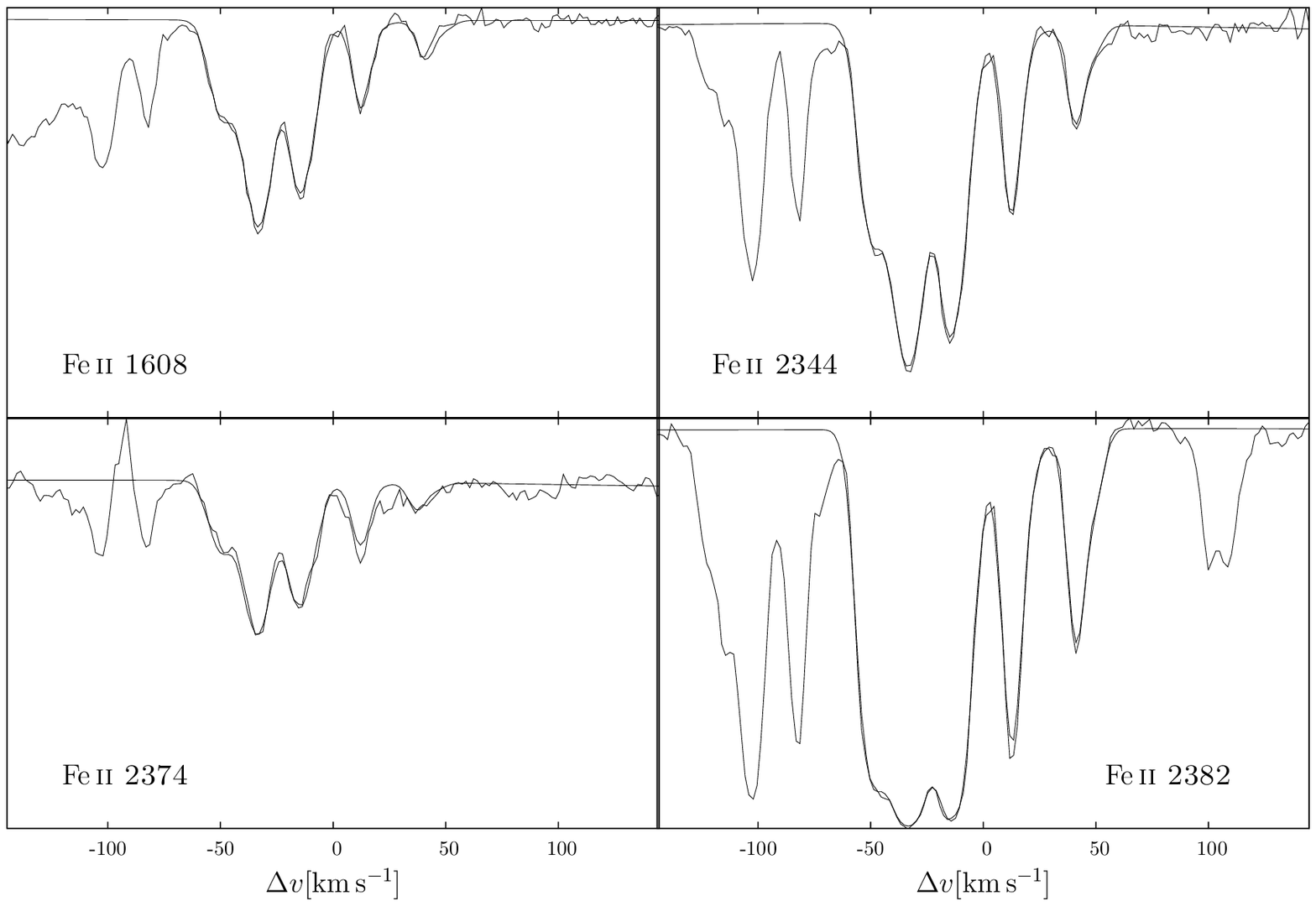}
\caption{PK2126-158, \(z=2.7684\)}
\label{fig:fits_pks2126_277}
\end{figure}
\begin{figure}
\includegraphics[bb=32 238 533 570,width=\columnwidth]{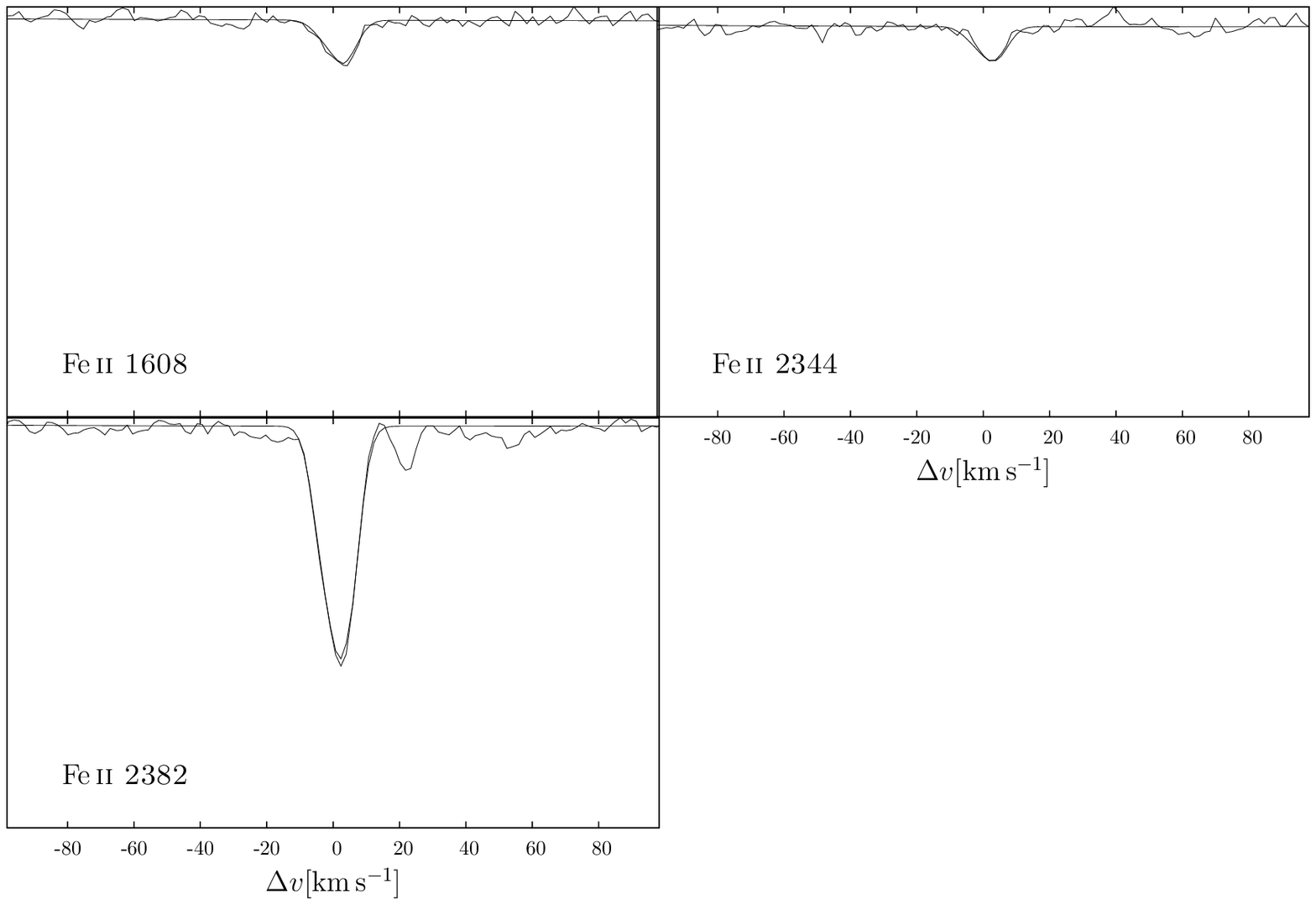}
\caption{Q0002-422, \(z=2.1678\)}
\label{fig:fits_q0002_217}
\end{figure}
\begin{figure}
\includegraphics[bb=32 238 533 570,width=\columnwidth]{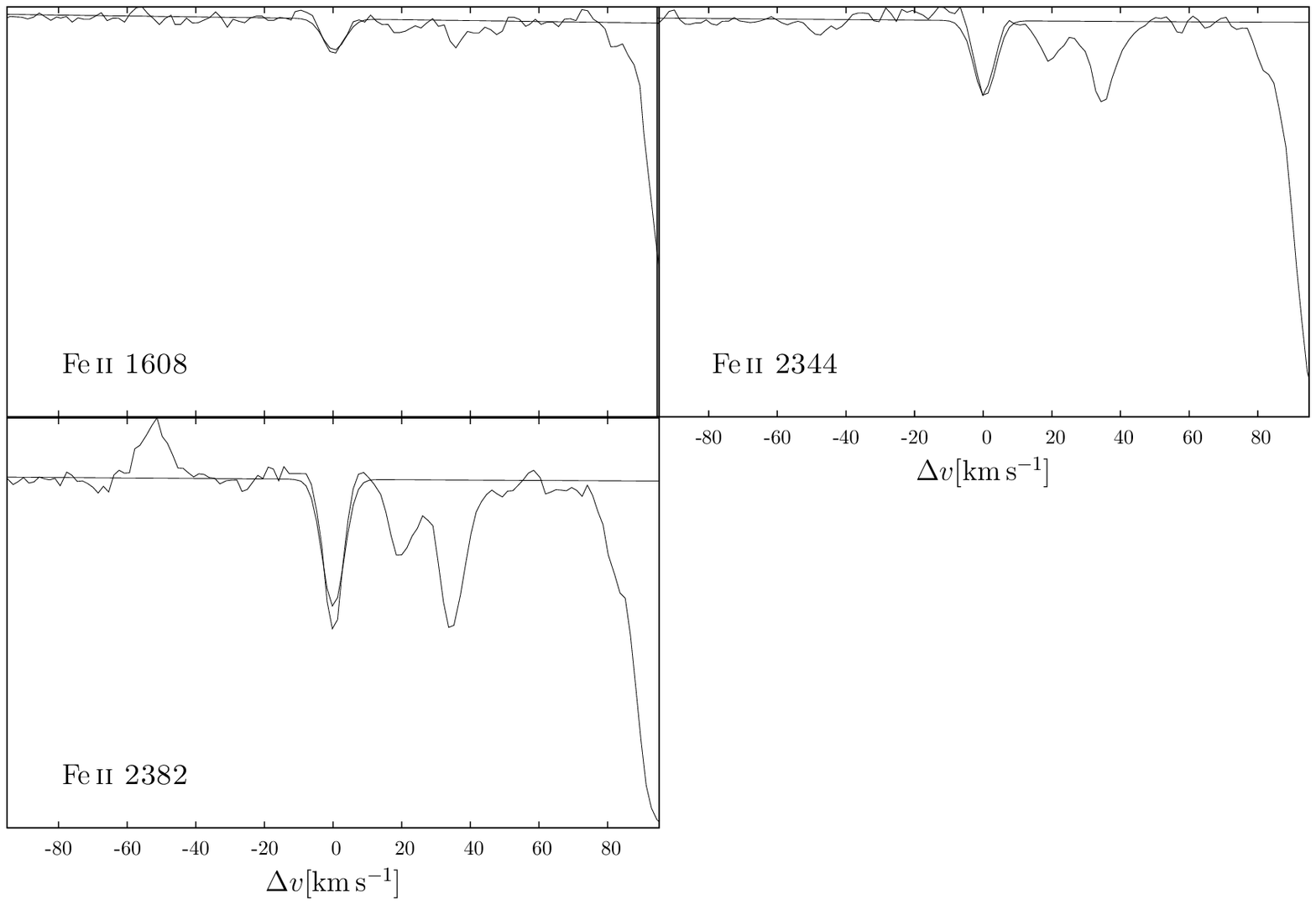}
\caption{Q0002-422, \(z=2.3006\)}
\label{fig:fits_q0002_230}
\end{figure}
\begin{figure}
\includegraphics[bb=32 238 533 570,width=\columnwidth]{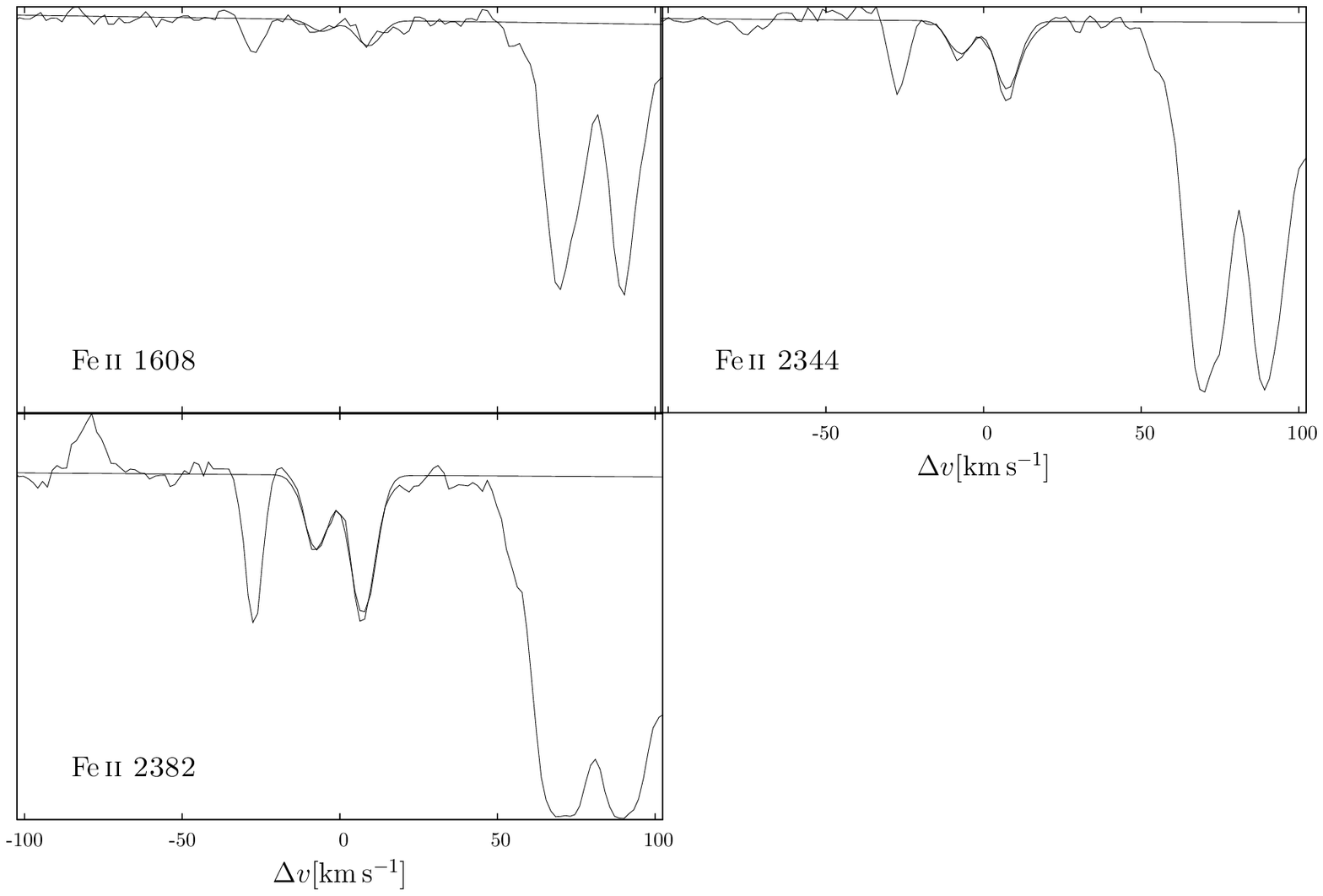}
\caption{Q0002-422, \(z=2.3008\)}
\label{fig:fits_q0002_230_2}
\end{figure}
\begin{figure}
\includegraphics[bb=32 238 533 570,width=\columnwidth]{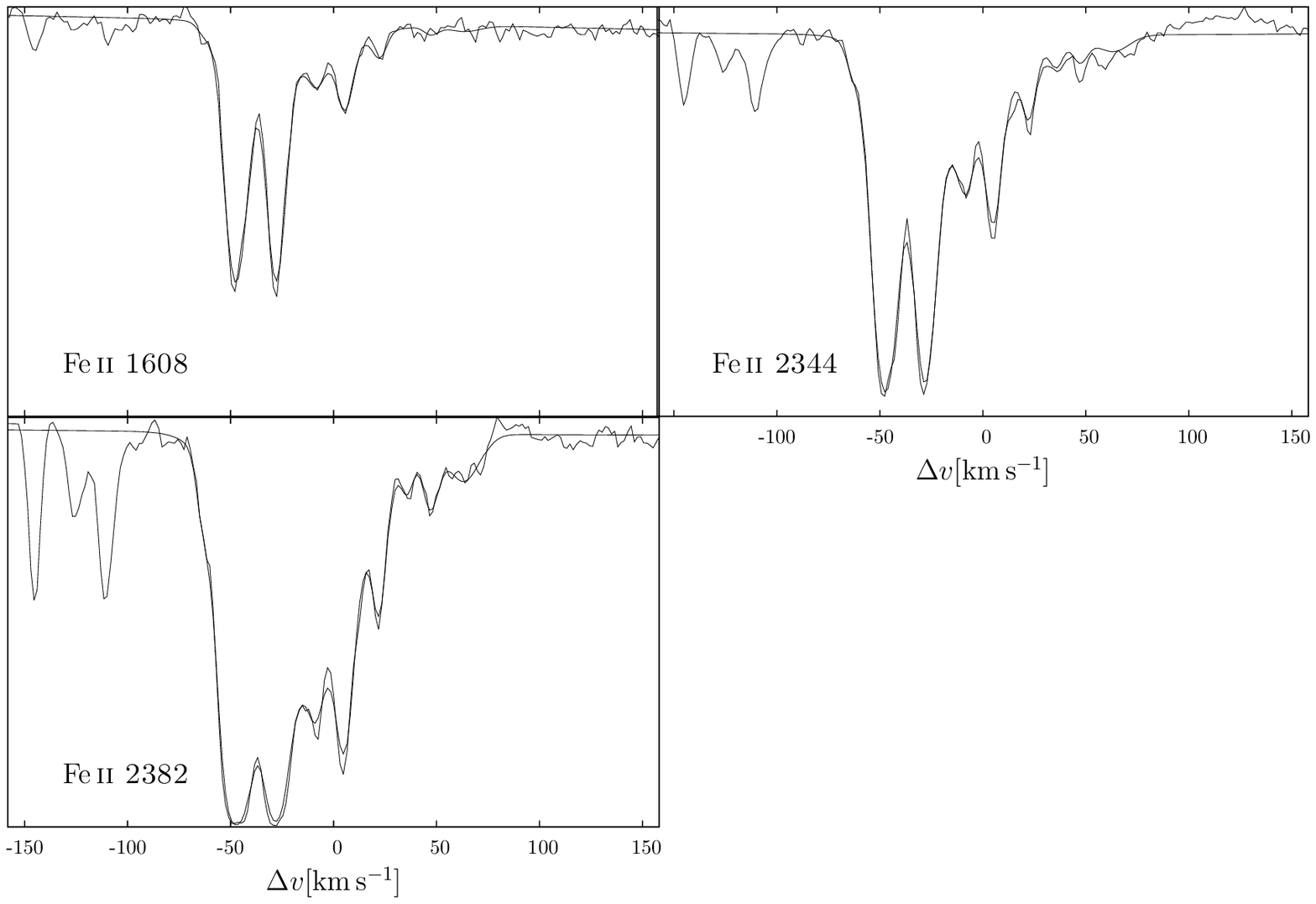}
\caption{Q0002-422, \(z=2.3015\)}
\label{fig:fits_q0002_230_3}
\end{figure}

\end{document}